\pdfoutput=1
\documentclass[cernpreprint,atlasdraft=false,texlive=2016,UKenglish,subfigure=true,texmf]{atlasdoc}
 
\usepackage[biblatex=true]{atlaspackage}
\usepackage{atlasbiblatex}
 
\usepackage{atlascontribute}
 
\usepackage[bsm=true]{atlasphysics}

\usepackage{amssymb}

\graphicspath{{logos/}{figures/}}
 
\usepackage{ANA-SUSY-2018-07-PAPER-defs}
 
\newcommand{\AtlasCoordFootnote}{
ATLAS uses a right-handed coordinate system with its origin at the nominal interaction point (IP)
in the centre of the detector and the $z$-axis along the beam pipe.
The $x$-axis points from the IP to the centre of the LHC ring,
and the $y$-axis points upwards.
Cylindrical coordinates $(r,\phi)$ are used in the transverse plane,
$\phi$ being the azimuthal angle around the $z$-axis.
The pseudorapidity is defined in terms of the polar angle $\theta$ as $\eta = -\ln \tan(\theta/2)$.
}
 
\usepackage{verbatim}
\usepackage{comment}

\AtlasTitle{Search for new phenomena with top quark pairs in final states with one lepton, jets, and missing transverse momentum in \(pp\) collisions at $\sqrt{s}=13~\mathrm{TeV}$ with the ATLAS detector}

\PreprintIdNumber{CERN-EP-2020-177}
 
\AtlasJournal{JHEP 04 (2021) 174}
\AtlasDOI{10.1007/JHEP04(2021)174}
 
\AtlasAbstract{
A search for new phenomena with top quark pairs in final states with one isolated electron or muon,
multiple jets, and large missing transverse momentum is performed. Signal regions are designed to
search for two-, three-, and four-body decays of the directly pair-produced supersymmetric partner of the top
quark (stop). Additional signal regions are designed specifically to search for spin-0 mediators that are
produced in association with a pair of top quarks and decay into a pair of dark-matter particles.
The search is performed using the Large Hadron Collider proton--proton collision dataset at a
centre-of-mass energy of $\sqrt{s}=13$~\TeV recorded by the ATLAS detector from 2015 to 2018, corresponding to
an integrated luminosity of {\SI{139}{\per \fb}}. No significant excess above the Standard Model background is
observed, and limits at 95\% confidence level are set in the stop--neutralino mass plane and as a
function of the mediator mass or the dark-matter particle mass. Stops are excluded up to 1200~GeV
(710~GeV) in the two-body (three-body) decay scenario. In the four-body scenario stops up to 640~GeV
are excluded for a stop--neutralino mass difference of 60~GeV. Scalar and pseudoscalar dark-matter
mediators are excluded up to 200~GeV
when the coupling strengths of the mediator to Standard Model and dark-matter particles are both equal to one
and when the mass of the dark-matter particle is 1~GeV.}

\AtlasRefCode{SUSY-2018-07}

\hypersetup{pdftitle={ATLAS document},pdfauthor={The ATLAS Collaboration}}
 
\begin{document}
 
\maketitle

\clearpage
 
% The next lines are included from the .//texfiles/introduction.tex input file
\section{Introduction}
\label{sec:intro}
This paper presents a search for new phenomena in events with top quark pairs, in a final state with exactly
one isolated charged lepton (electron or muon,\footnote{Electrons and muons from $\tau$-lepton decays are included.}
henceforth referred to as `lepton') from the decay of an on- or off-shell $W$ boson, jets, and a significant
amount of missing transverse momentum (\Ptmiss), the magnitude of which is denoted by \met. This experimental
signature may arise in Supersymmetry (SUSY)~\cite{Golfand:1971iw,Volkov:1973ix,Wess:1974tw,Wess:1974jb,Ferrara:1974pu,Salam:1974ig,Farrar:1978xj}
or in models with a spin-0 mediator produced in association with top quarks~\cite{DAmbrosio:2002vsn,Isidori_2012}
and subsequently decaying into a pair of dark matter (DM) particles.
 
SUSY extends the Standard Model (SM) by introducing a supersymmetric partner for each SM particle, the two having
identical quantum numbers except for a half-unit difference in spin. Searches for a light supersymmetric partner
of the top quark, referred to as the top squark or `stop', are of particular interest after the discovery of the
Higgs boson~\cite{HIGG-2012-27,CMS-HIG-12-028} at the Large Hadron Collider (LHC). Stops may largely cancel out
divergent loop corrections to the Higgs boson
mass~\cite{Dimopoulos:1981zb,Witten:1981nf,Dine:1981za,Dimopoulos:1981au,Sakai:1981gr,Kaul:1981hi,Barbieri:1987fn,deCarlos:1993yy},
and thus, supersymmetry may provide an elegant solution to the hierarchy
problem~\cite{Weinberg:1975gm,Gildener:1976ai,Weinberg:1979bn,Susskind:1978ms}. The superpartners of the left- and
right-handed top quarks, \stopL and \stopR, mix to form two mass eigenstates, \stopone and \stoptwo, where \stopone
is the lighter of the two. Significant mass splitting between the \stopone and \stoptwo particles is possible
due to the large top quark Yukawa coupling. A generic
$R$-parity-conserving\footnote{A multiplicative quantum number, referred to as $R$-parity, is introduced in SUSY models to conserve baryon and lepton number where $R$-parity is $1$ ($-1$) for all SM (SUSY) particles.}
minimal supersymmetric extension of the SM (MSSM)~\cite{Fayet:1976et,Fayet:1977yc,Farrar:1978xj,Fayet:1979sa,Dimopoulos:1981zb}
predicts pair production of SUSY particles and the existence of a stable lightest supersymmetric particle (LSP).
The mass eigenstates from the linear superposition of charged or neutral SUSY partners of the Higgs and
electroweak gauge bosons (higgsinos, winos and binos) are called charginos \chinoOneTwopm\ and neutralinos
\ninoOneTwoThreeFour. The lightest neutralino (\ninoone), assumed to be
the LSP, may provide a potential dark matter (DM) candidate because it is stable and only interacts
weakly with ordinary matter~\cite{Goldberg:1983nd,Ellis:1983ew}. This paper presents a search for direct pair production
of \stopone particles, with significant amount of \met, from the two weakly interacting LSPs that escape detection.
Scenarios with on- and off-shell production of $W$ bosons and top quarks in the stop decays are considered, leading to
two-, three- and four-body decays of the stop.
 
The search for a spin-0 mediator produced in association with top quarks and subsequently decaying into a pair of DM
particles is motivated by SM extensions which respect the principle of minimal flavour violation resulting in the
interaction strength between the spin-0 mediator and the SM quarks being proportional to the fermion masses via
Yukawa-type couplings.
 
Dedicated searches for direct \stopone pair production were recently reported by the
ATLAS~\cite{SUSY-2016-16,SUSY-2016-15,SUSY-2016-17,SUSY-2014-07} and
CMS~\cite{CMS-PAS-SUS-19-005,Sirunyan:2017wif,Sirunyan:2017xse,Sirunyan:2017leh,CMS-SUS-16-008,CMS-SUS-15-005,CMS-SUS-15-004,CERN-EP-2019-233}
Collaborations. Previous ATLAS and CMS searches extend the lower limit on \stopone masses at 95\% confidence level to
1.2~\TeV\ in the two-body decay scenario and up to $\sim$450~\GeV\ in the three-body decay scenario. Searches for
spin-0 mediators produced in association with heavy-flavour quarks and decaying into a pair of DM particles
have also been reported by the ATLAS~\cite{DMhfRun2,SUSY-2016-16} and CMS~\cite{CMS-EXO-16-005} Collaborations.

% End of text imported from the .//texfiles/introduction.tex input file
 
% The next lines are included from the .//texfiles/strategy.tex input file
\section{Signal models and search strategy}
\label{sec:strategy}
 
Two classes of physics models are targeted by this search, the production of \stopone pairs in simplified SUSY
models~\cite{Alwall:2008ve,Alwall:2008ag,Alves:2011wf} where the only light sparticles are \stopone and \ninoone,
and simplified benchmark models for DM production that assume the existence of a spin-0 mediator particle
that can be produced in association with two top quarks~\cite{DMhfRun2,Abercrombie:150700966} and decays
into a pair of DM particles $\chi\bar{\chi}$.
 
\begin{figure}[htbp]
\begin{center}
\subfigure[]{\includegraphics[width=0.30\textwidth]{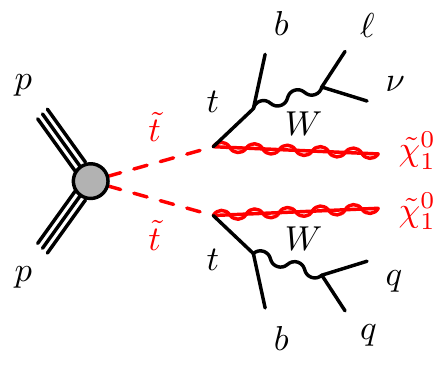}}
\subfigure[]{\includegraphics[width=0.30\textwidth]{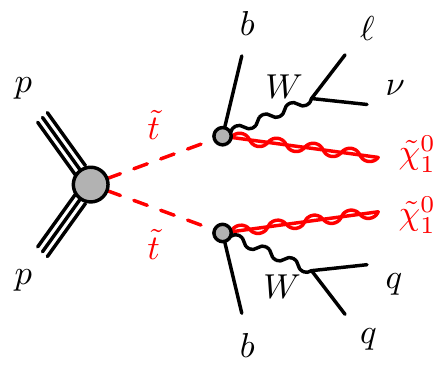}}
\subfigure[]{\includegraphics[width=0.30\textwidth]{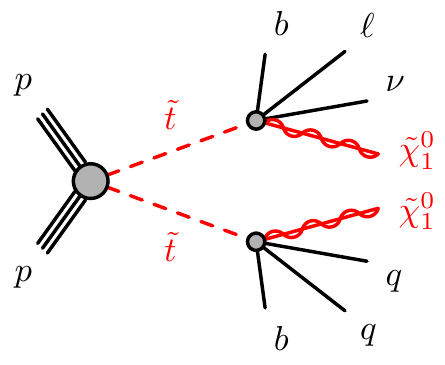}}
\caption{
Diagrams illustrating the stop decay modes, which are referred to as (a) \topLSP\ , (b) \threeBody\ and (c) \fourBody.
In these diagrams, the charge-conjugate symbols are omitted for simplicity. All the processes considered involve the production of a squark--antisquark pair.}
\label{fig:stopdiagrams}
\end{center}
\end{figure}

The experimental signatures of stop pair production can vary dramatically, depending on the mass-splitting between \stopone and \ninoone.
Figure~\ref{fig:stopdiagrams} illustrates the two-, three-
and four-body stop decays considered in this paper. As flavour-changing neutral
current processes are not considered, the dominant among the two-, three- or four-body stop decays is
assumed to have a 100\% branching ratio in a given $\Delta m_{\stopone,\ninoone}$ regime.
In the regime where $\Delta m_{\stopone,\ninoone} = m(\stopone)-m(\ninoone)$ is larger
than the top quark mass \mtop, the two-body decay \topLSP\ dominates. At smaller $\Delta m_{\stopone,\ninoone}$,
the three-body decay  \threeBody\ dominates as long as $\Delta m_{\stopone,\ninoone}$ is larger than the sum of
the $b$-quark and $W$ boson masses. At the smallest values of $\Delta m_{\stopone,\ninoone}$ the dominant decay
channel is the four-body decay \fourBody. The stop is always assumed to decay promptly.
 
The searches for stops presented in this paper use several signal regions dedicated to each of the decay channels
\topLSP, \threeBody\ and \fourBody. For instance, specific signal regions target the so-called compressed region where the stop undergoes a \topLSP\ decay but where $\Delta m_{\stopone,\ninoone} \approx \mtop$. The selections are optimised for given benchmark model points, and are
binned in key variables to retain sensitivity to the widest possible range of \stopone and \ninoone\ masses.
 
\begin{figure}[htbp]
\begin{center}
\includegraphics[width=0.30\textwidth]{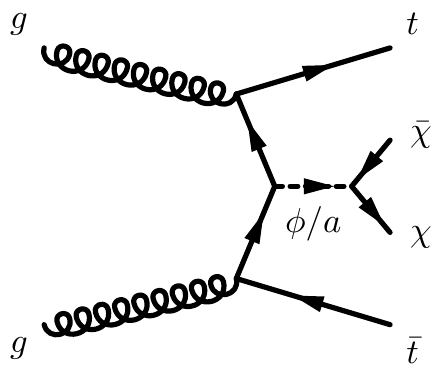}
\caption{A representative Feynman diagram for spin-0 mediator production. The $\phi$/$a$ is the scalar/pseudoscalar mediator, which decays into a pair of dark matter ($\chi$) particles.}
\label{fig:diagram_DM}
\end{center}
\end{figure}
 
The mediator-based DM scenarios consist of simplified models with a DM particle $\chi$ that is a SM singlet
and a single spin-0 mediator that couples $\chi$ to SM fermions. Both the scenarios
where the mediator is a scalar, $\phi$, or a pseudoscalar, $a$, are considered, as illustrated in Figure~\ref{fig:diagram_DM}.
These models have four parameters: the mass of the mediator, $m_{\mathrm{med}}$,
the DM mass, $m_{\mathrm{DM}}$, the DM--mediator coupling, $g_{\chi}$, and the coupling of the mediator to the SM
fermions, $g_{\mathrm{q}}$. In the models considered, the interaction strength between the mediator and SM particles is
proportional to the fermion masses via Yukawa-type couplings, and therefore final states involving top quarks dominate over those involving other fermions. Due to the associated production of top quarks with undetected DM particles in the same event, the mediator-based DM model predicts an excess of
\ttbar+\met final-state events above the SM expectation. A dedicated signal region common to both the scalar and
pseudoscalar models is developed. The signal region is binned in the azimuthal angle \dPMETlep\ between the
missing transverse momentum and the leading lepton, to retain maximum sensitivity to both the scalar and pseudoscalar
models and to a large range of mediator and DM particle masses.

The searches presented are based on eight dedicated analyses that target the various scenarios mentioned above.
Each of these analyses corresponds to a set of event selection criteria, referred to as a signal region (SR),
and is optimised to achieve three standard deviation expected sensitivity to the targeted benchmark model. Two techniques are employed
to define the SRs: `cut-and-count' and `shape-fit' methods. The former is based on counting events in a
single region of phase space, and is employed in the eight analyses. The latter is used in several SRs to improve the exclusion reach if no excess is observed in the cut-and-count signal regions, and employs SRs split into multiple bins in one or two key discriminating kinematic
variables. The shape-fit method exploits the varying signal-to-background ratios in different bins to
provide sensitivity to a wider range of new-particle masses than can be achieved by a single cut-and-count SR.
Including these background-rich regions in the single-bin discovery SRs would significantly reduce the sensitivity
to the targeted signatures.
 
The main background processes after the signal selections include \ttbar, $\ttbar+Z(\rightarrow \nu\bar{\nu})$,
$W$+jets and the associated production of a single top quark and a $W$ boson ($Wt$). Backgrounds from these SM processes are estimated by exploiting dedicated control regions (CRs) enriched
in these processes.
The backgrounds are normalised to data by applying a likelihood
fit simultaneously to the SR and associated CRs,
making the analysis more robust against potential mis-modelling in simulated events and reducing the uncertainties in the background normalisation.
Before looking at the data in 
the signal regions, the background modelling
and the normalisation procedure are tested in a series of validation regions (VRs) by applying the normalisation factors determined by a background-only fit in the CRs.
A background-only fit to the CRs and SRs then provides a statistical test that quantifies the existence and extent of a
potential excess of events in data in the SRs. In the absence of an excess, exclusion limits are set on the associated model
parameters by using the theoretical cross-sections. An overview of the signal regions and the benchmark models for
optimisation is presented in Table~\ref{tab:signalRegionOverview}.

\begin{table}[h]
\begin{center}
\caption{Signal scenarios, benchmark models and signal regions. For each SR, the table lists
the analysis technique used for exclusion limits. The last column points to the section where the signal region is defined. For \tNhigh\ no exclusion technique is defined. The \tNmed\ shape-fit also covers the \tNhigh-like phase space.}
\vspace{3mm}
\renewcommand{\arraystretch}{1.5}
{\scriptsize
\begin{tabular}{ l  c  l  c  c }
\toprule
Signal scenario &    Benchmark              & Signal Region & Exclusion technique &      Section \\
\midrule
\topLSP         & $m(\stopone,~\ninoone) = (800, 400)~\GeV$   & \tNmed      & shape-fit of \met and \mt & \ref{subsec:SR:tNbulk} \\
\topLSP         & $m(\stopone,~\ninoone) = (950, 1)~\GeV$~~~~     & \tNhigh     & --                         & \ref{subsec:SR:tNbulk} \\
\topLSP         & $m(\stopone,~\ninoone) = (225, 52)~\GeV$~~    & \tNdiaglow  & cut-and-count             & \ref{subsec:SR:tNcompressed} \\
\topLSP         & $m(\stopone,~\ninoone) = (500, 327)~\GeV$   & \tNdiaghigh & cut-and-count             & \ref{subsec:SR:tNcompressed} \\
\threeBody     & $m(\stopone,~\ninoone) = (500, 380)~\GeV$   & \bWN        & shape-fit in RNN score    & \ref{subsec:SR:3B} \\
\fourBody       & $m(\stopone,~\ninoone) = (450, 400)~\GeV$   & \bffNbtag   & shape-fit in \lepPtoverMET\ and $\Delta \phi (\vec{p}_{\mathrm{T}}^{b\textrm{-jet}},\Ptmiss)$ & \ref{subsec:SR:4B} \\
\fourBody       & $m(\stopone,~\ninoone) = (450, 430)~\GeV$   & \bffNsoftb  & shape-fit in \lepPtoverMET  & \ref{subsec:SR:4B} \\
\midrule
Spin-0 mediator & $m(\phi/a,~\chi) = (20, 1)~\GeV$~~~~~~~~      & \DM         & shape-fit in \dPMETlep      & \ref{subsec:SR:DM} \\
\bottomrule
\end{tabular}
}
\label{tab:signalRegionOverview}
\end{center}
\end{table}
% End of text imported from the .//texfiles/strategy.tex input file
 
% The next lines are included from the .//texfiles/detector.tex input file
\section{ATLAS detector and data collection}
\label{sec:atlasdetector}
 
The ATLAS detector~\cite{PERF-2007-01} at the LHC is a multipurpose particle detector with almost \(4\pi\)
coverage in solid angle around the interaction point.\footnote{\AtlasCoordFootnote The transverse momentum, \pt,
is defined in the $x$--$y$ plane.}
It consists of an inner tracking detector (ID) surrounded by a superconducting solenoid providing a \SI{2}{\tesla}
axial magnetic field, electromagnetic and hadronic calorimeters, and a muon spectrometer (MS), which is based on three
large air-core toroidal superconducting magnets consisting of eight coils each. The ID provides charged-particle tracking in the range $|\eta| < 2.5$.
During the LHC shutdown between Run~1 (2010--2012) and Run~2 (2015--2018), a new innermost layer of silicon pixels was
added~\cite{ATLAS-TDR-19,PIX-2018-001}, which improves the track impact parameter resolution, vertex position resolution and
$b$-tagging performance~\cite{ATL-PHYS-PUB-2016-012}.
High-granularity electromagnetic and hadronic calorimeters provide energy measurements up to \(|\eta| = 4.9\). The
electromagnetic calorimeters, as well as the hadronic calorimeters in the endcap and forward regions, are sampling
calorimeters with liquid argon as the active medium and lead, copper, or tungsten absorbers. The hadronic calorimeter
in the central region of the detector is a sampling calorimeter with scintillator tiles as the active medium and steel absorbers.
The MS surrounds the calorimeters and has three layers of precision tracking chambers with coverage up
to $|\eta|=2.7$ and fast detectors for triggering in the region $|\eta|<2.4$.
A two-level trigger system~\cite{Aaboud:2016leb} is used to select events. The first-level trigger is hardware-based, followed by a software-based trigger system.
 
The results in this paper utilise the full Run 2 data sample collected from 2015 to 2018 at a centre-of-mass energy of $\sqrt{s} = 13$~\TeV. The average number of simultaneous $pp$ interactions per bunch crossing, referred to as `pile-up', in the recorded data is approximately 34.
After the application of beam, detector and data-quality requirements, the total integrated luminosity is \ourLumi.
The uncertainty in the combined 2015--2018 integrated luminosity is 1.7\%. It is derived from the calibration of the luminosity scale using $x$--$y$ beam-separation scans, following a methodology similar to that detailed in Ref.~\cite{DAPR-2013-01}, and using the LUCID-2 detector for the baseline luminosity measurements~\cite{LUCID2}.
 
All events were recorded with triggers that accepted events with \met\ above a given threshold.
The \met\ triggers relied on energy measurements in the calorimeter and on several algorithms based on
cells, jets or topological clusters in addition to two methods for correcting for the effects of pile-up.
The triggers were fully efficient for events passing an offline-reconstruction requirement of $\met > 230$~\GeV.
% End of text imported from the .//texfiles/detector.tex input file
 
% The next lines are included from the .//texfiles/datasets.tex input file
\section{Simulated event samples}
\label{sec:simulation}
 
Samples of Monte Carlo (MC) simulated events are used for the description of the SM background
processes and to model the signals.
Details of the simulation samples used, including the matrix
element (ME) event generator and parton distribution function (PDF) set, the parton shower (PS) and
hadronisation model, the set of tuned parameters (tune) for the underlying event (UE) and the
order of the cross-section calculation, are summarised in Table~\ref{tab:mc_samples}.
 
\begin{table}[htbp]
\footnotesize
\centering
\caption{Overview of the nominal simulated samples. The cross-sections of top, single-top and SUSY
samples were calculated at next-to-next-to-leading order (NNLO) with the resummation of soft gluon
emission at next-to-next-to-leading-logarithm (NNLL) accuracy. The $V$+jets background samples were
calculated at NNLO. The cross-sections of other background and DM samples were calculated at
next-to-leading order (NLO).
}
\vspace{3mm}
\footnotesize
\makebox[\textwidth][c]{
\footnotesize
\begin{tabular}{  l  lllll }
\toprule
Process & ME event generator & ME PDF  & PS and & UE tune   & Cross-section\\
&           & & hadronisation &  & calculation \\
\midrule
$\ttbar$                    & \POWHEGBOXV{v2}~\cite{Alioli:2010xd}         & NNPDF3.0~\cite{Ball:2014uwa} & \PYTHIA 8~\cite{Sjostrand:2007gs} & A14~\cite{ATL-PHYS-PUB-2014-021} & NNLO+NNLL~\cite{Czakon:2013goa,Czakon:2012pz,Czakon:2012zr,Baernreuther:2012ws,Cacciari:2011hy,Czakon:2011xx} \\
Single-top                  & & & & & \\
\quad $t$-channel           & \POWHEGBOXV{v1}~                             & NNPDF3.0                     & \PYTHIA 8                         & A14                              & NNLO+NNLL~\cite{Kidonakis:2011wy} \\
\quad $s$- and $Wt$-channel & \POWHEGBOXV{v2}~                             & NNPDF3.0                     & \PYTHIA 8                         & A14                              & NNLO+NNLL~\cite{Kidonakis:2010ux,Kidonakis:2010tc} \\
$V$+jets $(V=W/Z)$          & \SHERPA~2.2.1~\cite{Gleisberg:2008ta}   & NNPDF3.0                     & \SHERPA                           & Default                          & NNLO~\cite{Catani:2009sm}\\
Diboson                     & \SHERPA~2.2.1--2.2.2                  & NNPDF3.0                     & \SHERPA                           & Default                          & NLO\\
Multiboson                  & \SHERPA~2.2.1--2.2.2                  & NNPDF3.0                     & \SHERPA                           & Default                          & NLO\\
$\ttbar+V $                 & \MGaMC~2.3.3~\cite{Alwall:2014hca}      & NNPDF3.0                     & \PYTHIA 8                         & A14                              & NLO~\cite{Alwall:2014hca}\\
SUSY signal                 & \madgraph~2.6.2~\cite{Alwall:2014hca} & NNPDF2.3~\cite{Ball:2012cx}  & \PYTHIA 8                         & A14                              & NNLO+NNLL~\cite{Beneke:2009ye,,Beenakker:2016lwe}\\
DM signal                   & \madgraph~2.6.2                       & NNPDF3.0                     & \PYTHIA 8                         & A14                              & NLO~\cite{dMtt_xsec1,dMtt_xsec2} \\
\bottomrule
\end{tabular}
}
\label{tab:mc_samples}
\end{table}
 
The samples produced with \MGMCatNLOV~\cite{Alwall:2014hca} and
\POWHEGBOX~\cite{Alioli:2010xd,Re:2010bp,Frixione:2007nw,Frederix:2012dh,Alioli:2009je}
used {\textsc{EvtGen}} v1.6.0~\cite{EvtGen} for the modelling of $b$-hadron decays. The signal samples
were all processed with a fast simulation~\cite{SOFT-2010-01}, whereas all background samples were
processed with the full simulation of the ATLAS detector~\cite{SOFT-2010-01} based on
$\GEANT4$~\cite{Agostinelli:2002hh}. All samples were produced with varying numbers of minimum-bias
interactions generated by \PYTHIAV{8} with the A3 tune~\cite{ATL-PHYS-PUB-2016-017} and overlaid on the
hard-scattering event to simulate the effect of multiple $pp$ interactions in the same or nearby bunch
crossings. The number of interactions per bunch crossing was reweighted to match the distribution in data.

The nominal \ttbar~sample and single-top sample cross-sections were calculated at NNLO with the resummation
of soft gluon emission at NNLL accuracy and were generated with \POWHEGBOX (at NLO accuracy)
interfaced to \PYTHIAV{8} for parton showering and hadronisation. Additional $t\bar{t}$ samples were
generated with \MGMCatNLOV\unskip (at NLO accuracy)+\PYTHIAV{8} and
\POWHEGBOX\unskip+\HERWIGV{7}~\cite{Bahr:2008pv,Bellm:2015jjp} for modelling comparisons and the
evaluation of systematic uncertainties~\cite{ATL-PHYS-PUB-2017-007}. The \ttbar and $Wt$ processes have
identical $WWbb$ final states and can interfere.
Additional \ttbar, $Wt$ and $WWbb$ samples were generated as multi-leg processes at LO with
\madgraph\ and used to estimate the systematic uncertainty from the interference modelling.
The \tNmed\ and \tNhigh\ regions
receive significant contributions from both \ttbar and $Wt$ in a phase space where the interference is significant.
Techniques used to model the interference such as diagram subtraction (DS) and diagram removal
(DR)~\cite{Frixione:2008yi} were shown to provide predictions bracketing the
data~\cite{TOPQ-2017-05}, but can lead to large uncertainties.
Both schemes are investigated in this paper, but the DR scheme is ultimately
used for the nominal $Wt$ sample.
 
The \Wjets\ and \Zjets\ samples were generated with \SHERPAV{2.2.1}~\cite{Gleisberg:2008ta,ATL-PHYS-PUB-2017-006}
with up to two partons at NLO and up to four partons at leading order (LO). Diboson and multiboson~\cite{ATL-PHYS-PUB-2017-005}
events were generated with \SHERPAV{2.2.1} and 2.2.2. For dibosons, the events include up to one parton at NLO and up
to three partons at LO. For triboson processes, up to two extra partons were considered at LO.
The \SHERPA samples used matrix elements from \textsc{Comix}~\cite{Gleisberg:2008fv} and \textsc{OpenLoops}~\cite{Cascioli:2011va},
which were merged with the \SHERPA parton
shower~\cite{Schumann:2007mg} using the {\textsc{ME+PS@NLO}} prescription~\cite{Hoeche:2012yf}.
The \Wjets\ and \Zjets\ events were further normalised to the NNLO cross-sections~\cite{Catani:2009sm}.
 
The $\ttbar+V $ samples were generated with \MGMCatNLOV\unskip (at NLO accuracy) interfaced to \PYTHIAV{8} for
parton showering and hadronisation. The corresponding MC tune and generator comparisons can be found
in Ref.~\cite{ATL-PHYS-PUB-2016-005}.

The SUSY samples were generated at LO with \MADGRAPHV{2.6.2} including up to two extra partons,
and interfaced to \PYTHIAV{8} for parton showering and hadronisation. For the \topLSP\ samples,
the stop was decayed in \PYTHIAV{8} using only phase-space considerations and not the full ME.
Since the decay products in the generated event samples did not preserve
spin information, a polarisation reweighting was applied following Refs.~\cite{stopPol1,stopPol2}.
A value of $\cos\theta_t = 0.553$ was assumed, corresponding to a \stopone composed mainly of \stopR ($\sim$70\%).
For the \threeBody\ and \fourBody\ samples the stops were decayed with \MADSPIN~\cite{Artoisenet:2012st},
interfaced to \PYTHIAV{8} for the parton showering. \MADSPIN emulates kinematic distributions
such as the mass of the $bW^{(\ast)}$ system to a good approximation without calculating the full ME.
 
The signal cross-sections for stop pair production were calculated to approximate next-to-next-to-leading
order in the strong coupling constant, adding the resummation of soft gluon emission at
next-to-next-to-leading-logarithm accuracy (approximate
NNLO+NNLL)~\cite{Beenakker:2016lwe,Beenakker:1997ut,Beenakker:2010nq,Beenakker:2016gmf}. The nominal
cross-section and its uncertainty were derived using the PDF4LHC15\_mc PDF set, following the
recommendations of Ref.~\cite{Butterworth:2015oua}. The stop pair production cross-section varies from approximately 200~fb
at $m_{\stopone}=$~600~\GeV to about 2~fb at $m_{\stopone}=$~1150~\GeV.
 
Signal events for the spin-0 scalar and pseudoscalar mediator models were generated at LO with up to one
additional parton with \MADGRAPHV{2.6.2} interfaced to \PYTHIAV{8} for parton showering and hadronisation.
In the DM sample generation the couplings of the mediator to the DM and SM particles ($g_{\chi}$ and $g_{\mathrm{q}}$)
were set to one.
When interpreting the experimental results, a single common coupling $g=g_{\chi}=g_{\mathrm{q}}$ is always assumed. Coupling values of $g=1$ as well as $g<1$ are considered. The kinematics of the mediator decay were found to not depend strongly on the values of the couplings;
however, the particle kinematic distributions are sensitive to the scalar or pseudoscalar nature of the
mediator and to the mediator and DM particle masses. The cross-sections were computed at
NLO~\cite{dMtt_xsec1,dMtt_xsec2} and decrease significantly when the mediator is produced off-shell.
The production cross-section varies from approximately 26~pb to 130~fb over a scalar mediator mass range of 10 to 200~\GeV
and from approximately 600~fb to 120~fb over a pseudoscalar mediator mass range of 10 to 200~\GeV.
 
% End of text imported from the .//texfiles/datasets.tex input file
 
% The next lines are included from the .//texfiles/eventreconstruction.tex input file
\section{Event reconstruction}
\label{sec:event_reconstruction}
 
Events selected in the analysis must satisfy a series of beam, detector and data-quality criteria.
The primary vertex, defined as the reconstructed vertex with the highest $\sum_\text{tracks} \pt^2$,
must have at least two associated tracks with $\pt > 500$~\MeV.
 
Depending on the quality and kinematic requirements imposed, reconstructed physics objects are
labelled as either {\textit{baseline}} or {\textit{signal}}, where the latter is a subset of the
former, with tighter selection criteria applied. Baseline objects are used when classifying overlapping
selected objects and to compute the
missing transverse momentum. Background contributions from \ttbar\ and $Wt$ production where both
$W$ bosons decay leptonically, referred to as dileptonic \ttbar\ or $Wt$ events, are suppressed by
vetoing events with more than one baseline lepton. Signal objects are used to construct kinematic
and discriminating variables necessary for the event selection.
 
Electrons are identified as energy clusters formed in the electromagnetic calorimeter matched to tracks in the ID.
Baseline electrons are required to have $\pT > 4.5$~\GeV\ and $|\eta| < 2.47$, and to
satisfy `LooseAndBLayer' likelihood identification criteria that follow the methodology
described in
Ref.~\cite{PERF-2017-01}.
Furthermore, their longitudinal
impact parameter ($z_0$), defined as the distance along the beam direction between the
primary vertex and the track's point of closest approach to the beam axis, must satisfy
$|z_0 \sin \theta| < 0.5$~mm where $\theta$ is the polar angle of the track.
Signal electrons must satisfy all the baseline requirements and have a
transverse impact parameter ($d_0$) that satisfies $|d_0|/\sigma_{d_0} < 5$, where $\sigma_{d_0}$
is the uncertainty in $d_0$. Furthermore, signal electrons are required to be isolated.
The isolation is defined as the sum of the transverse energy or momentum reconstructed
in a cone of size $\Delta R = \sqrt{(\Delta\eta)^2+(\Delta\phi)^2}$ around the electron, excluding the energy of
the electron itself. The isolation criteria rely on both track- and calorimeter-based information
with a fixed requirement on the isolation energy divided by the electron's $\pt$. Electrons that satisfy
the signal identification criteria, including the loose isolation, are called loose electrons.
In addition, tight electrons must satisfy both a tight electron likelihood identification criterion and a tight isolation criterion.
 
Muon candidates are reconstructed from combined tracks that are formed from ID and MS tracks, or
stand-alone MS tracks. Baseline muons up to $|\eta|=2.7$ are used, and are required to have
$\pt>4$~\GeV, a longitudinal impact parameter $| z_0 \sin \theta| < 0.5$~mm, and to satisfy the
`Medium' identification criterion~\cite{PERF-2015-10}. Signal muons must satisfy all baseline requirements
and in addition have a transverse impact parameter satisfying $|d_0|/\sigma_{d_0} < 3$. Tight signal muons must
satisfy tight isolation criteria, similar to those used for tight signal electrons, but with a
fixed requirement on track-based isolation energy divided by the muon's $\pt$. A category of loose signal
muons is also defined, which requires the `Loose' identification criterion~\cite{PERF-2015-10} and satisfies a looser isolation criterion.
 
Dedicated efficiency scale factors are derived from $Z\rightarrow \ell\bar{\ell}$ and
$J/\psi\rightarrow \ell\bar{\ell}$ data samples to correct the simulations for minor mis-modelling of electron
and muon identification, impact parameter and isolation selections.
The \pt\ threshold of signal leptons is 25~\GeV\ for electrons and muons in all signal regions
except for signal regions dedicated to \fourBody, where electrons with $\pt > 4.5$~\GeV\ and muons
with $\pt > 4$~\GeV\ are used.
 
Jet candidates are built from topological clusters~\cite{PERF-2014-07,PERF-2011-03} in the calorimeters
using the anti-$k_t$ algorithm~\cite{Cacciari:2008gp} with a jet radius parameter $R=0.4$ implemented
in the FastJet package~\cite{Cacciari:2011ma}. Jets are corrected for contamination from pile-up using
the jet area method~\cite{Cacciari:2007fd,Cacciari:2008gn,PERF-2014-03} and are then calibrated to
account for the detector response~\cite{PERF-2012-01,PERF-2016-04}. Jets in data are further calibrated
according to \textit{in situ} measurements of the jet energy scale~\cite{PERF-2016-04}. Baseline jets
are required to have $\pt > 20$~\GeV. Signal jets are required to have $|\eta| < 2.5$ and $\pt > 25$~\GeV\ in
all signal regions, except in the four-body signal regions, where the $\pt$ threshold of signal jets is
20~\GeV. Furthermore, signal jets with $\pT < 120$~\GeV\ and $|\eta| < 2.5$ must satisfy
track-based criteria designed to reject jets originating from pile-up~\cite{PERF-2014-03}. Events containing
a signal jet that does not satisfy specific jet-quality requirements (`jet cleaning') are rejected to suppress detector noise and non-collision backgrounds~\cite{DAPR-2012-01,ATLAS-CONF-2015-029}.
The number of signal jets in an event is denoted \Njet. In addition to these jet candidates, the same anti-$k_t$ algorithm is used to define larger radius (large-$R$) jets to provide discriminating variables for the reconstruction of top quarks, as described in Section~\ref{sec:vars}.
 
Jets identified as containing $b$-hadrons are referred to as $b$-tagged jets. Their identification
is performed using the MV2c10 $b$-tagging algorithm, which examines quantities such as the impact
parameters of associated tracks and characteristics of reconstructed secondary
vertices~\cite{Aad:2015ydr,FTAG-2018-01}. The algorithm is used at a working point that provides a
77\% $b$-tagging efficiency in simulated $\ttbar$ events, and corresponds to a rejection factor of
about 130 for jets originating from gluons and light-flavour quarks (light jets) and about 6 for
jets induced by charm quarks. Corrections derived from data control samples are applied to account
for differences between data and simulation in the efficiency and mis-tag rate of the
$b$-tagging algorithm. The number of $b$-tagged jets in an event is denoted \Nbjet. Since MV2c10
is only applicable to baseline jets with $\pt>20$~\GeV, it is not sensitive to low-\pt $b$-hadrons.
The presence of low-\pt $b$-hadrons, below $20$~\GeV, is instead inferred using
a soft $b$-tagging algorithm, which does not rely on the presence of a jet in the calorimeter, but requires
the presence of secondary vertices~\cite{ATLAS-CONF-2019-027}. This technique is used to gain
sensitivity to the \fourBody\ signal in the regime with $\Delta m_{\stopone,\ninoone}$ lower than
$\sim$40~\GeV. The number of secondary vertices in an event is denoted \NSV. Corrections derived
from dedicated \ttbar and \Wjets\ control regions are applied to the soft $b$-tagging
efficiencies to account for differences between data and simulation.
 
Jets and associated tracks are also used to identify hadronically decaying $\tau$-leptons using
the `Loose' identification criterion described in Refs.~\cite{ATLAS-CONF-2017-029,PERF-2014-06},
which has a 85\% (75\%) efficiency for reconstructing $\tau$-leptons decaying into one (three)
charged pions. The hadronic $\tau$-lepton decay candidates are required to have one or three associated tracks,
with total electric charge opposite to that of the signal electron or muon, $\pt>20$~\GeV, and
$|\eta|<2.5$. The $\tau$-lepton candidate \pT\ requirement is applied after a dedicated energy
calibration~\cite{HIGG-2014-09,PERF-2013-06}.
 
To avoid labelling the same detector signature as more than one object, an overlap removal
procedure is applied. Given a set of baseline objects, the procedure checks for overlap based on
either a shared track, ghost-matching~\cite{Cacciari:2008gn}, or a minimum  distance $\DeltaRy$
between pairs of objects.\footnote{Rapidity $y \equiv 1/2\ln \left[(E+p_{z})/(E-p_{z})\right]$ is used instead of
pseudorapidity ($\eta$) when computing the distance $\DeltaRy$ between objects in the overlap removal procedure.}
First, if a baseline lepton and a baseline jet are separated by
$\DeltaRy<0.2$, then the lepton is retained and the jet is discarded. Second, if a baseline jet and
a baseline lepton are separated by $\DeltaRy<0.4$, then the jet is retained and the lepton is
discarded, to minimise the contamination from jets misidentified as leptons.
For the remainder of the paper, all baseline and signal objects are those that have survived the
overlap removal procedure.
 
The missing transverse momentum \Ptmiss\ is reconstructed as the negative vector sum of the transverse
momenta of baseline electrons, muons, jets, and a soft term built from high-quality tracks that
are associated with the primary vertex but not with the baseline physics
objects~\cite{PERF-2016-07,ATLAS-CONF-2018-023}. Photons and hadronically decaying $\tau$-leptons
are not explicitly included but enter either as jets, electrons, or via the soft term.
% End of text imported from the .//texfiles/eventreconstruction.tex input file
 
% The next lines are included from the .//texfiles/variables.tex input file
\section{Discriminating variables}
\label{sec:vars}
 
The backgrounds contributing to a final state with one isolated lepton, jets and \met are primarily
semileptonic \ttbar events with one of the $W$ bosons decaying leptonically, and \Wjets\ events with a leptonic
decay of the $W$ boson. Both backgrounds can be efficiently reduced by requiring the transverse mass of the
event, \mt, to be significantly larger than the $W$ boson mass. The transverse mass is defined as
$\mt=\sqrt{2p_{\text{T}}^{\ell}\met[1-\cos(\Delta\phi)]}$, where $\Delta\phi$ is the azimuthal angle between
the lepton and missing transverse momentum directions and $\pt^{\ell}$ is the transverse momentum of the charged lepton.
Other discriminating variables used to distinguish signal from several categories of background events are
described below.

\subsection{Dileptonic \ttbar reconstruction}
 
The \mtTwo variable~\cite{Lester:1999tx} is a generalisation of the transverse mass, applied to signatures
where two particles are not directly detected. The variable \mtTwoTau~\cite{SUSY-2013-15} is a variant of
\mtTwo developed to identify and remove \ttbar events where one $W$ boson decays into a hadronically decaying $\tau$-lepton candidate.
In this case the `$\tau$-jet' is used as the visible particle for one top branch and the observed electron or muon
for the other top branch. For \ttbar events where one $W$ boson decays leptonically and the other into a hadronically decaying $\tau$-lepton,
\mtTwoTau\ has an endpoint at the $W$ boson mass.
 
Events with dileptonic decays of \ttbar\ pairs, where one lepton is not identified, constitute a significant background. The
lost lepton can lead to significant missing transverse momentum and
\mtTwo\ above the $W$ boson mass.
The topness variable~\cite{Graesser:2007topness} quantifies how well an event can be reconstructed under a dileptonic top
hypothesis and is defined as the logarithm of the minimum of the following quantity $\topness$:
\begin{eqnarray}
{\cal S}(p_{Wx},p_{Wy},p_{Wz},p_{vz}) & = & \frac{ [m_{W}^{2}-(p_{\ell}+p_{\nu})^{2}]^{2}  } {a_{W}^4 } +
\frac{[m_{t}^{2}-({p_{b}}_{1}+p_{\ell}+p_{\nu})^{2}]^{2}}{ a_{t}^4 } +  \nonumber \\
& &
\frac{[m_{t}^{2}-({p_{b}}_{2}+p_{W})^{2}]^{2}}{a_{t}^4} +
\frac{[4m_{t}^2 - (\Sigma_i p_i)^2]^2}{ a_{\mathrm{CM}}^4},  \nonumber
\label{eq:topness}
\end{eqnarray}
when minimised with respect to $p_{W}$ and $p_{\nu}$ with the constraint $\Ptnu + \Ptw = \Ptmiss$.
The quantity $p_{W}$ represents the four-momentum vector of the $W$ boson for which the lepton was not reconstructed and is thus completely
invisible. The quantities $p_{\ell}$ and $p_{\nu}$ are the lepton and neutrino four-momentum vectors from the $W$ boson whose lepton was
identified. Finally, ${p_{b}}_{i}$ refer to the two $b$-jets. The sum in the last term runs over
the five assumed final-state particles. If the event contains two $b$-tagged jets, the two permutations are
tested in the minimisation. If the event has a single $b$-tagged jet, then permutations where the second $b$-jet can be either of
the two leading momentum untagged jets are tested during the minimisation. The values of resolution parameters $a_W, a_t$ and
$a_{\mathrm{CM}}$ are constants taken from Ref.~\cite{Graesser:2007topness}.
 
\subsection{Reconstruction of hadronic top decays}
Signal events contain one hadronic top decay $t\to q\bar{q}'b$, while such decays are absent from the dileptonic
\ttbar background. Therefore, reconstructing the hadronic top quark decay can provide additional discrimination
against dileptonic \ttbar events. A recursive reclustering jet algorithm searches for large-radius jets with
radius parameter $R$ corresponding to the radius \( R(\pT) = 2 \times \mtop / \pT \)
expected from a hadronic top quark decay $t \to q \bar{q}' b$~\cite{SUSY-2016-16}. The algorithm is based on the
anti-$k_t$ algorithm using signal jets as inputs and
with initial radius parameter $R_{0}=3.0$. If a reclustered large-radius jet is significantly narrower
than the radius expected from a hadronic top quark decay of that \pT, it is discarded. The radius of the
remaining reclustered jets is iteratively reduced until the radius approximately matches the radius expected
from a hadronic top quark decay. Surviving reclustered jets constitute hadronic top candidates. If more
than one hadronic top candidate is found, the candidate whose mass \mtoprecl\ is closest to \mtop\ is retained.
 
A second hadronic top quark candidate algorithm is employed that fully reconstructs the direction of both
the leptonically and the hadronically decaying top quarks, denoted \tlep\ and \thad\ respectively. This
algorithm is applied to events with at least four jets and one $b$-tagged jet. The \mTopChi\ variable is
defined as the invariant mass of the triplet of signal jets (one of which must be $b$-tagged) most compatible with
\mtop, taking into account the jet momentum and energy resolution. The component of the \Ptmiss\ perpendicular
to \tlep\ in the \ttbar\ rest frame, \perpmet, is small in semileptonic top quark decays since \Ptmiss\ tends to align with the leptonically decaying top quark.
 
\subsection{Backgrounds with mismeasured missing momentum}
In some signal regions, additional suppression against backgrounds with mismeasured missing momentum, arising from mismeasured jets, is required.
This additional rejection is provided by  \(
\HTmissSig = (|\HTmissVec|-M)/\sigma_{|\HTmissVec|} \),
where \HTmissVec\ is the negative vectorial sum of the momenta of the signal jets and signal lepton~\cite{SUSY-2013-15}.
The denominator is computed from the per-event jet energy uncertainties, while the lepton resolution is neglected.
The offset parameter $M$ is a characteristic scale of the background processes and is fixed at 100~\GeV.
 
\subsection{Variables for compressed \topLSP}
\label{subsec:compressedvariables}
To discriminate stop pair production from SM \ttbar production, in the phase space dominated by the decay
\topLSP\ in the compressed regime $\Delta m_{\stopone,\ninoone} \approx \mtop$, events are reconstructed according
to both the stop and semileptonic \ttbar hypotheses. These techniques are employed in the \tNdiaglow\ and \tNdiaghigh\ SRs.
 
The reconstruction of the event under the semileptonic \ttbar hypothesis starts by searching for the hadronically decaying
top quark candidate through the minimisation of the loss function
\begin{equation*}
L_{t}=\frac{(\mW^{\mathrm{cand}}-\mW)^2}{m_W} + \frac{(m_{t_{\mathrm{had}}}^{\mathrm{cand}}-\mtop)^2}{\mtop}
\end{equation*}
with \mW\ and \mtop\ being the experimentally known $W$ boson and top quark masses. The $W$ boson candidate
mass $\mW^{\mathrm{cand}}$ is either the mass of a single large anti-$k_{t}$ jet with radius 1.0 or 1.2 or the
invariant mass of two anti-$k_{t}$ jets with radius 0.4. The hadronically decaying top quark candidate
$t_{\mathrm{had}}$ is either one of the large-$R$ jets or the $W$ boson candidate plus a $b$-tagged jet. The jet permutation with the minimum loss
function is considered as the candidate
for the hadronic top. The visible part of the leptonically decaying top quark candidate
($t_{\mathrm{lep}^{\mathrm{vis}}}$) four-momentum vector is determined by adding the four-momentum vectors of the remaining highest-\pT
$b$-tagged jet and the signal lepton.
 
The reconstruction of the event under the stop hypothesis relies on the collinear
approximation~\cite{An:2015uwa,Macaluso:2015wja}, in which the top quark and the neutralino from the stop decay
are collinear. This approximation is valid for compressed \topLSP\ models ($\Delta m_{\stopone,\ninoone} \approx \mtop$), where the requirement of a high-\pt\ initial-state radiation (ISR) jet in the event forces the momentum of the \stopone to be much larger than the sum of the top and \ninoone masses.
 
With this approximation and a given value of the parameter $\alpha=m_{\ninoone}/m_{\stopone}$,
the four-momentum vector \pfvecnu$(\alpha)$ of the neutrino can be calculated from the missing transverse energy and
the measured momenta of the hadronic and visible leptonic top quark candidates under the assumption that the longitudinal
neutrino momentum $p_z$ is zero. The resulting \pfvecnu$(\alpha)$ is then used to
compute the leptonically decaying $W$ boson's transverse mass $\mt^{\alpha}$ and the difference in \mt between
the calculation under the hypothesis of a \ttbar event and under the signal hypothesis,
$\Delta \mt^{\alpha} = \mt - \mt^{\alpha}$.
 
The \tNdiaglow\ SR is optimised to probe the previously unexcluded region around the point with a stop mass of 200~\GeV and neutralino mass of 27~\GeV ~\cite{SUSY-2016-16} which corresponds to $\alpha=0.135$. Therefore this region uses the variable \dmtalpha\ with $\alpha$ fixed to 0.135. For other compressed regions, which are targeted by the \tNdiaghigh\ SR, $\alpha$ can be determined dynamically by minimising the loss function
 
\begin{equation*}
L_{\alpha} = \frac{[m(\ell+\nu)-\mW]^2}{\mW} + \frac{[m(\tlepvis+\nu)-\mtop]^2}{\mtop}
\end{equation*}
 
where $m(\ell+\nu)$ is the invariant mass of the lepton and the neutrino, and $m(\tlepvis+\nu)$ is the invariant
mass of the leptonic top candidate and the neutrino. Using the approximation $\alpha=m_{\ninoone}/(m_{\ninoone}+m_{t_{\mathrm{had}}})$ and the measured value of $m_{t_{\mathrm{had}}}^{\mathrm{cand}}$, the values of \dmtalpha\ and the mass of the \ninoone\ at the minimum of the loss function can be determined. These variables are labelled \dmtdyn\ and \mchidyn\ respectively.

Although the neutrino momentum under the collinear approximation is fully known for a given value of $\alpha$,
there is an ambiguity as to how the remaining missing transverse momentum is split between the two neutralinos.
To resolve this, the following loss function, which compares the reconstructed leptonic and hadronic \stopone masses
with a given \stopone mass hypothesis, $m_{\stopone}$, is defined and used in the \tNdiaglow\ SR:
\begin{equation*}
L_{\stopone} = \frac{\left(m_{\stopone}^{\mathrm{had}}-m_{\stopone}\right)^2}{m_{\stopone}} + \frac{\left(m_{\stopone}^{\mathrm{lep}}-m_{\stopone}\right)^2}{m_{\stopone}}
\end{equation*}
A minimisation of this loss function, again under the assumption that $\alpha = 0.135$, is performed with respect to the angles between each neutralino momentum vector and each of the two top quarks.
The mass \mstoplep, which denotes the leptonic \stopone mass at the minimum of this loss function, takes lower and more peaked values for compressed \topLSP models than for the SM top quark backgrounds. Finally, the ratio \xone\ of the hadronic top quark momentum to the parent stop momentum is also used to discriminate between the stop signal and the background. Since it is computed as a projection, \xone\ can take negative values for background processes, or if the collinear assumption does not hold.

% End of text imported from the .//texfiles/variables.tex input file
 
% The next lines are included from the .//texfiles/signal_regions.tex input file
\section{Signal regions}
\label{sec:srs}
 
A preselection that exploits the basic characteristics of the signals is applied: the presence of a signal lepton,
$b$-tagged jets and missing transverse momentum. The preselection is designed to have very high efficiency for
the signal and to remove the most trivial backgrounds. To cover signals with both high-momentum decay
products such as in \topLSP\ and low-momentum decay products such as in \fourBody\ , `soft-lepton' and
`hard-lepton' preselections are defined and are presented in Table~\ref{tab:preselection}. All regions require
$\met>230$~\GeV to ensure that the trigger was fully efficient. To reject multijet events
with mismeasured jet momenta, a minimum azimuthal angular distance is required between the missing transverse
momentum direction and the two leading jets, $\Delta\phi(j_{1,2},\Ptmiss)>0.4$.
 
The signal regions are then optimised using simulated event samples to maximise the expected $Z$ significance~\cite{2008NIMPA.595..480C,Li1983:ApJ272317L}
for the benchmark signals.\footnote{Significance $Z$ of observing $n$ events for a prediction of
$b\pm\sigma$ is defined as\\ $Z = \sqrt{2\left\{n\ln{\left[\frac{n(b+\sigma^2)}{b^2+n\sigma^2}\right]}-\frac{b^2}{\sigma^2}\ln{\left[1+\frac{\sigma^2(n-b)}{b(b+\sigma^2)}\right]}\right\}}$
when $n\ge b$, or \\ $Z = -\sqrt{2\left\{n\ln{\left[\frac{n(b+\sigma^2)}{b^2+n\sigma^2}\right]}-\frac{b^2}{\sigma^2}\ln{\left[1+\frac{\sigma^2(n-b)}{b(b+\sigma^2)}\right]}\right\}}$ when $n<b$.}
A set of benchmark signal models, selected to cover the various stop and spin-0 mediator models, is used for optimisation.
The optimisation is performed using an iterative algorithm, considering all discriminating variables and accounting for
statistical and systematic errors in the evaluation of the discovery significance. An overview of the signal regions and
the benchmark models for optimisation is presented in Table~\ref{tab:signalRegionOverview}. The SRs are not designed to be
orthogonal. The final exclusion limits are obtained by selecting at each point of the model parameter space the SR with the
best expected sensitivity.
 
\begin{table}[t]
\begin{center}
\caption{Preselection criteria used for the hard-lepton signal regions (left) and the soft-lepton signal regions (right).}
\label{tab:preselection}
\vspace{3mm}
\renewcommand{\arraystretch}{1.1}
\begin{tabular}{ l r  c c }
\toprule
\multicolumn{2}{l}{Selection}                   & \textbf{hard-lepton}                     & \textbf{soft-lepton} \\
\midrule
Trigger &                                         & \multicolumn{2}{c}{\met\ trigger}  \\
Data quality &                                    & \multicolumn{2}{c}{jet cleaning, primary vertex} \\
Second-lepton veto &                              & \multicolumn{2}{c}{no additional baseline leptons} \\
\midrule
\multicolumn{2}{l}{Number of leptons, tightness} & $= 1$ `loose' lepton                     & $= 1$ `tight' lepton \\
Lepton $\pt$ & [\si{\GeV}]                       & $>25$                                    & $>4\ (4.5)$ for $\mu\  (e)$ \\
Number of jets  (jet \pt) &                       & $\ge4~(>25 \ \GeV)$                       & $\ge1$~($>200\ \GeV$) or $\ge2$~($>20\ \GeV$)  \\
\met\ & [\si{\GeV}]                               & \multicolumn{2}{c}{$>230$}\\
$\Delta\phi(j_{1,2},\Ptmiss)$ &  [rad]            & \multicolumn{2}{c}{$> 0.4$} \\
\Nbjet &                                          & $\ge1$                                   & -- \\
\mt\ & [\si{\GeV}]                                & $>30$                                    & -- \\
\mtTwoTau\ & [\si{\GeV}]                          & $>80$                                    & -- \\
\bottomrule
\end{tabular}
\end{center}
\end{table}

\subsection{\topLSP}
\label{subsec:SR:tNbulk}
Two signal regions, \tNmed\ and \tNhigh, are designed for models with $\Delta m_{\stopone,\ninoone}$ significantly
larger than \mtop, and rely on large missing momentum and energetic jets. Selections on \mt, \HTmissSig,
\perpmet ~and topness are dictated by the need to suppress the three main backgrounds, namely \Wjets, \ttbar, and \ttV.
The presence of a hadronic top quark candidate with  $\mtoprecl > 150$~\GeV\ is required primarily to ensure orthogonality
with the control regions. The \tNmed\ and \tNhigh\ definitions are given in
Table~\ref{tab:tnbulk_sr}. A common exclusion region is defined by performing a two-variable shape-fit on the
\tNmed\ signal region, if no excess is observed in the single-bin discovery signal regions.
The binning is designed to maximise the excluded parameter space in the $m_{\stopone}$--$m_{\ninoone}$ plane.
The two variables chosen for the binning are the two discriminating variables
that best distinguish between \tNmed\ and \tNhigh, namely  \met and \mt. The resulting six bins are given in
Table~\ref{tab:tnbulk_sr}.
 
\begin{table}[ht]
\begin{center}
\caption{Event selections defining the signal regions \tNmed\ and \tNhigh.}
\vspace{3mm}
\renewcommand{\arraystretch}{1.1}
\begin{tabular}{ l r   c  c }
\toprule
\multicolumn{2}{l}{Selection}    & \tNmed               & \tNhigh                \\
\midrule
\multicolumn{2}{l}{Preselection} & \multicolumn{2}{c}{hard-lepton preselection} \\
\midrule
\Njet, \Nbjet &                  & $\geq(4,1)$          & $\geq(4,1)$         \\
Jet $\pT$ & [\si{\GeV}]          & $>$ (100, 90, 70, 50)  & $>$ (120, 50, 50, 25) \\
\midrule
\met & [\si{\GeV}]               & $>230$                & $>520$               \\
\perpmet & [\si{\GeV}]           & $>400$                & --                   \\
\HTmissSig &                     & $>16$                 & $>25$                \\
\mt & [\si{\GeV}]                & $>220$                & $>380$               \\
Topness &                        & $>9$                  & $>8$                 \\
\mtoprecl & [\si{\GeV}]           & \multicolumn{2}{c}{$>$ 150}                 \\
\deltaRbl &                       & $<2.8$                & $<2.6$               \\
\midrule
Exclusion technique &             & \multicolumn{2}{c}{Based on shape-fit in \met and \mt in \tNmed}  \\
&             & \multicolumn{2}{l}{ \met $\in$ [230, 400], \mt $>$ 220}  \\
&             & \multicolumn{2}{l}{ \met $\in$ [400, 500], \mt $>$ 220}  \\
Bin boundaries & [\si{\GeV}]      & \multicolumn{2}{l}{ \met $\in$ [500, 600], \mt $\in$ [220, 380]}  \\
&             & \multicolumn{2}{l}{ \met $\in$ [500, 600], \mt $>$ 380}  \\
&             & \multicolumn{2}{l}{ \met $>$ 600, \mt $\in$ [220, 380]}  \\
&             & \multicolumn{2}{l}{ \met $>$ 600, \mt $>$ 380}  \\
\bottomrule
\end{tabular}
\label{tab:tnbulk_sr}
\end{center}
\end{table}

\subsection{Compressed \topLSP}
\label{subsec:SR:tNcompressed}
 
The kinematics of the decay \topLSP\ in the region where $\Delta m_{\stopone,\ninoone} \approx \mtop$ differ significantly from
the two signal regions defined above, and the stop signal is kinematically very similar to the dominant \ttbar background. This region of
parameter space is referred to as the diagonal region. Two dedicated signal regions, \tNdiaglow\ and \tNdiaghigh, are designed
to target scenarios on the diagonal for low-mass and high-mass stops respectively. The sensitivity of the \tNdiaglow\ SR is
such that it is expected to be able to exclude scenarios with $\Delta m_{\stopone,\ninoone} = \mtop$ and $m(\stopone)$ between
200 and 250~\GeV. Both the \tNdiaglow\ and \tNdiaghigh\ signal regions rely on the presence of a high-\pT
ISR jet, which serves to boost the di-stop system. The signal region definitions are shown in Table~\ref{tab:tn_diag_selection}
and are used both for exclusion and for discovery.
 
\begin{table}
\centering
\caption{Event selections defining the signal regions \tNdiaglow\ and \tNdiaghigh.}
\label{tab:tn_diag_selection}
\begin{tabular}{ l l  c c }
\toprule
\multicolumn{2}{l}{Selection}			& \tNdiaglow	& \tNdiaghigh \\
\midrule
\multicolumn{2}{l}{Preselection}    		& \multicolumn{2}{c}{hard-lepton preselection without $\tau$-lepton veto} \\
\midrule
\multicolumn{2}{l}{\Njet, \Nbjet}                 & \multicolumn{2}{c}{$>$ (4, 1)} \\
Jet $\pT$ & [\si{\GeV}]                              & \multicolumn{2}{c}{$>$ (400, 40, 40, 40)} \\
\midrule
\mt & [\si{\GeV}]                                    & $>150$           & $>110$ \\
\met & [\si{\GeV}]                                   & --            & $>400$ \\
$m_\mathrm{T2}$ & [\si{\GeV}]                        & --            & $<360$ \\
\midrule
\dmtalpha & [\si{\GeV}]                              & $>40$        & -- \\
\dmtdyn & [\si{\GeV}]	                        & --            & $>60$ \\
\mstoplep & [\si{\GeV}]                            & $<600$       & -- \\
\mchidyn & [\si{\GeV}]                             & $>5$         & [220, 595] \\
\multicolumn{2}{l}{\xone}                       & --            & $>-0.2$ \\
\midrule
Exclusion technique &                               & \multicolumn{2}{c}{cut-and-count} \\
\bottomrule
\end{tabular}
\end{table}

\subsection{\threeBody\ }
\label{subsec:SR:3B}
 
The signal region for the decay \threeBody\ is labelled \bWN\ and defined using an optimised two-step machine
learning (ML) approach, applied to events preselected according to the hard-lepton preselection criteria and
additionally satisfying $\mt > 110$~\GeV.
The background mostly consists of \ttbar, which has strong similarities to the signal in this region of phase space.
For this reason the ML technique is selected.
The jet multiplicity in signal events varies significantly due to the potential presence of
ISR jets and fluctuations in the number of low-energy jets reconstructed from the hadronically decaying
$W$ boson. To deal with the variable number of signal jets, the first step of the ML procedure is
to use a recurrent neural network (RNN) that has the ability to extract information from sequences of variable
length~\cite{Goodfellow:2016}. The RNN uses a long short-term memory (LSTM) algorithm~\cite{Hochreiter:1997}
and takes the four-momentum vectors of the jets as inputs. The LSTM output becomes the input of the second step,
made up of a shallow neutral network (NN) with a single hidden layer and an output corresponding to the signal
probability. The RNN and NN are trained simultaneously in one step. The NN uses the following discriminating
variables as input: output of the RNN, \met, \mt, the
azimuthal $\phi$ angle of \Ptmiss, the azimuthal angle \dPMETlep\ between the lepton and \Ptmiss, the invariant
mass $m_{\ell\mathrm{b}}$ of the lepton and the $b$-tagged jet, the transverse momentum of the $b$-tagged jet,
the lepton four-momentum vector, \Njet\ and \Nbjet.
 
Before training, the hard-lepton preselection and the additional selection $\mt > 110$~\GeV are applied. The
size of the training sample is a crucial aspect for the performance of any ML method. Generating fully simulated
signal samples with adequate sample sizes after the hard-lepton preselection and $\mt>110$~\GeV is
computationally expensive. To overcome this difficulty, signal events without detector simulation were used for
the training to enhance the number of signal events by two orders of magnitude. Fully simulated SM background events
were available in sufficiently large numbers to be used directly for the training. For the signal, the generated
events are `smeared' using a dedicated procedure to emulate the effects of detector simulation and reconstruction.
Parameterisations for reconstruction and identification efficiencies are obtained from dedicated ATLAS
measurements and applied to jets, leptons and $b$-tagged jet identification. Particle-level electron, muon and
jet four-momentum vectors are smeared according to their respective $\pt$, $\eta$ and identification working point.
The \met is recomputed from all smeared objects. The kinematic distributions of all input variables after smearing
are found to have fair agreement with distributions after full event reconstruction. The output score of the ML
classifier, denoted \NNscore, shows good agreement between smeared samples and fully simulated samples after
full event reconstruction. The classifier output also shows a good agreement between simulation and data.
The smeared samples are used only for the training, while signal and background predictions are
obtained with the samples described in Section~\ref{sec:simulation}.
 
The discovery signal region is defined by selecting events with $\NNscore > 0.9$. The exclusion limits are
obtained by performing a shape-fit using ten bins in \NNscore, with bin boundaries
$\{0.65, 0.7, 0.75, 0.8, 0.82, 0.84,$ $0.86, 0.88, 0.90, 0.92, 1.0\}$. The \ttbar background in
the first three bins is reduced by applying an additional selection, namely $\mt > 150$~\GeV.
The selections that define the \bWN\ signal region are presented in Table~\ref{tab:3b_sr}.
\begin{table}[htpb]
\centering
\caption{Event selections defining the signal region \bWN, along with its CR and VR.}
\begin{tabular}{ l r  c c c }
\toprule
\multicolumn{2}{l}{Selection}            & \multicolumn{1}{c}{\bWN}                              & \multicolumn{1}{c}{\bWN-TCR}  & \multicolumn{1}{c}{\bWN-TVR}  \\
\midrule
\multicolumn{2}{l}{Preselection}         & \multicolumn{3}{c}{hard-lepton preselection}                  \\
\midrule
\Njet, \Nbjet          &                   & \multicolumn{3}{c}{$\geq (4,1)$}                              \\
Jet \pt                & [\si{\GeV}]        & \multicolumn{3}{c}{> (25, 25, 25, 25)}                          \\
\mt                    & [\si{\GeV}]        & $>110$                             & $>150$           & $>150$ \\
\midrule
\NNscore               &                   & $>0.9$                             & $\in [0.4, 0.6]$ &  $\in [0.60, 0.65]$ \\
\midrule
\multicolumn{2}{l}{Exclusion technique}  & \multicolumn{3}{c}{shape-fit in \NNscore} \\ 
Bin boundaries         &                   & \multicolumn{3}{c}{$\{0.65, 0.7, 0.75, 0.8, 0.82, 0.84, 0.86, 0.88, 0.90, 0.92, 1.0\}$} \\ 
&                   & \multicolumn{3}{c}{and \mt $>150$~\GeV if \NNscore $<0.8$}  \\ 
\bottomrule
\end{tabular}
\label{tab:3b_sr}
\end{table}

\subsection{\fourBody\ }
\label{subsec:SR:4B}
The four-body decay \fourBody\ occurs when $\Delta m_{\stopone,\ninoone}$ is smaller than the $W$ boson mass.
In this scenario, the decay products have low momenta and often fall below the standard jet and lepton reconstruction \pT thresholds.
It is therefore necessary to apply a soft-lepton preselection and require the presence of a high-momentum
ISR jet, with $\pT > 200$~\GeV, to boost the momenta of the final-state particles. A first four-body signal
region, labelled as \bffNbtag, is optimised by requiring the presence of at least one $b$-tagged jet. The
background in the \bffNbtag\ signal region mostly consists of \ttbar events.
Because the $b$-tagged jets are required to have $\pT > 20$~\GeV, \bffNbtag\ is not sensitive to
$\Delta m_{\stopone,\ninoone}$ below $\sim$40~\GeV. For this reason a second signal region, labelled as \bffNsoftb, is
defined. This region does not rely on $b$-tagged jets but instead requires a soft $b$-tag identified by the presence
of a secondary vertex. The dominant background processes in this region are \ttbar\ and \Wjets.
The \bffNbtag\ signal region also exploits the correlation between the ISR jet \pT and \met by
cutting on the \CTtwo\ variable defined by \( \CTtwo = \mathrm{min}(\met,\pT^{\mathrm{ISR}}-25~\GeV)\).
The key variable used at the last stage of the selection is the ratio of the lepton's transverse momentum to the missing
transverse momentum, \lepPtoverMET, which has small values for the \fourBody\ signal and large
values for the backgrounds. The exact definitions of the four-body signal regions are given in Table~\ref{tab:4b_sr}.
For exclusion limits, the last selection, namely on \lepPtoverMET, is replaced by a shape-fit. In the
\bffNsoftb, the shape-fit is performed in five bins of the variable \lepPtoverMET\ with bin boundaries
$\{0,0.015,0.025,0.04,0.06,0.08\}$. In the \bffNbtag\ signal region the shape-fit is performed in two variables, namely
three bins in \lepPtoverMET\ with bin boundaries $\{0,0.03,0.06,0.1\}$ and two bins in
$\Delta\phi(\pT^{b\textrm{-jet}},\Ptmiss)$ with bin boundaries $\{0,0.8,1.5\}$.
 
\begin{table}[htpb]
\centering
\caption{Event selections defining the signal regions \bffNsoftb\ and \bffNbtag.}
\begin{tabular}{ l r  c c }
\toprule
\multicolumn{2}{l}{Selection}            & \bffNsoftb                         & \bffNbtag     \\
\midrule
\multicolumn{2}{l}{Preselection}         & \multicolumn{2}{c}{soft-lepton preselection}  \\
\midrule
\Njet &                                    & $\geq 1 $                          & $\geq 2 $ \\
Jet \pt & [\si{\GeV}]                       & \multicolumn{2}{c}{$>200$}     \\
\Nbjet &                                   & =0                                 & $\geq 1 $ \\
$b$-jet \pt & [\si{\GeV}]                   & --                                 & $<50$ \\
\NSV &                                     & $\geq 1$                           & -- \\
\mt & [\si{\GeV}]                           & \multicolumn{2}{c}{$>90$}  \\
\midrule
\met & [\si{\GeV}]                          & $>250$                             & -- \\
\dPMETlep                  & [rad]         & $<2.0$                             & -- \\
\midrule
\CTtwo & [\si{\GeV}]                        & --                                 & $>400$ \\
$\Delta\phi(\pT^{b\textrm{-jet}},\Ptmiss)$ & [rad] & --                         & $<1.5$\\
\midrule
\lepPtoverMET &                            & $<0.04$                           & $<0.05$ \\
\midrule
Exclusion technique &                      &  shape-fit in \lepPtoverMET       & shape-fit in \lepPtoverMET \\
&                      &                                    & and $\Delta\phi(\pT^{b\textrm{-jet}},\Ptmiss)$ \\
Bin boundaries in \lepPtoverMET & & $\{0,0.015,0.025,0.04,0.06,0.08\}$ & $\{0,0.03,0.06,0.1\}$ \\
Bin boundaries in $\Delta\phi(\pT^{b\textrm{-jet}},\Ptmiss)$ & [rad] & & $\{0,0.8,1.5\}$ \\
\bottomrule
\end{tabular}
\label{tab:4b_sr}
\end{table}
 
\subsection{Dark matter}
\label{subsec:SR:DM}
 
The dominant background to the search for spin-0 mediator models is the $\ttbar+V $ process. The optimisation
of this signal region favours a selection with at least
two $b$-tagged jets and a leading $b$-tagged jet with $\pT>80$~\GeV. The distribution of \dPMETlep\
differentiates the scalar and pseudoscalar models from each other and also from the background. The
resulting \DMscalar\ and \DMpseudoscalar\ signal region definitions are given in Table~\ref{tab:dm_sr}.
In addition to the selection criteria optimised for discovery described above, the exclusion sensitivity is maximised
by relying on a shape-fit in the region \DMscalar\ with the binning in \dPMETlep\ given in Table~\ref{tab:dm_sr}.
 
\begin{table}[ht]
\begin{center}
\caption{Event selections defining the DM signal regions.}
\vspace{3mm}
\renewcommand{\arraystretch}{1.1}
\begin{tabular}{ l r  c  c }
\toprule
\multicolumn{2}{l}{Selection}       & \DMscalar   & \DMpseudoscalar \\
\midrule
\multicolumn{2}{l}{Preselection}    & \multicolumn{2}{c}{hard-lepton preselection} \\
\midrule
\multicolumn{2}{l}{\Njet, \Nbjet}   & \multicolumn{2}{c}{$\geq(4,2)$} \\
Jet $\pT$ & [\si{\GeV}]                & \multicolumn{2}{c}{$>(80, 60, 30, 25)$} \\
$b$-tagged jet \pT & [\si{\GeV}]       & \multicolumn{2}{c}{$>(80, 25)$}  \\
\midrule
\met & [\si{\GeV}]                   & \multicolumn{2}{c}{$>230$} \\
\HTmissSig &                        & \multicolumn{2}{c}{$>15$}    \\
\mt & [\si{\GeV}]                    & \multicolumn{2}{c}{$>180$}   \\
Topness &                             & \multicolumn{2}{c}{$>8$}     \\
\mtoprecl & [\si{\GeV}]                & \multicolumn{2}{c}{$>150$} \\
\minDeltaPhi, $i\in[1,4]$ & [\si{rad}] & \multicolumn{2}{c}{$>0.9$}\\
\midrule
\dPMETlep    &  [\si{rad}]            & $>1.1$ & $>1.5$ \\
\midrule
Exclusion technique &                 & \multicolumn{2}{c}{Based on shape-fit in \dPMETlep} \\
Bin boundaries in \dPMETlep &         & \multicolumn{2}{c}{$\{1.1, 1.5, 2.0, 2.5, \pi \}$} \\
\bottomrule
\end{tabular}
\label{tab:dm_sr}
\end{center}
\end{table}
% End of text imported from the .//texfiles/signal_regions.tex input file
 
\clearpage
 
% The next lines are included from the .//texfiles/backgrounds.tex input file
\section{Backgrounds}
\label{sec:bkg}
 
Data can be used to constrain the normalisation of the most significant background processes. To this end,
control regions (CRs) are defined by minimally modifying the SR selections to suppress the signal while
enhancing the fraction of the targeted background process. The CRs are then incorporated into a simultaneous
likelihood fit to constrain the background process normalisations in the signal region. The ratio of
the number of background events of a given process in the SR to those in a CR is estimated in MC background
samples but is allowed to deviate from that ratio within dedicated MC modelling systematic uncertainties.
Less significant background processes, such as diboson production and \Zjets, are estimated directly from MC
simulation since they typically represent only a few percent of the signal region yields. CRs are defined to
normalise \ttbar (TCR), \Wjets\ (WCR), single-top (STCR) and \ttZ\ (TZCR). Whether a control region is defined
for a given background and signal region depends on the relative contribution of the process to the SR yield.
 
To validate the background estimates from the CRs, validation regions (VRs) are introduced for
\ttbar (TVR) and \Wjets\ (WVR). The VRs are disjoint from both the SRs and CRs. The TZCR is designed
to be as close as possible to the signal region in order to obtain the most precise estimate of the large
\ttZ\ background, and thus does not leave space between the SR and the CR to introduce a VR for this process.
Background normalisations, referred to as normalisation factors (NF), determined in the CRs are applied to the VRs and compared with the data.
The VRs are not included in the final simultaneous fit, but provide a statistically independent test of the
background estimates in background-dominated regions.
 
The CRs and VRs are designed to minimise potential contamination from signal processes. The signal
contamination is generally well below 10\%, but in some TCRs and TVRs, for models close to the
previously excluded region of parameter space, it can reach approximately 15\%. The signal
contributions to the CRs are not included in the background-only fits but are taken into account
in the exclusion fits described in Section~\ref{sec:results_limits}. The CRs and VRs used for each
SR are summarised in Table~\ref{tab:CROverview}. If a process is not normalised via a control
region then it is estimated directly from MC simulation and theoretical cross-sections.
 
\begin{table}[h!]
\begin{center}
\caption{Summary of the control and validation regions used ($\checkmark$) for each signal region.}
\vspace{3mm}
\renewcommand{\arraystretch}{1.5}
\begin{tabular}{ l  c  c  c  c  c  c c }
\toprule
Signal Region & Signal Scenario & TCR        & WCR        & STCR       & TZCR       & TVR         & WVR \\
\midrule
\tNmed        & \topLSP         & $\checkmark$ & $\checkmark$ & $\checkmark$ & $\checkmark$ & $\checkmark$  & $\checkmark$ \\
\tNhigh       & \topLSP         & \checkmark & \checkmark & \checkmark & \checkmark & \checkmark  & \checkmark \\
\tNdiaglow    & \topLSP         & \checkmark & --         & --         & --         & \checkmark  & --  \\
\tNdiaghigh   & \topLSP         & \checkmark & --         & --         & --         & \checkmark  & -- \\
\midrule
\bWN          & \threeBody      & \checkmark & --         & --         & --         & \checkmark  & -- \\
\midrule
\bffNbtag     & \fourBody       & \checkmark & \checkmark & --         & --         & \checkmark  & \checkmark \\
\bffNsoftb    & \fourBody       & \checkmark & \checkmark & --         & --         & \checkmark  & \checkmark \\
\midrule
\DM           & spin-0 mediator & \checkmark & --         & --         & \checkmark & \checkmark  &  -- \\
\bottomrule
\end{tabular}
\label{tab:CROverview}
\end{center}
\end{table}

\subsection{Control and validation regions for \topLSP and spin-0 mediator signals}
The dominant background process in the \tNmed, \tNhigh\ and \DM\ signal regions is \ttZ, and therefore each of these SRs
uses a dedicated TZCR. The TZCRs aim at capturing \ttZ\ events where the $Z$ boson decays into two electrons or muons,
and thus is kinematically similar to the \ttZ\ background in the signal regions where the $Z$ boson decays into a pair
of neutrinos. This CR is built by selecting events with three leptons (electrons or muons), one pair of which must be
of opposite charge and same flavour with an invariant mass within 10~\GeV of the $Z$ boson mass. The exact definitions
of the TZCRs follow the definitions of the \tNmed, \tNhigh\ and \DM\ SRs in terms of the number of jets, $b$-tagged
jets and jet \pT thresholds.
A modified missing momentum variable, \metll, is defined, where the leptons associated with the $Z$ boson decay are
considered invisible. The \metll is the magnitude of the vector with components \ensuremath{\vec{\tilde{p}}_\mathrm{x,y}^{\mathrm{miss}}} derived
from the $x,~y$ components \ensuremath{\vec{p}_\mathrm{x,y}^{\mathrm{miss}}} of \Ptmiss\ introduced in
Section~\ref{sec:event_reconstruction}. The components \ensuremath{\vec{\tilde{p}}_\mathrm{x,y}^{\mathrm{miss}}} are obtained as follows:
\ensuremath{\vec{\tilde{p}}_\mathrm{x,y}^{\mathrm{miss}}}=\ensuremath{\vec{p}_\mathrm{x,y}^{\mathrm{miss}}}+\ensuremath{\vec{p}_\mathrm{x,y}^{\mathrm{l2}}}
+\ensuremath{\vec{p}_\mathrm{x,y}^{\mathrm{l3}}}, where \ensuremath{\vec{p}_\mathrm{x,y}^{\mathrm{l2}}} and \ensuremath{\vec{p}_\mathrm{x,y}^{\mathrm{l3}}}
are the $x,~y$ components of the momenta of the leptons that make up the $Z$ boson candidate.
The TZCRs require \metll $>$230~\GeV. The remaining SR selections are not applied to the TZCRs, in order to retain
a large enough event sample.
 
The \Wjets\ and dileptonic \ttbar processes are significant in \tNmed\ and \tNhigh, and therefore dedicated CRs,
WCR and TCR, are employed. The \DM\ signal region also employs a TCR but does not require a WCR due to the
smaller size of the \Wjets\ background. These CRs have the same requirements on the number of jets, the number of $b$-tagged jets and
the jet \pT thresholds as listed in Tables~\ref{tab:tnbulk_sr} and~\ref{tab:dm_sr} for their respective signal regions.
Table~\ref{tab:tN_DM_cr_vr} presents the definitions of the TCRs, WCRs and VRs, by showing which selections differ
from the \tNmed\ and \tNhigh\ SRs definitions. Neither \Wjets\ nor dileptonic \ttbar processes yield
hadronic top decays, so a veto on the presence of a hadronic reclustered top candidate is used to ensure
orthogonality with the signal regions. The number of events in TCR and WCR is increased by relaxing several
selections compared with the SR selections. The \HTmissSig\ selection is lowered to 10 for both \tNmed\ and
\tNhigh, and to 13 for \DM. In addition, \perpmet\ is lowered to 300~\GeV for \tNmed, while \met is lowered
to 450~\GeV for \tNhigh. In the \DM\ signal region the requirement on \minDeltaPhi is lowered to 0.6.
 
The topness and \mt selections are used to differentiate between WCR and TCR. In the WCR, \mt is required to be in
the range 30$--$90~\GeV, compatible with the presence of a semileptonic $W$ decay, but incompatible with dileptonic \ttbar
because of the topness selection. In the TCR, the topness selection of the SR is inverted, thus selecting events
compatible with dileptonic \ttbar, while $\mt > 120$~\GeV\ is required, as larger values are favoured by
the presence of two leptonically decaying $W$ bosons. The TCR dedicated to the \DM\ signal region has the same
\mt selection as its signal region, $\mt > 180$~\GeV. The purity of the WCR is further improved by using only
positively charged leptons, exploiting the lepton charge asymmetry in \Wjets\ events from $pp$ collisions.
 
To validate the dileptonic \ttbar background normalisation, a TVR dominated by \ttbar production is designed.
The TVRs for \tNmed\ and \tNhigh\ have the same selections as the corresponding TCR, except for the veto
on the presence of a hadronic reclustered top quark candidate, which is replaced by a selection requiring the presence
of such a hadronic top quark, with a mass $\mtoprecl > 150$~\GeV.
 
The validation of the \Wjets\ background for \tNhigh\ is performed with a WVR with the same selection as \tNhigh\
but requiring the presence of a hadronic reclustered top candidate with $\mtoprecl > 150$~\GeV and $\HTmissSig > 25$, in
order to be closer to the SR. The WVR for \tNmed\ is defined starting from the WCR selections, but replacing several
selections with those used in its SR: $\met > 400$~\GeV, $\HTmissSig > 16$ and the presence of a hadronic top quark,
with a mass $\mtoprecl > 150$~\GeV.
 
\begin{table}[htpb]
\centering
\caption{Event selections defining the CRs and VRs in \tNmed, \tNhigh\ and \DM\ relative to their respective signal regions.
Only variables for which the selection criteria in the CRs or VRs differ from those in the SRs are listed.}
\begin{tabular}{ l r  c c c c }
\toprule
\multicolumn{2}{l}{Selection}  & \tNmed   & \tNmed-TCR (-TVR)  & \tNmed-WCR (-WVR)  & \tNmed-STCR\\
\midrule
\mtoprecl     & [\si{\GeV}]       & $>150$    & veto ($>150$)       & veto ($>150$)       & veto \\
\HTmissSig  &                  & $>16$     & $>10$               & $>10$ ($>16$)        & $>10$ \\
\perpmet    & [\si{\GeV}]       & $>400$    & $>300$              & $>300$ ($>400$)       & 350 \\
\mt           & [\si{\GeV}]       & $>220$    & $>120$              & $\in [30, 90]$     & $\in [30, 120]$ \\
Topness       &                  & $>9$     & $<9$                & $>9$                & $>10$ \\
\deltaRbl     &                  & $<2.8$    & --                 & --                 & -- \\
\dRbb         &                  & --       & --                 & $<1.4$              & $>1.4$ \\
Lepton charge &                  & --       & --                 & $>0$                & -- \\
\Nbjet        &                  & $\geq 1$ & $\geq 1$           & $\geq 1$           & $\geq 2$ \\
\midrule
\multicolumn{2}{l}{Selection}  & \tNhigh  & \tNhigh-TCR (-TVR) & \tNhigh-WCR (-WVR) & \tNhigh-STCR\\
\midrule
\mtoprecl     & [\si{\GeV}]       & $>150$    & veto ($>150$)       & veto ($>150$)       & veto \\
\HTmissSig  &                  & $>25$     & $>10$               & $>$10 ($>25$)        & $>10$ \\
\met        & [\si{\GeV}]       & $>520$    & $>450$              & $>450$              & $>450$ \\
\mt           & [\si{\GeV}]       & $>380$   & $>120$             & $\in [30, 90]$     & $\in [30, 120]$ \\
Topness       &                  & $>8$     & $<8$                & $>8$                & $>10$ \\
\deltaRbl     &                  & $<2.6$    & --                 &  --                 & -- \\
\dRbb         &                  & --       & --                 & $<1.4$              & $>1.4$ \\
Lepton charge &                  & --       & --                 & $>0$                & -- \\
\Nbjet        &                  & $\geq 1$ & $\geq 1$           & $\geq 1$           & $\geq 2$ \\
\midrule
\multicolumn{2}{l}{Selection}  & \DM      & \DM-TCR (-TVR) &  & \\
\hline
\mtoprecl     & [\si{\GeV}]       & $>150$    & veto ($>150$)   &  & \\
\HTmissSig  &                  & $>15$     & $>13$ ($>15$)    &  & \\
Topness       &                  & $>8$     & $<8$            &  & \\
\minDeltaPhi  & [\si{rad}]       & $>0.9$    & $>0.6$          &  & \\
\bottomrule
\end{tabular}
\label{tab:tN_DM_cr_vr}
\end{table}
 
The \DM\ SR contains only a small fraction of \Wjets\ events due to the requirement of two $b$-tagged jets, and
therefore only a TVR is considered. It is constructed from the \DM\ SR definition, but with the topness selection
inverted, and to increase the number of events and limit signal contamination, the selection on \minDeltaPhi is relaxed to 0.6.

\begin{figure}[ht] \centering
\subfigure[]{\includegraphics[width=0.45\textwidth]{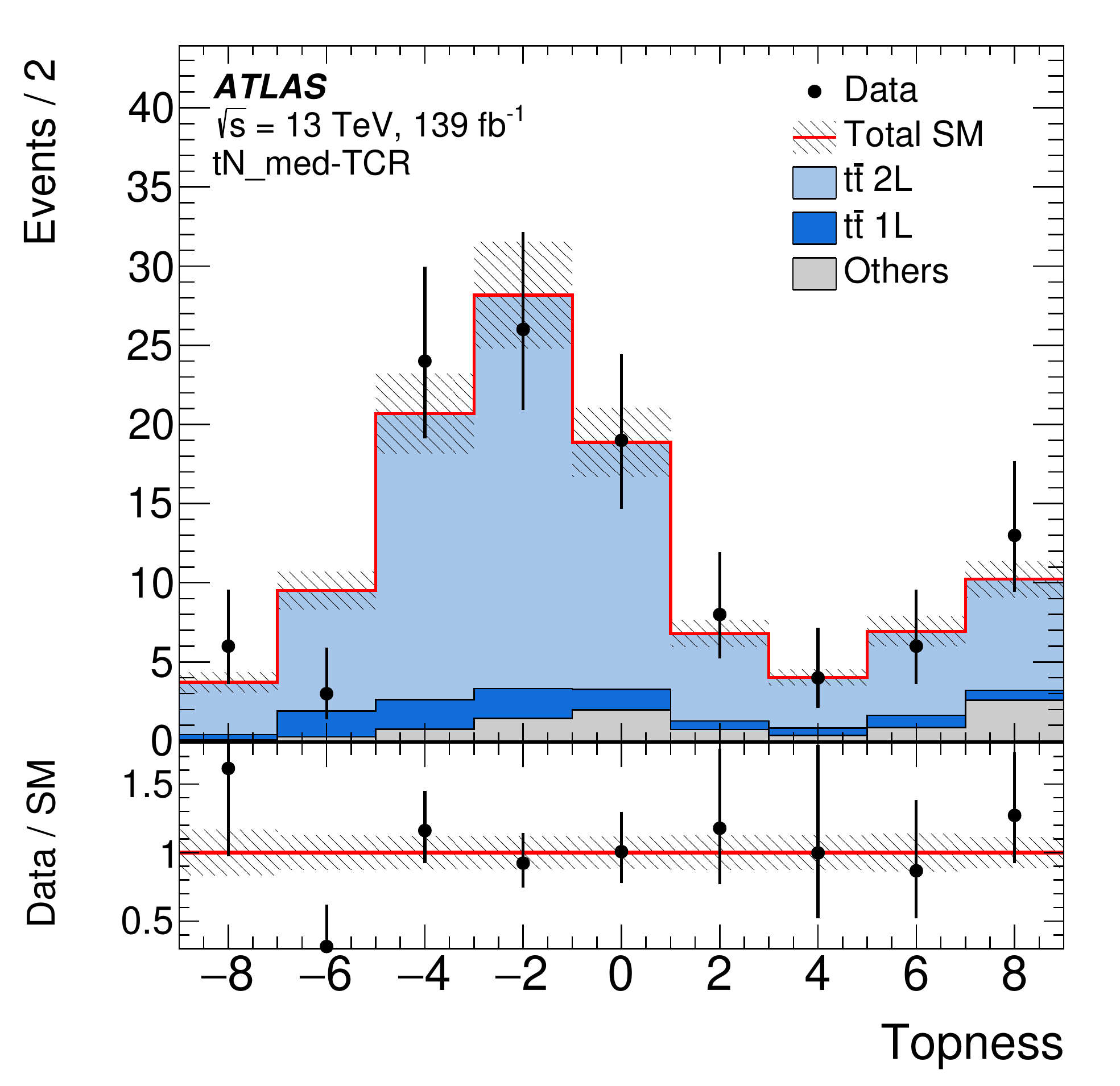}}
\subfigure[]{\includegraphics[width=0.45\textwidth]{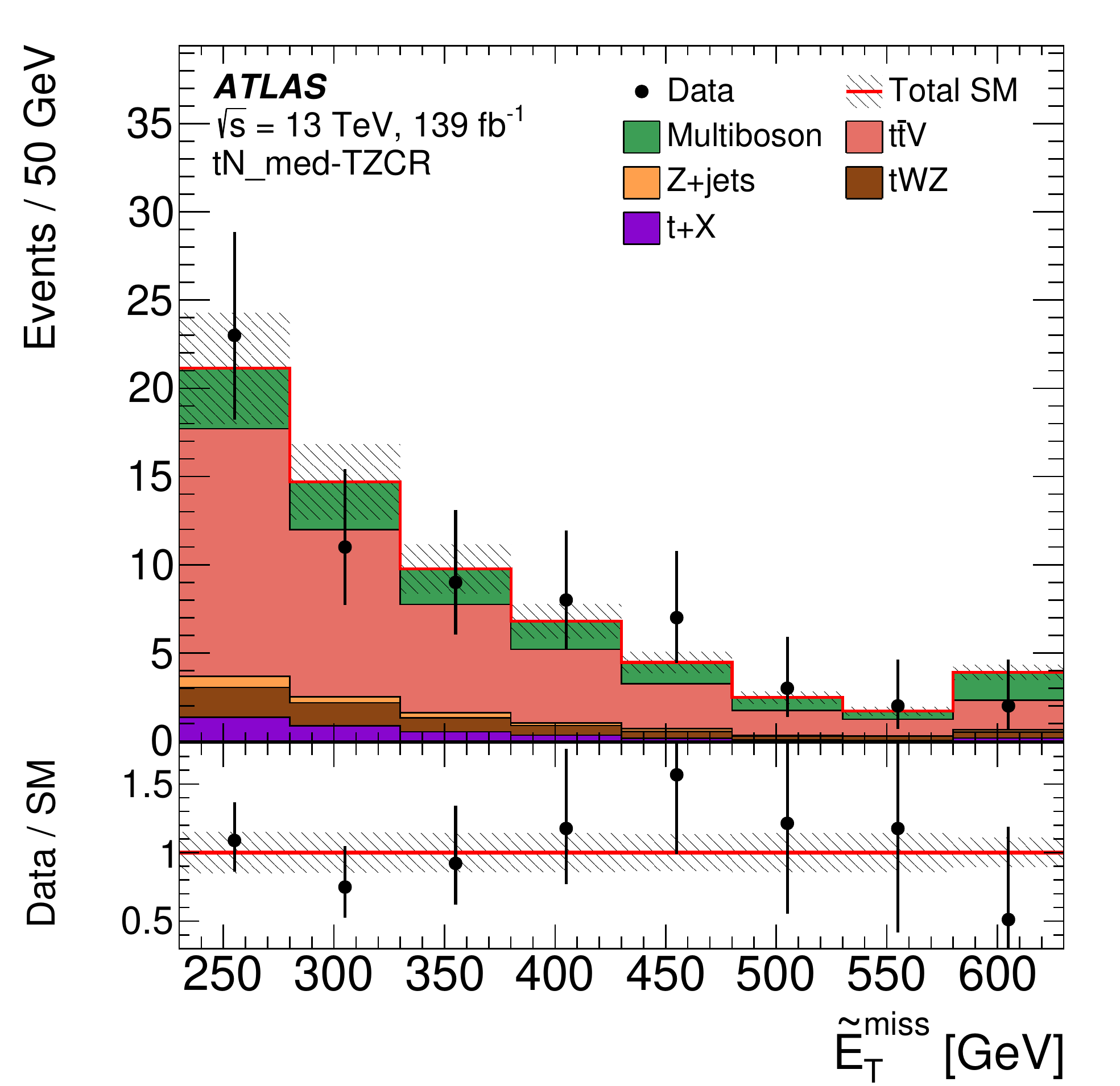}}
\subfigure[]{\includegraphics[width=0.45\textwidth]{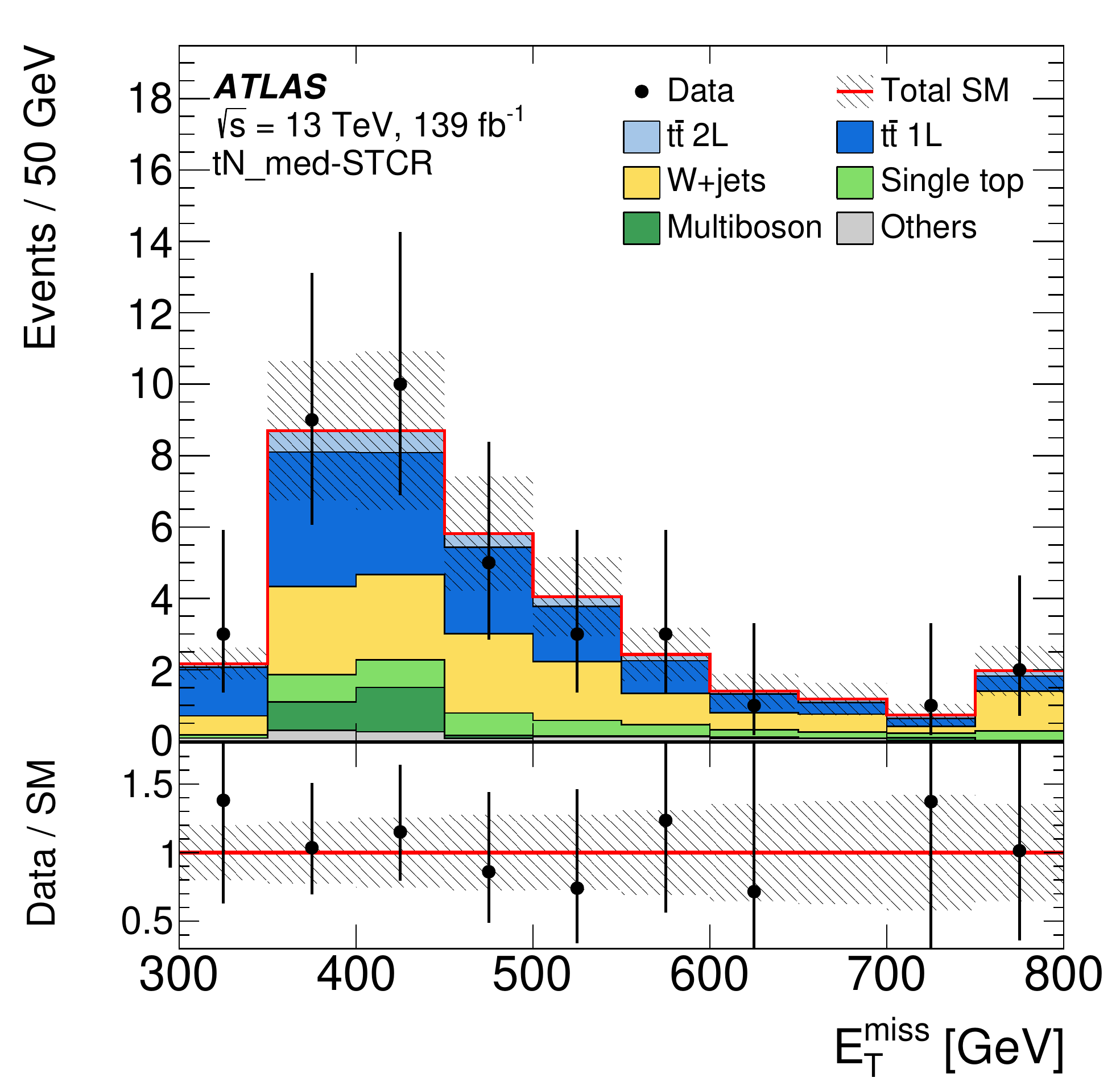}}
\subfigure[]{\includegraphics[width=0.45\textwidth]{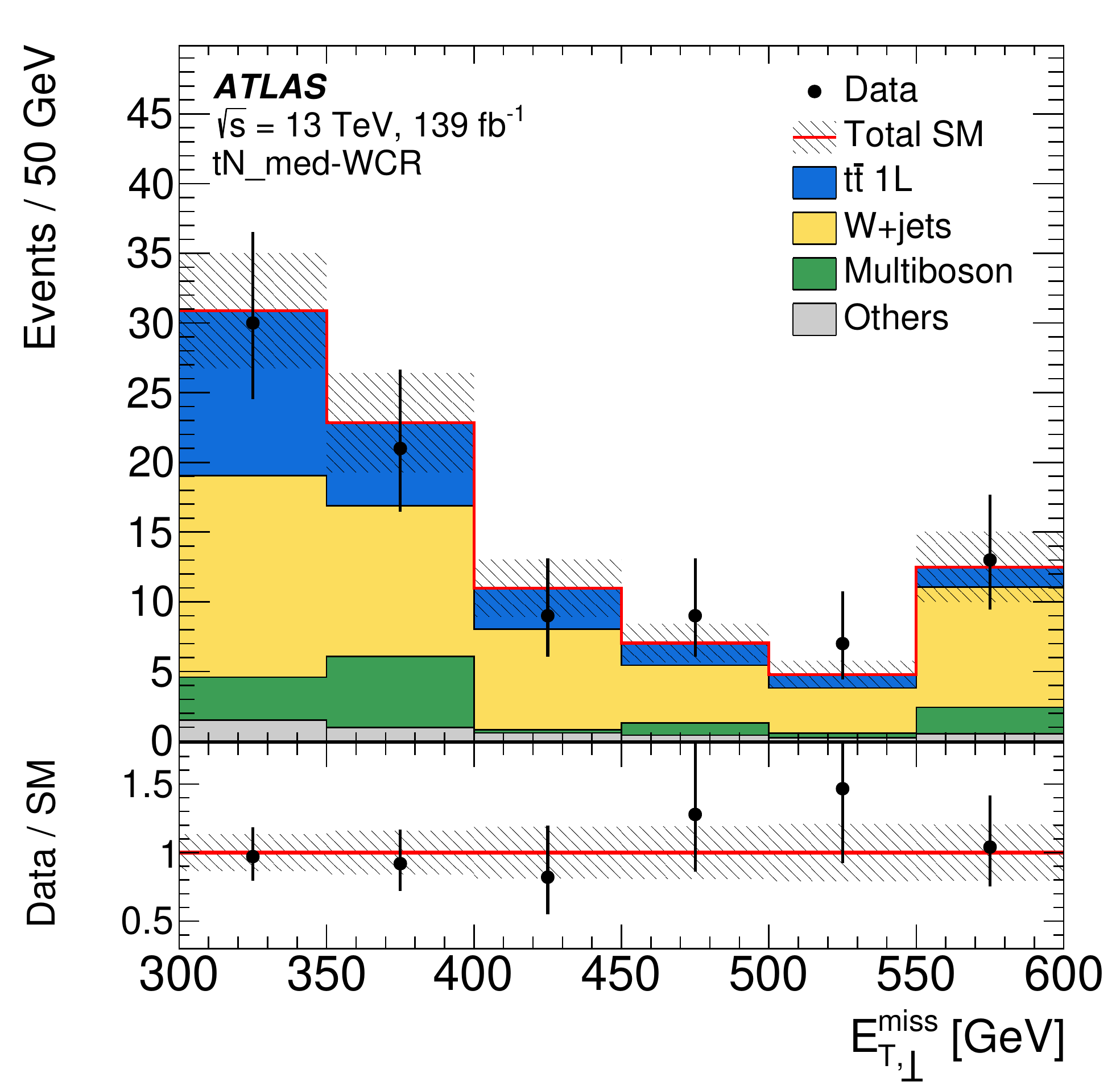}}
\caption{Selected kinematic distributions in \tNmed\ CRs: (a) topness in the TCR, (b) \metll in the TZCR, (c) \met in the STCR, (d) \perpmet\ in the WCR. The distributions shown are post-fit, i.e.\ each background is scaled by a normalisation factor obtained from a background-only likelihood fit to the CRs (see Table~\ref{tab:yields}). The hatched area around the total SM prediction and the hatched band in the Data/SM ratio include all statistical and systematic uncertainties. The last (first) bin contains overflows (underflows).}
\label{fig:bkg:tN_med}
\end{figure}

\begin{figure}[ht] \centering
\subfigure[]{\includegraphics[width=0.45\textwidth]{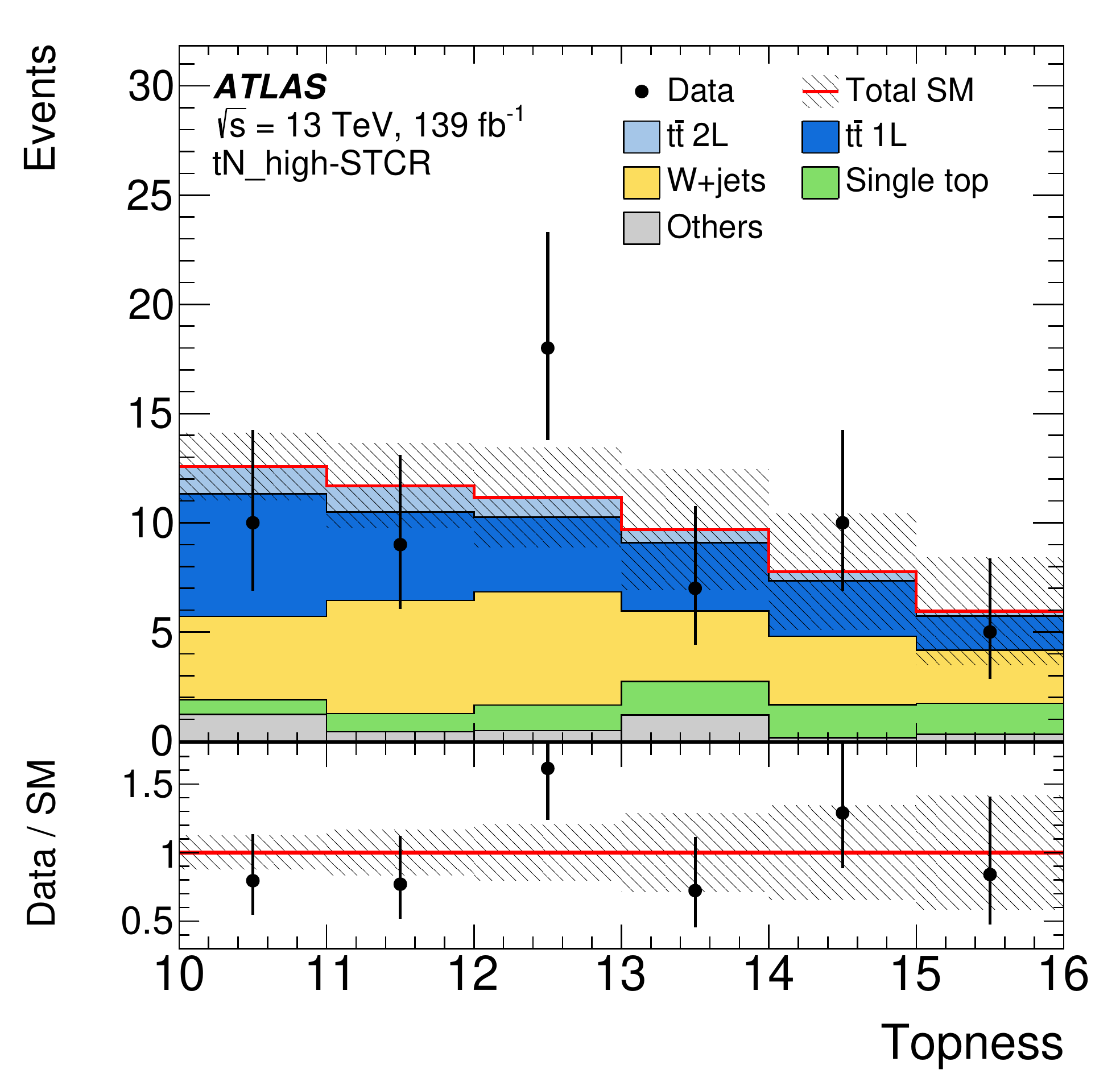}}
\subfigure[]{\includegraphics[width=0.45\textwidth]{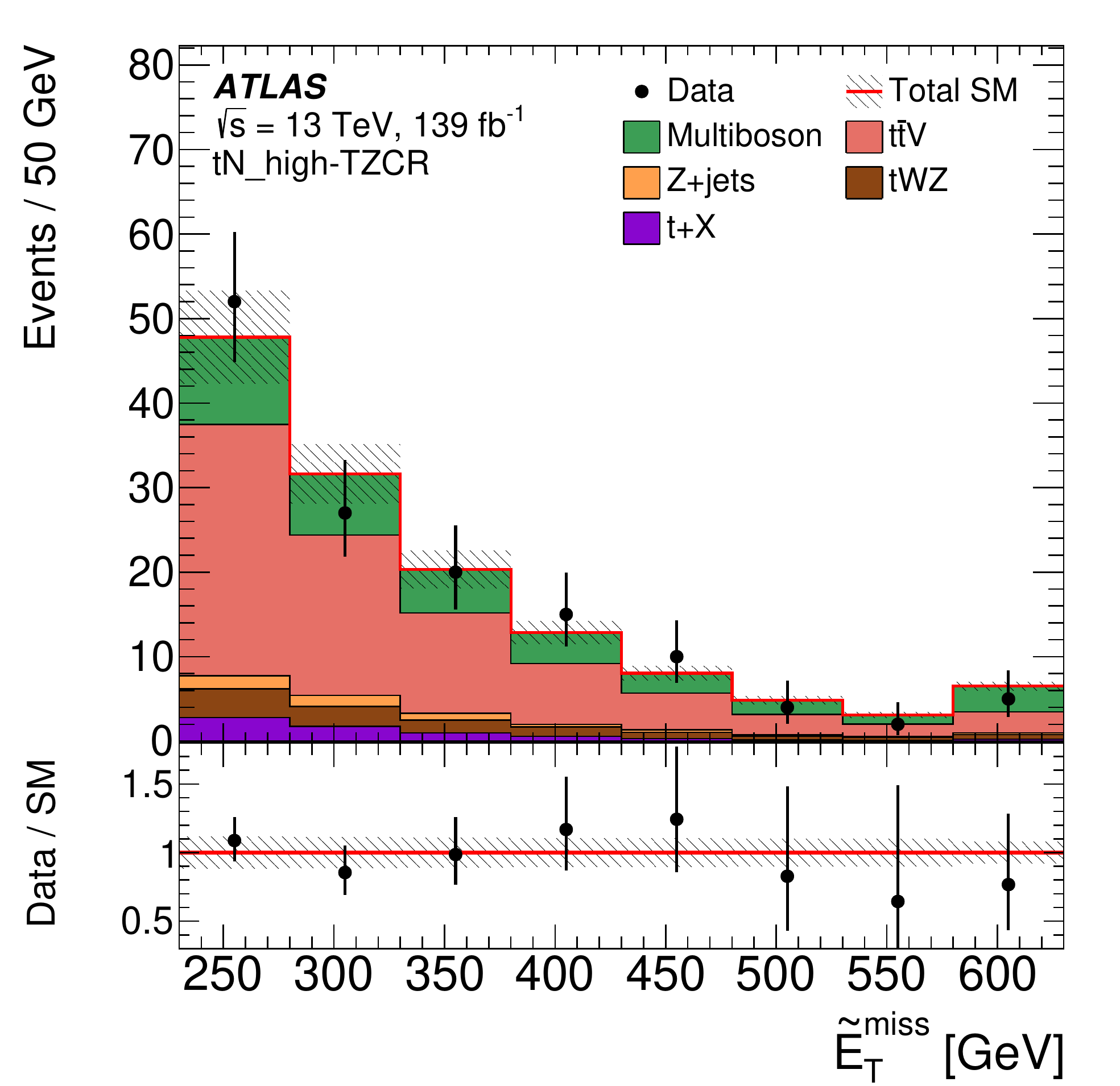}}
\subfigure[]{\includegraphics[width=0.45\textwidth]{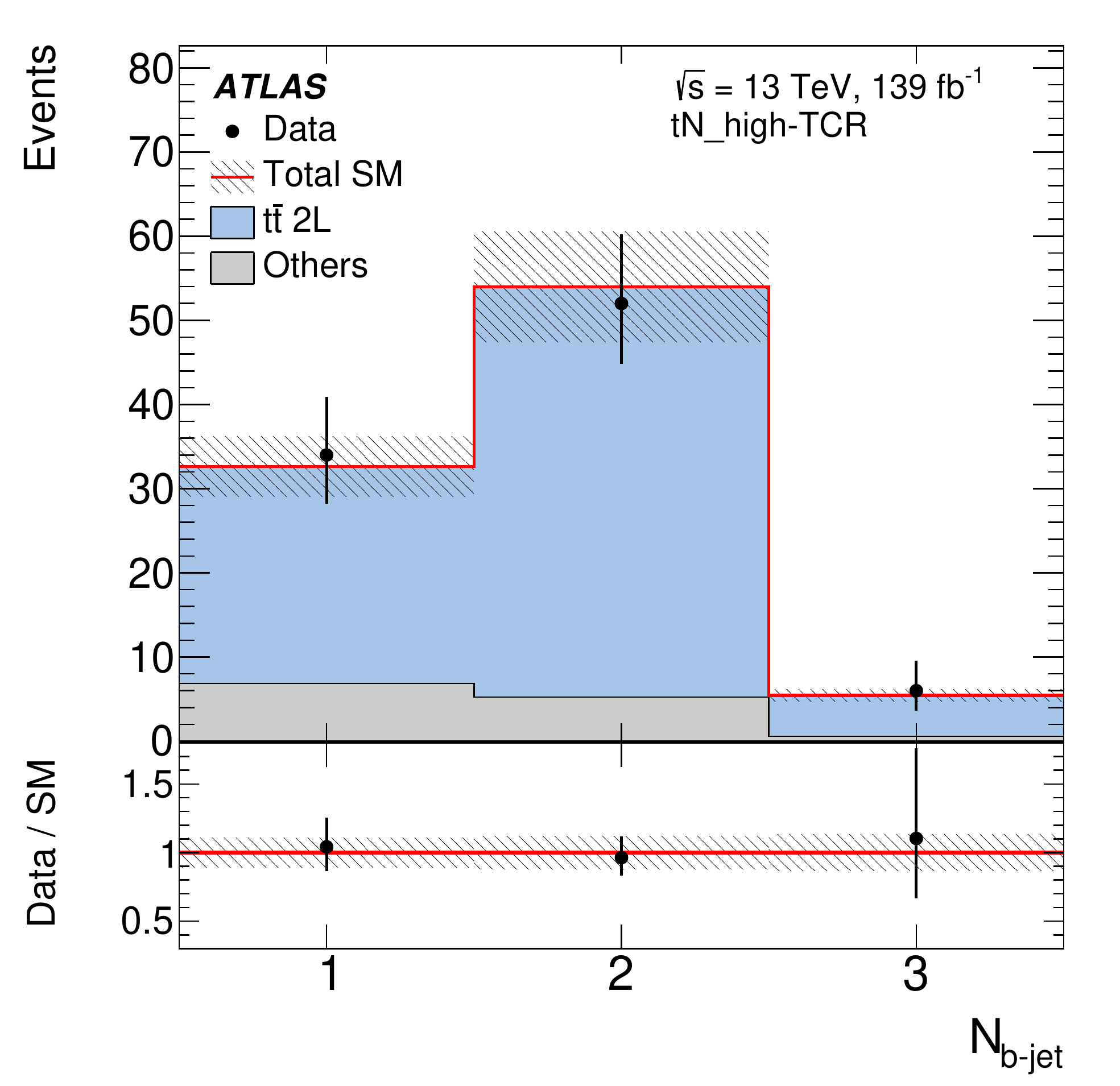}}
\subfigure[]{\includegraphics[width=0.45\textwidth]{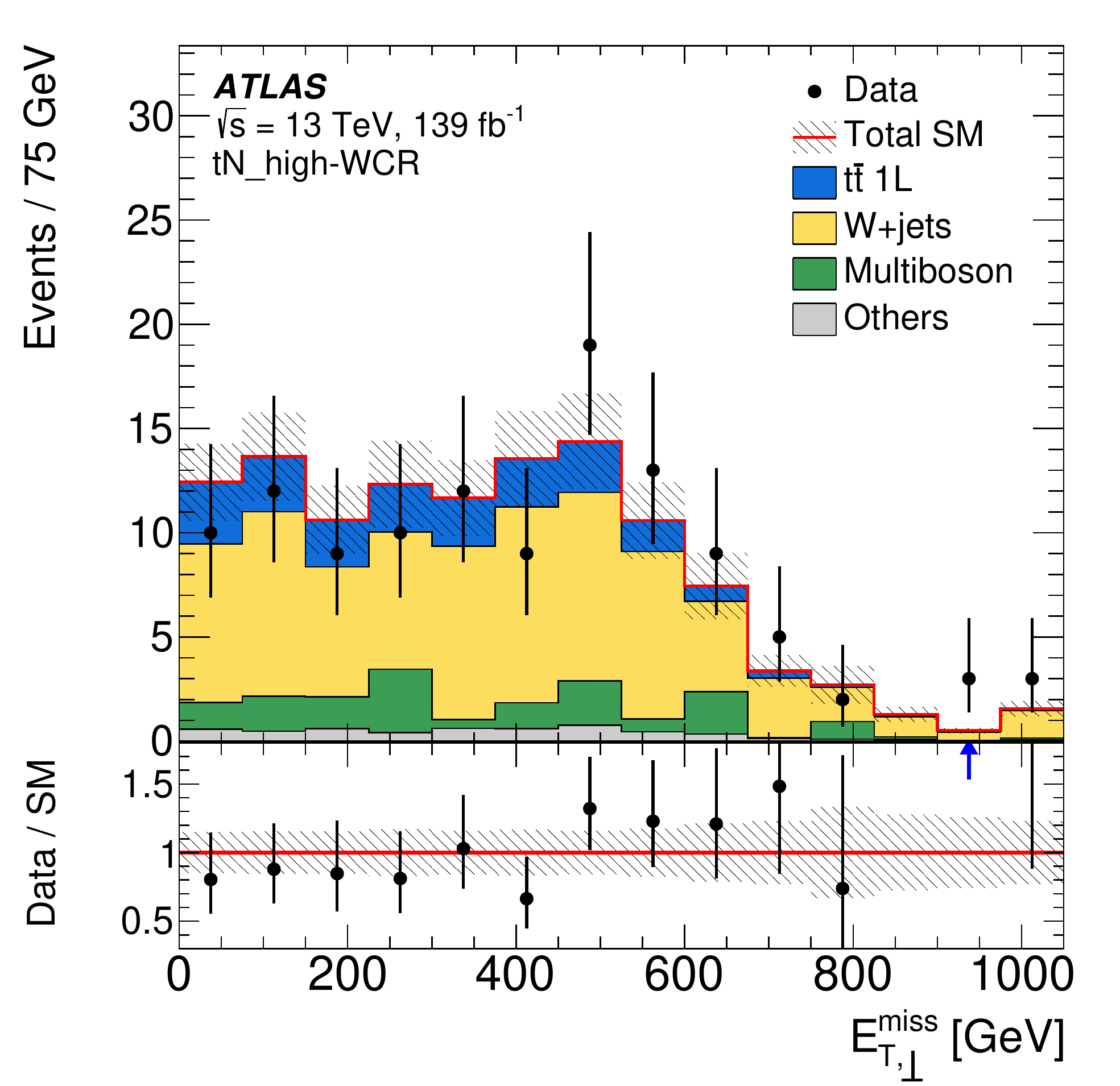}}
\caption{Selected kinematic distributions in \tNhigh\ CRs: (a) topness in the STCR, (b) \metll in the TZCR, (c) \Nbjet\ in the TCR, (d) \perpmet\ in the WCR. The distributions shown are post-fit, i.e.\ each background is scaled by a normalisation factor obtained from a background-only likelihood fit to the CRs (see Table~\ref{tab:yields}). The hatched area around the total SM prediction and the hatched band in the Data/SM ratio include all statistical and systematic uncertainties. The last (first) bin contains overflows (underflows).}
\label{fig:bkg:tN_high}
\end{figure}
 
\begin{figure}[ht] \centering
\subfigure[]{\includegraphics[width=0.45\textwidth]{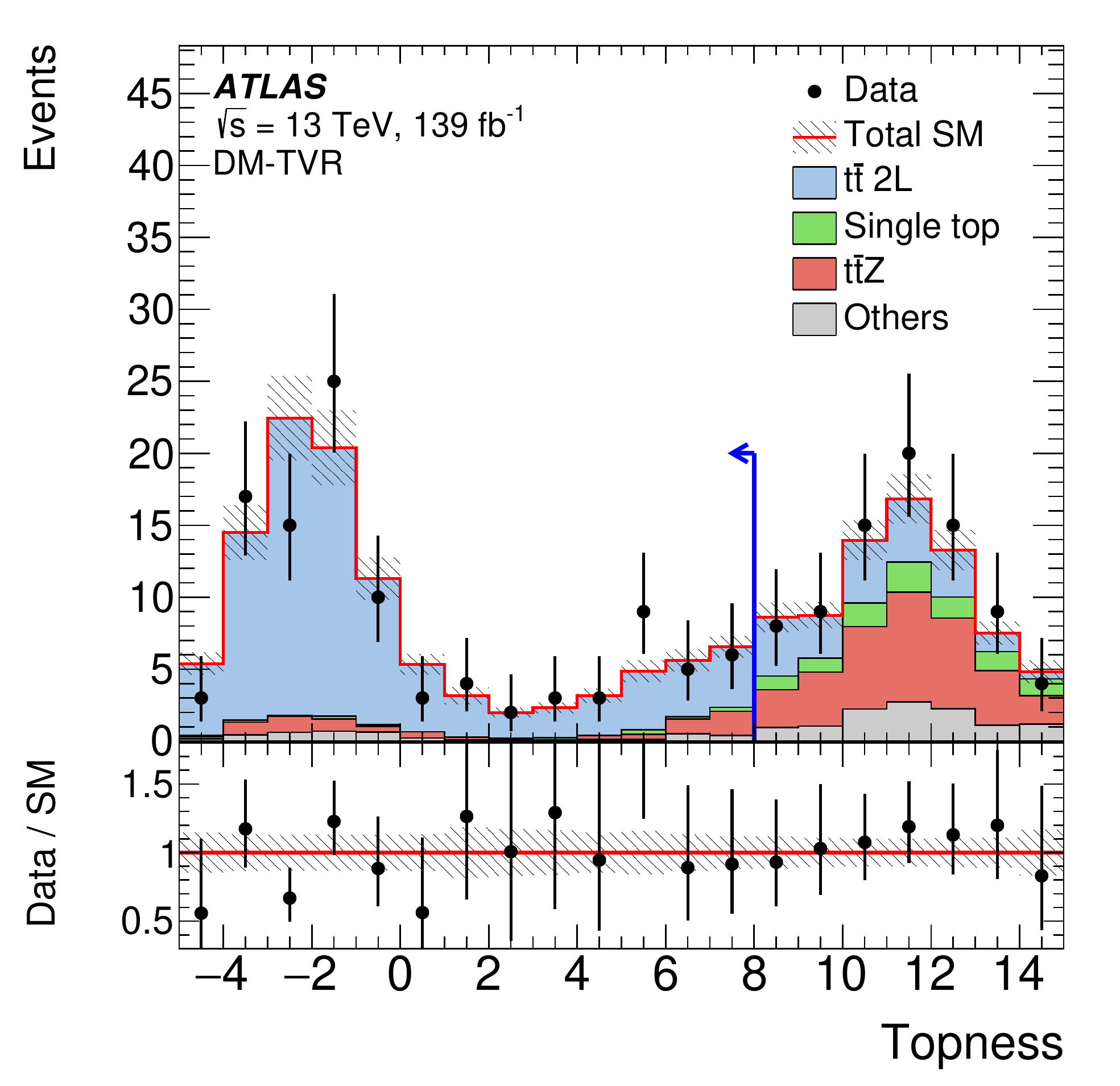}}
\subfigure[]{\includegraphics[width=0.45\textwidth]{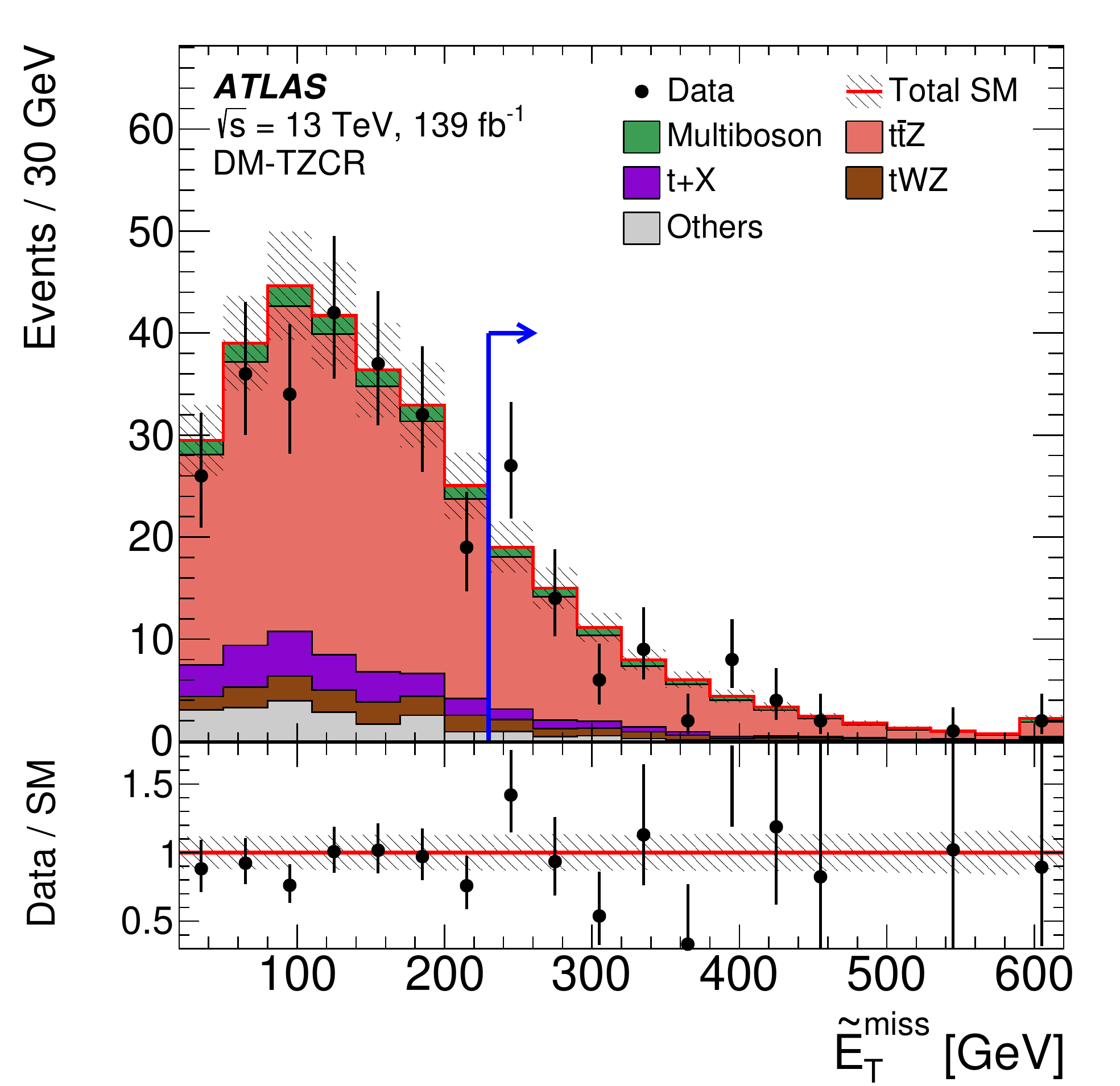}}
\subfigure[]{\includegraphics[width=0.45\textwidth]{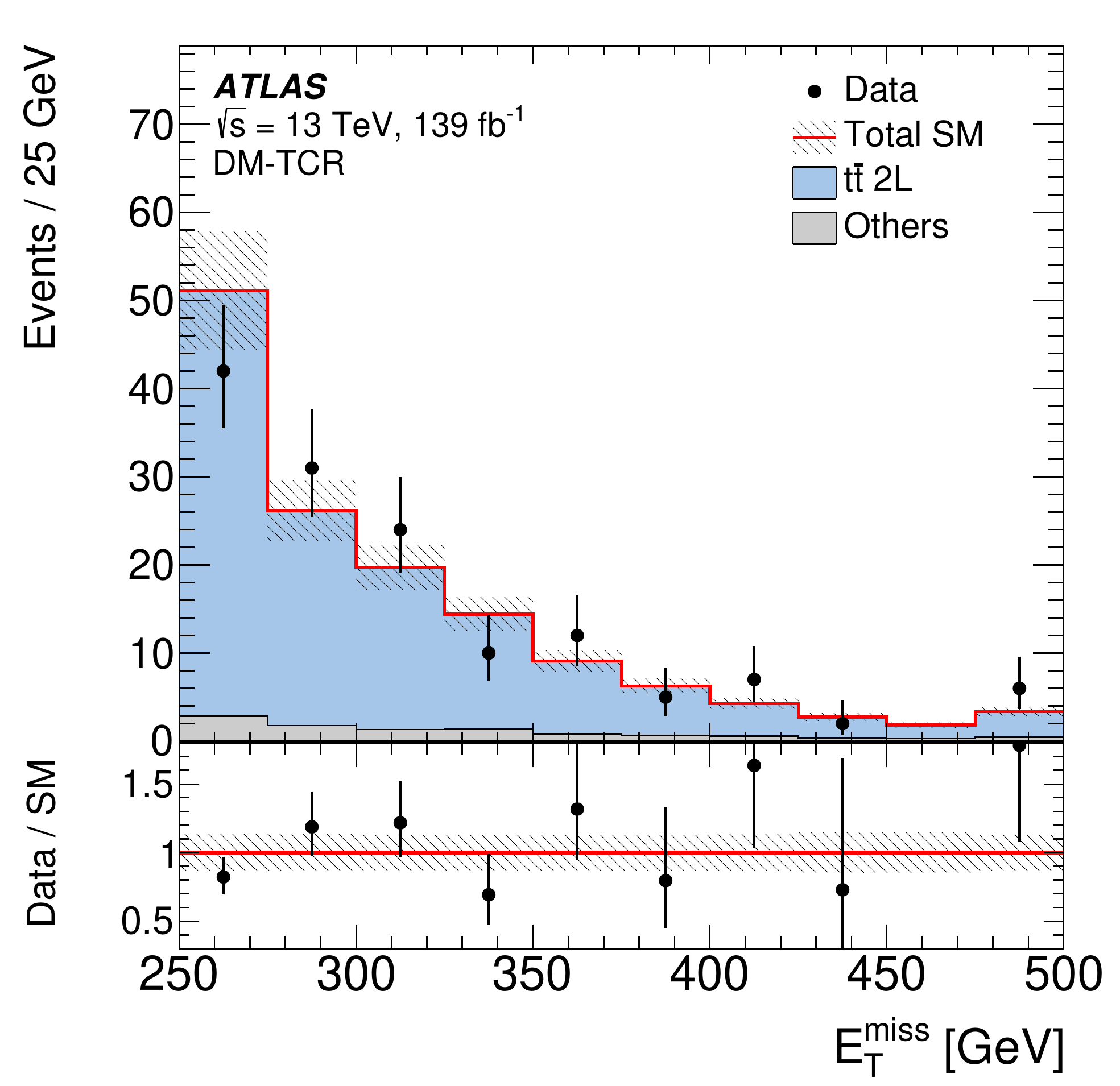}}
\subfigure[]{\includegraphics[width=0.45\textwidth]{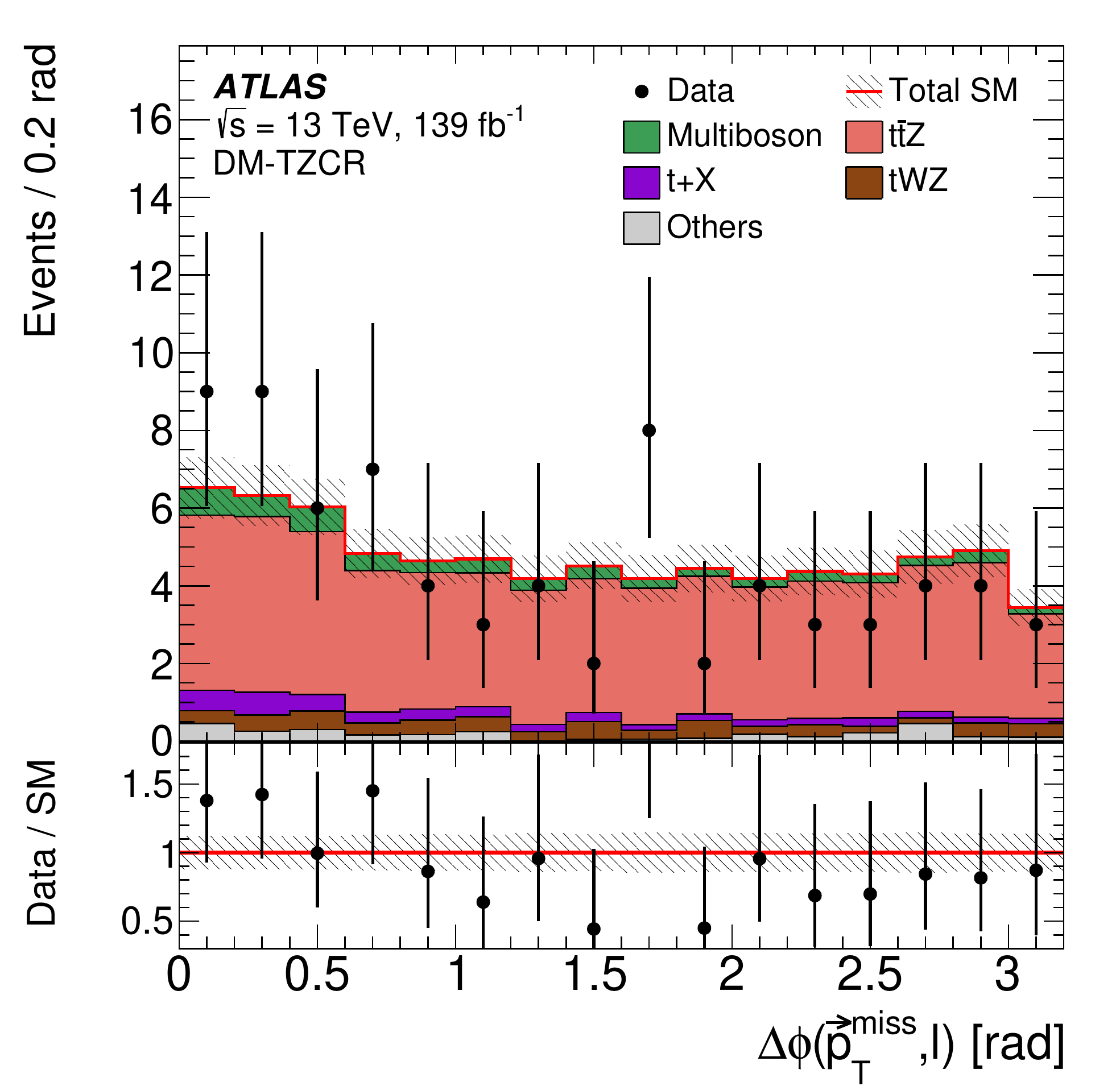}}
\caption{Selected kinematic distributions in \DM\ CRs and VRs: (a) topness in the TVR before applying the topness selection, (b) \metll in the TZCR before applying the \metll selection, (c) \met in the TCR, (d) \dPMETlep\ in the TZCR. For distributions where the requirement on the displayed variable is removed an arrow indicates the final selection on that variable. The distributions shown are post-fit, i.e.\ each background is scaled by a normalisation factor obtained from a background-only likelihood fit to the CRs (see Table~\ref{tab:yields}). The hatched area around the total SM prediction and the hatched band in the Data/SM ratio include all statistical and systematic uncertainties.
The last (first) bin contains overflows (underflows).}
\label{fig:bkg:DM}
\end{figure}
 
\FloatBarrier

The \tNmed\ and \tNhigh\ definitions permit the construction of a STCR with enough data events for comparison
with the DS- and DR-based MC predictions. The STCR is defined with selections
close to those of the WCR, but requires a second $b$-tagged jet, $30 < \mt < 120~\GeV$ and the distance
$\Delta R(b_{\mathrm{1}},b_{\mathrm{2}})$ between the two $b$-tagged jets to be larger than 1.4. To ensure
orthogonality with the WCR, events with two $b$-tagged jets inside the WCR must have
$\Delta R(b_{\mathrm{1}},b_{\mathrm{2}})<1.4$. It is found that the DS and DR scheme predictions bracket the observed
number of events in the STCR data, with a large discrepancy between the two predictions. The largest discrepancy is
observed in the STCR associated with the \tNmed\ SR. The data-to-prediction ratio in the STCR is
0.1$^{+0.3}_{-0.1}$ with the DR scheme and $1.5\pm{1.3}$ with the DS scheme.
 
The availability of the STCR allows the normalisation of the single-top background to be constrained from data.
The fit to the STCR is performed with both the DS and DR MC schemes, and the resulting two predictions for single-top
in the STCR and in the SRs are compatible within uncertainties. Therefore, once the STCR is used to constrain the single-top
normalisation, the choice of the DS or DR scheme is found to have a negligible impact on the single-top prediction in
the SR. In accordance with Ref.~\cite{TOPQ-2017-05}, the DR scheme is used for the default $Wt$ sample.
 
Figures~\ref{fig:bkg:tN_med}, \ref{fig:bkg:tN_high} and~\ref{fig:bkg:DM} compare data and prediction in CRs and VRs
for several variables used in the \topLSP and \DM\ SRs. Good agreement is observed between data and prediction,
within uncertainties.
 
\subsection{Control and validation regions for compressed \topLSP}
The definitions of the control and validation regions for the compressed two-body decays rely on dedicated
discriminating variables introduced in Section~\ref{subsec:compressedvariables}. The dominant background process
in both \tNdiaglow\ and \tNdiaghigh\ is \ttbar\ production. Each of these regions has its own dedicated TCR. The
TCRs build upon the same \Njet, \Nbjet, \mt and hard-lepton preselection as the SRs. In
both TCRs the selection on the \pT of the leading jet is lower than in the SR, and is chosen to be in the
range 200--360~\GeV for \tNdiaglow\ and in the range 200--440~\GeV\ for \tNdiaghigh. In the TCR associated with \tNdiaglow,
\dmtalpha\ is required to be below zero to avoid signal contamination. In addition, to increase the number of events,
the selection on \mstoplep\ is removed and the selection on \mt is lowered to 110~\GeV. The TCR associated with \tNdiaghigh\
requires \dmtdyn\ below 30~\GeV to ensure orthogonality with the SR and limit signal contamination. To increase
the number of events in the TCR associated with \tNdiaghigh, the \met selection is lowered to 350~\GeV and selections
on \mchidyn\ and \mtTwo\ are removed.
 
The top background normalisation in the two-body compressed region is validated using one VR for each of the regions
\tNdiaglow\ and \tNdiaghigh. The TVR corresponding to \tNdiaglow\ is identical to the TCR except for the
leading jet \pt, required to be above 400~\GeV. The TVR associated with \tNdiaghigh\ is identical to the TCR, but
requires the leading jet \pT to be larger than 440~\GeV. Table~\ref{tab:tn_diag_cr_vr} summarises the definitions
of the TCR and TVR for \tNdiaglow\ and \tNdiaghigh\ relative to the SR definitions.
 
\begin{figure}[ht] \centering
\subfigure[]{\includegraphics[width=0.45\textwidth]{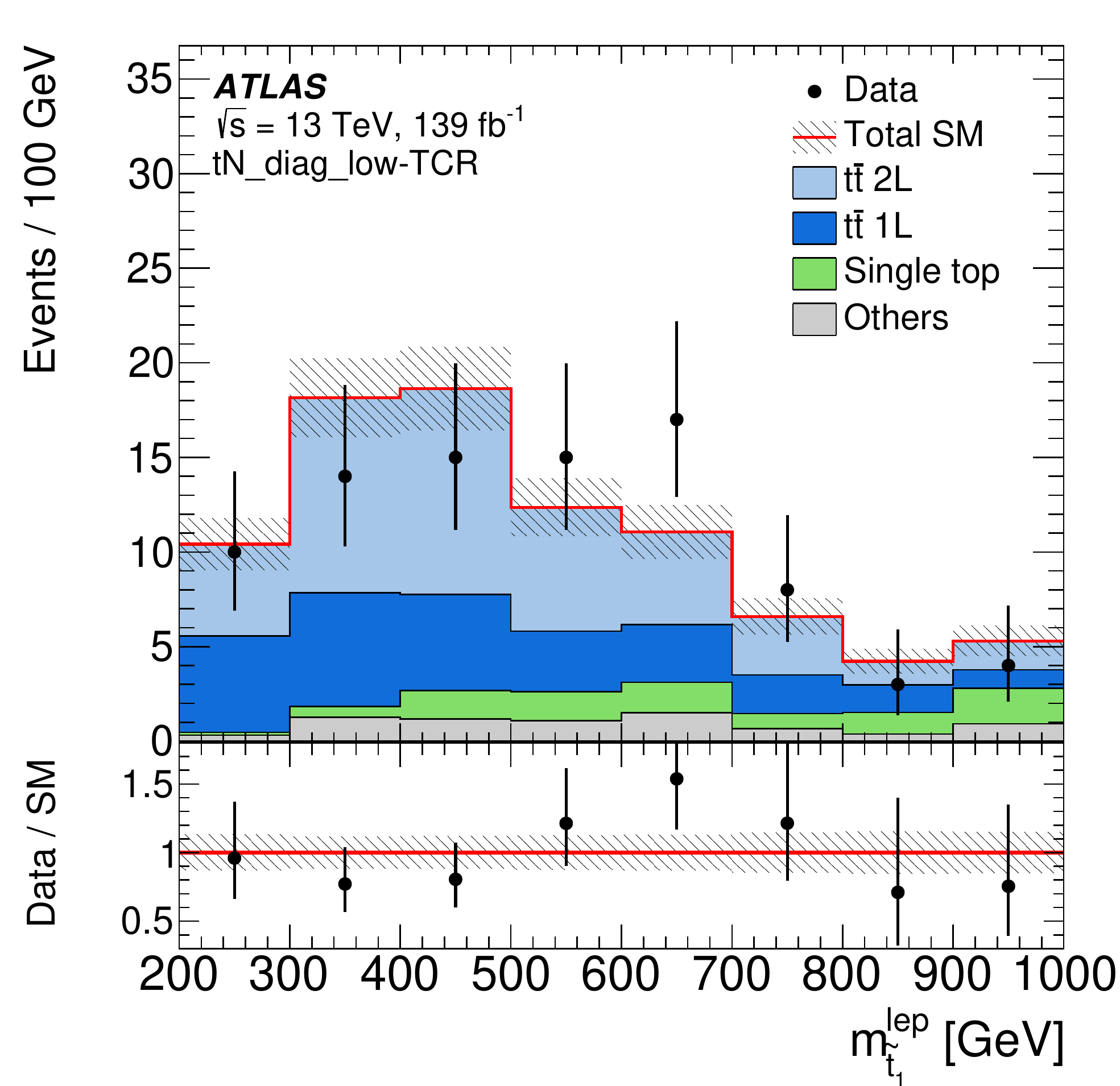}}
\subfigure[]{\includegraphics[width=0.45\textwidth]{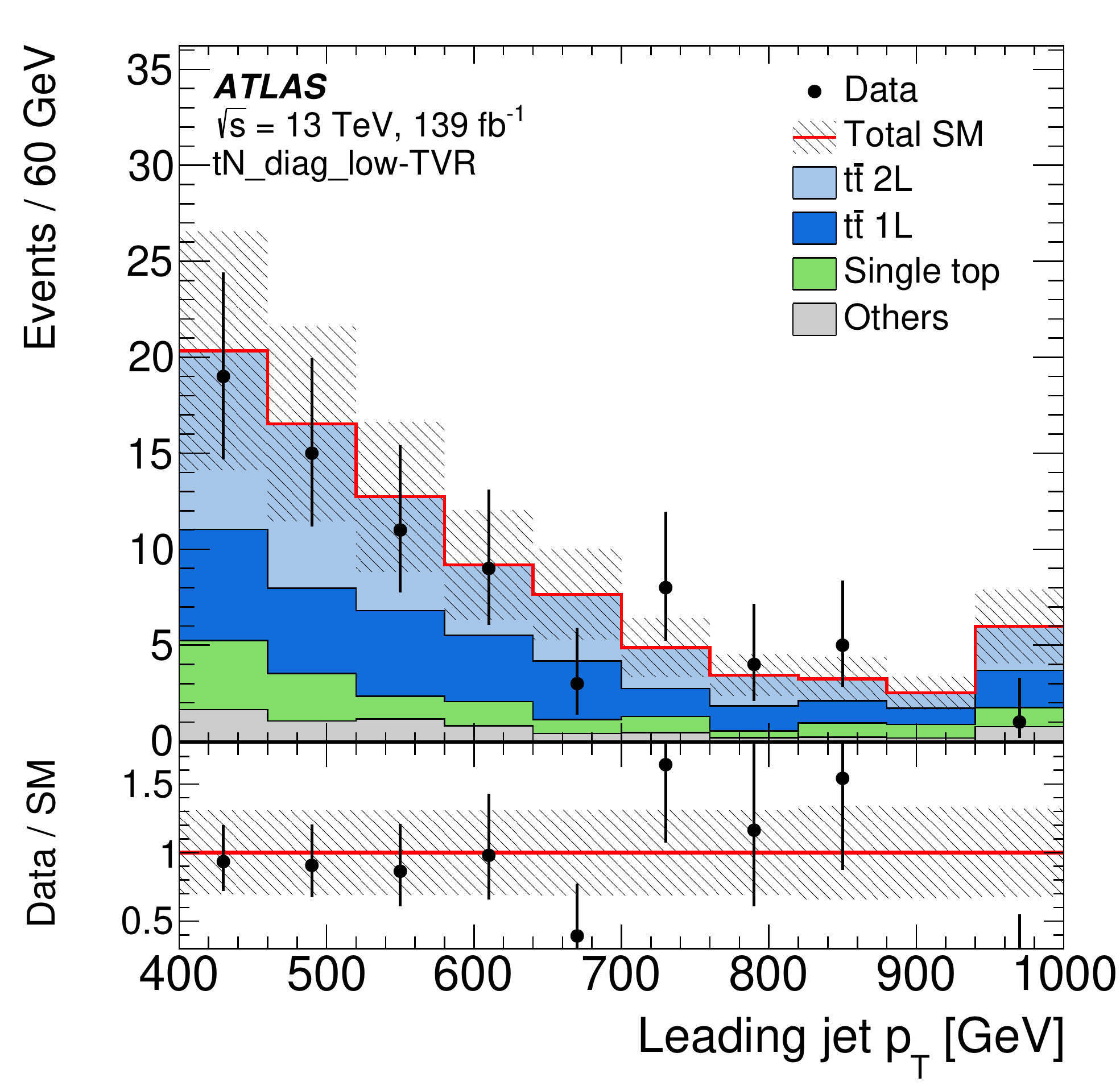}}
\subfigure[]{\includegraphics[width=0.45\textwidth]{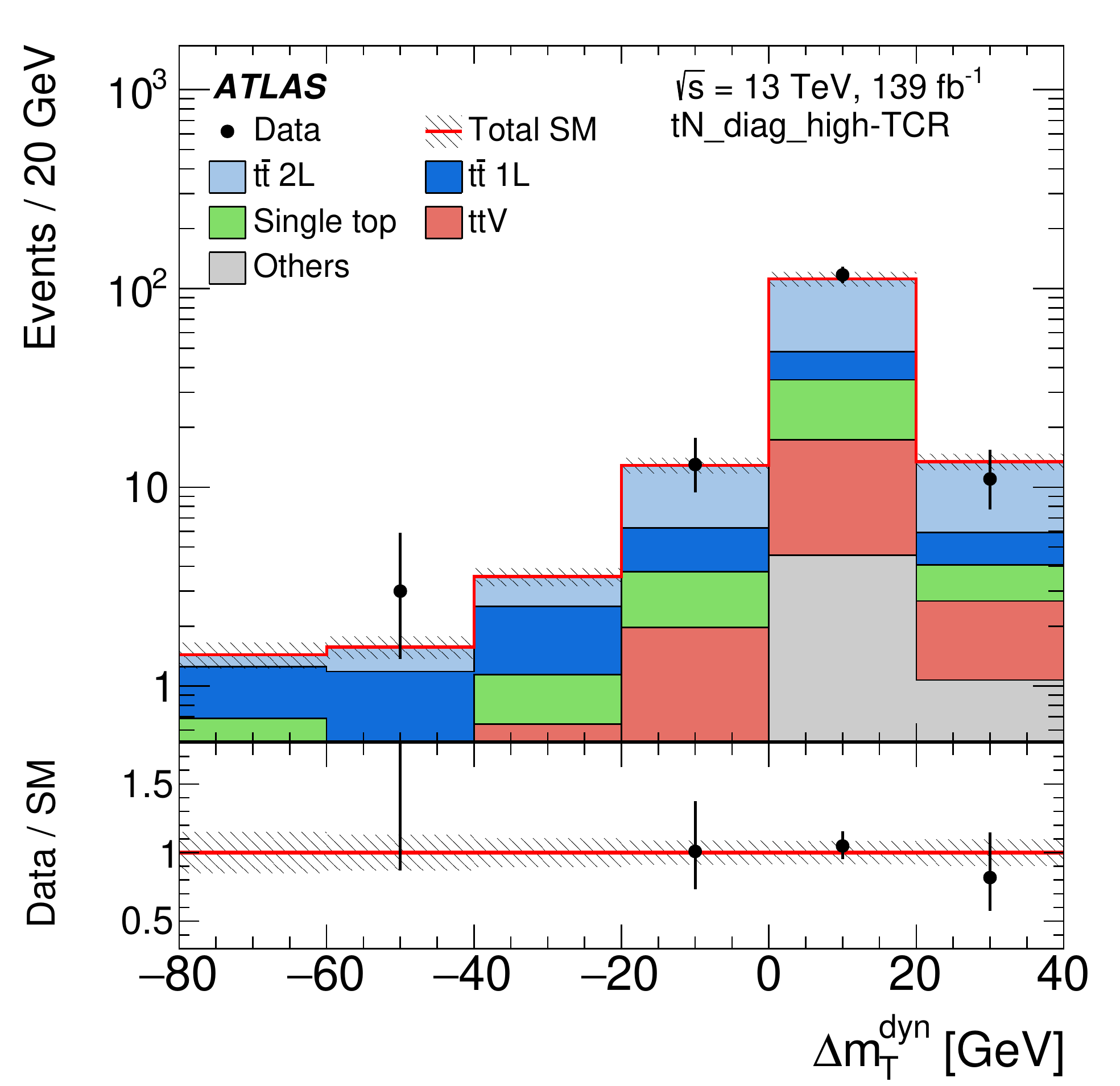}}
\subfigure[]{\includegraphics[width=0.45\textwidth]{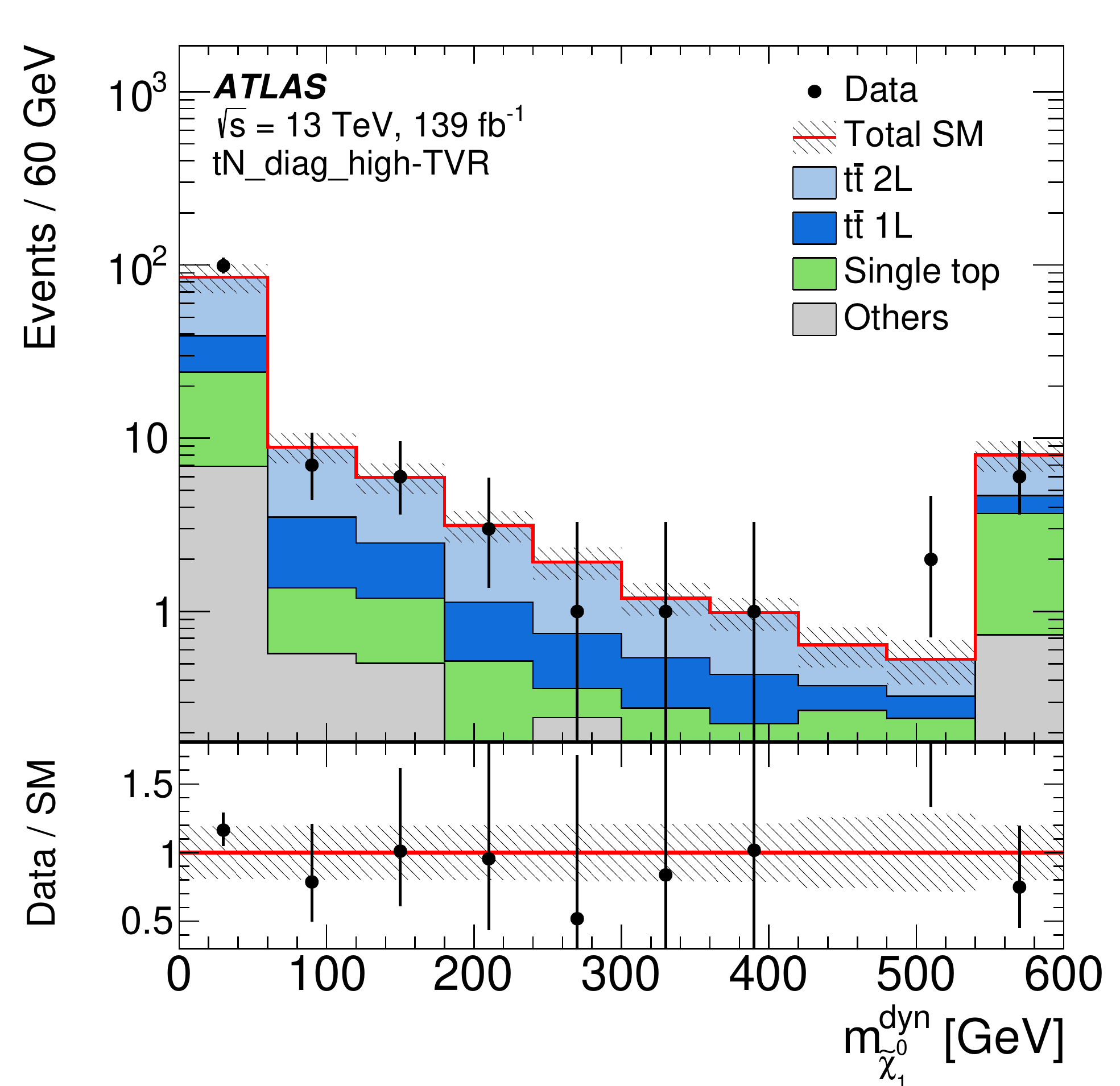}}
\caption{Selected kinematic distributions in \tNdiaglow\ and \tNdiaghigh\ TCRs and TVRs: (a) \mstoplep\ in the \tNdiaglow\ TCR, (b) leading jet \pt in the \tNdiaglow\ TVR, (c) \dmtdyn\ in the \tNdiaghigh\ TCR, (d) \mchidyn\ in the \tNdiaghigh\ TVR. The distributions shown are post-fit, i.e.\ each background is scaled by a normalisation factor obtained from a background-only likelihood fit to the CRs (see Table~\ref{tab:yields}). The hatched area around the total SM prediction and the hatched band in the Data/SM ratio include all statistical and systematic uncertainties. The last (first) bin contains overflows (underflows).}
\label{fig:bkg:tNdiag}
\end{figure}

Figure~\ref{fig:bkg:tNdiag} compares data and prediction in CRs and VRs for several variables used
in the \tNdiaglow\ and \tNdiaghigh\ SRs. Good agreement is observed between data and prediction, within
uncertainties.

\begin{table}
\centering
\caption{Event selections defining the \tNdiaglow\ and \tNdiaghigh\ TCR and TVR relative to their respective signal regions.}
\label{tab:tn_diag_cr_vr}
\begin{tabular}{ l l  c c c}
\toprule
\multicolumn{2}{l}{Selection}			& \tNdiaglow	& \tNdiaglow-TCR & \tNdiaglow-TVR \\
\midrule
Leading jet $\pT$ & [\si{\GeV}]                      & $>400$        & [200, 360]       & $>400$ \\
\mt & [\si{\GeV}]                                    & $>150$        & \multicolumn{2}{c}{$>110$} \\
\dmtalpha & [\si{\GeV}]                              & $>40$        & \multicolumn{2}{c}{$<0$} \\
\mstoplep & [\si{\GeV}]                            & $<600$       & \multicolumn{2}{c}{--} \\
\midrule
\multicolumn{2}{l}{Selection}			& \tNdiaghigh	& \tNdiaghigh-TCR & \tNdiaghigh-TVR \\
\midrule
Leading jet $\pT$ & [\si{\GeV}]                      & $>400$        & [200, 440]      & $>440$ \\
\met & [\si{\GeV}]                                   & $>400$        & \multicolumn{2}{c}{$>350$} \\
$m_\mathrm{T2}$ & [\si{\GeV}]                        & $<360$        & \multicolumn{2}{c}{--} \\
\dmtdyn & [\si{\GeV}]	                            & $>60$         & \multicolumn{2}{c}{$<30$}\\
\mchidyn & [\si{\GeV}]                             & [220, 595]    & \multicolumn{2}{c}{--} \\
\bottomrule
\end{tabular}
\end{table}

\FloatBarrier
 
\subsection{Control and validation regions for \threeBody\ }
More than 80\% of the \bWN\ SR yield consists of dileptonic \ttbar events, while other background components range
between 2\% and 5\% of the SR yield. Therefore, the \threeBody\ SR requires only a TCR. In addition to
the hard-lepton preselection, the \bWN\ SR requires $\NNscore > 0.9$ and $\mt > 100$~\GeV. To
prevent overlap with the signal region, the TCR requires a lower interval of the output score, namely
\NNscore\ $\in$ [0.4, 0.6]. To preserve the same background composition as the signal region, the
selection on \mt is tightened to 150~\GeV.
 
\begin{figure}[ht] \centering
\subfigure[]{\includegraphics[width=0.45\textwidth]{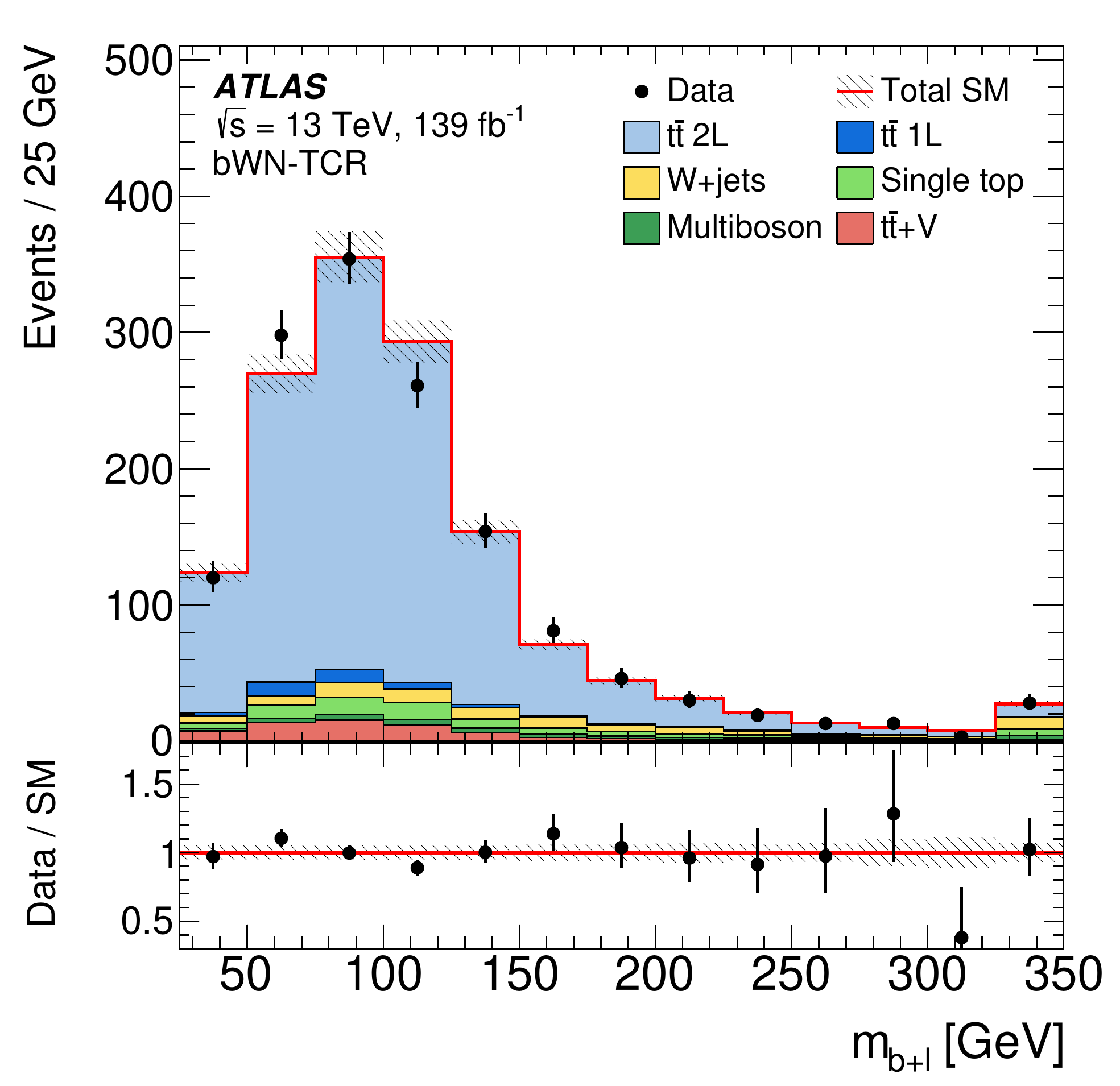}}
\subfigure[]{\includegraphics[width=0.45\textwidth]{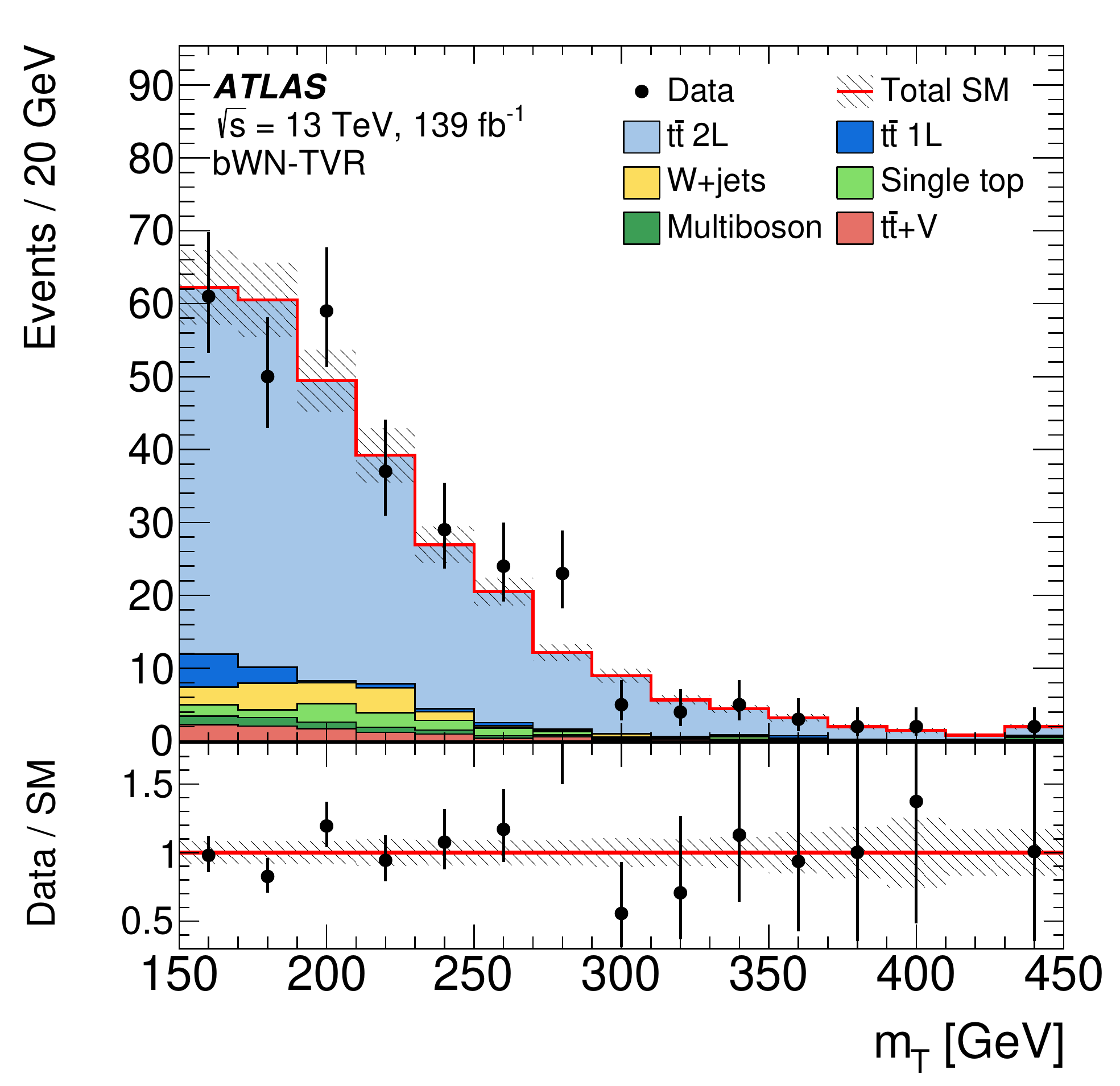}}
\caption{Selected distributions in the \bWN\ CR and VR: (a) $m_{\mathrm{b}+\ell}$ in the TCR, (b) \mt in the TVR. The distributions shown are post-fit, i.e.\ each background is scaled by a normalisation factor obtained from a background-only likelihood fit to the CRs (see Table~\ref{tab:yields}). The hatched area around the total SM prediction and the hatched band in the Data/SM ratio include all statistical and systematic uncertainties. The last (first) bin contains overflows (underflows).}
\label{fig:bkg:bWN}
\end{figure}

The TVR for the \threeBody\ signal is designed by selecting \NNscore\ $\in$ [0.6, 0.65], in between the
\NNscore\ ranges used for the CR and the SR. In addition, $\mt > 150$~\GeV is required to enhance the
contribution of semileptonic \ttbar events, representative of the background in the SR. The selections
that define the TCR and TVR for \bWN\ are summarised in Table~\ref{tab:3b_sr}.
Figure~\ref{fig:bkg:bWN} compares data and prediction in the CR and VR associated with \bWN. Good agreement is
observed between data and prediction, within uncertainties.
 
\FloatBarrier
 
\subsection{Control and validation regions for \fourBody\ }
In the signal region \bffNsoftb\ the largest background is \Wjets\ followed by approximately equal amounts of
dileptonic and semileptonic \ttbar. A \Wjets\ control region, WCR, is built upon the same selections as the SR,
but to enhance the fraction of \Wjets\ events, \lepPtoverMET\ is required to be in the interval [0.16, 0.32] while
the lepton charge is required to be positive. A \ttbar control region, TCR, is designed to have approximately equal amounts
of the two types of \ttbar backgrounds. The TCR builds on the same selections as \bffNsoftb, but it requires the
presence of at least one $b$-tagged jet in order to increase the fraction of \ttbar\ and to ensure orthogonality with
the \bffNsoftb\ SR. The TCR also requires \lepPtoverMET\ to be in the interval [0.12, 0.25] and that the lepton
charge be negative for orthogonality with the WCR.
 
The validation region TVR associated with \bffNsoftb\ is defined with the same selections as the TCR,
except for the ratio \lepPtoverMET, which is required to be in the interval [0.08, 0.12]. The WVR for \bffNsoftb\ has the same
selections as the WCR, except \lepPtoverMET\ is required to be in the interval [0.08, 0.16]. Table~\ref{tab:4b_vr_cr}
shows the selection differences between the \bffNsoftb\ SR and the associated CRs and VRs.
 
The soft $b$-tagging efficiency and mis-tag rate depend on the track multiplicity, kinematics of the $b$-hadrons,
and the $b$-hadron fragmentation. Two highly populated soft $b$-tagging regions enriched in
\ttbar and \Wjets\ are defined, where it is found that the track multiplicity differs between data and MC simulations.
In each region a weight is defined as a function of the track multiplicity to reweight the
MC simulation to the data. After reweighting, good agreement is found between the data and the simulation
in a range of secondary vertex variables such as vertex mass, vertex momentum, vertex distance to the primary
vertex and vertex track multiplicity. The `after-to-before-reweighting' ratios of efficiencies and mis-tag rates define
scale factors that are derived separately for \SHERPA and \PYTHIA~8. The largest discrepancy after reweighting
is found in the \Wjets\ region and is of the order of 20\%. A corresponding 20\% systematic uncertainty in the
soft $b$-tagging scale factors is introduced.
 
\begin{table}[htpb]
\centering
\caption{Event selections defining the CRs and VRs in \bffNsoftb\ and \bffNbtag\ relative to their respective signal regions.}
\begin{tabular}{ l r  c c c }
\toprule
\multicolumn{2}{l}{Selection}                     & \bffNsoftb                         & \bffNsoftb-TCR (-TVR)                 &  \bffNsoftb-WCR (-WVR) \\
\midrule
\Nbjet        &                                     &                               $=0$ & $\geq 1$                              & =0 \\
Lepton charge &                                     & --                                 & $<0$                                  & $>0$ \\
\lepPtoverMET &                                     & $<0.04$                           & $\in [0.12,0.25]$ ($\in [0.08,0.12]$) & $\in [0.16,0.32]$ ($\in [0.08,0.16]$) \\
\midrule
\multicolumn{2}{l}{Selection}                     & \bffNbtag                          & \bffNbtag-TCR (-TVR)                  &  \bffNbtag-WCR (-WVR) \\
\midrule
$b$-jet \pt   & [\si{\GeV}]                          & $<50$                              & $>100$ ($\in [50,100]$)               & --       \\
\mt           & [\si{\GeV}]                          & $>90$                              & $>110$                                & $>$90    \\
\lepPtoverMET &                                     & $<0.05$                            & $>0.05$                               & $>0.05$  \\
$\Delta\phi(\pT^{b\textrm{-jet}},\Ptmiss)$ & [\si{rad}] & $<1.5$                             & $<1.5$                                & $>2.3$ ($\in [1.5, 2.3]$) \\
Lepton charge &                                     & --                                 & --                                    & $>0$     \\
\bottomrule
\end{tabular}
\label{tab:4b_vr_cr}
\end{table}
 
In the signal region \bffNbtag, the dominant background process is dileptonic \ttbar, representing almost half
of all background events, followed by \Wjets. To ensure orthogonality with the SR and limit signal contamination,
the TCR control region has the same selections as the SR, but requires the leading $b$-tagged jet \pT to be above 100~\GeV,
$\mt > 110$~\GeV and $\lepPtoverMET > 0.05$. Compared with the SR, the WCR removes the upper bound on the leading $b$-tagged jet
\pT, requires $\Delta\phi(\pT^{b\textrm{-jet}},\Ptmiss)>2.3$, $\lepPtoverMET > 0.05$ and a positive lepton charge.
 
The TVR associated with the \bffNbtag\ signal region has the same selections as TCR but the $b$-tagged jet \pT is required to
be in the interval [50, 100]~\GeV, and is thus between the signal region and control region. The WVR has the same selections
as the WCR except for the angle $\Delta\phi(\pT^{b\textrm{-jet}},\Ptmiss)$, which is required to be in the intermediate
range between the WCR and the SR, namely the interval [1.5, 2.3].
 
\begin{figure}[ht] \centering
\subfigure[]{\includegraphics[width=0.45\textwidth]{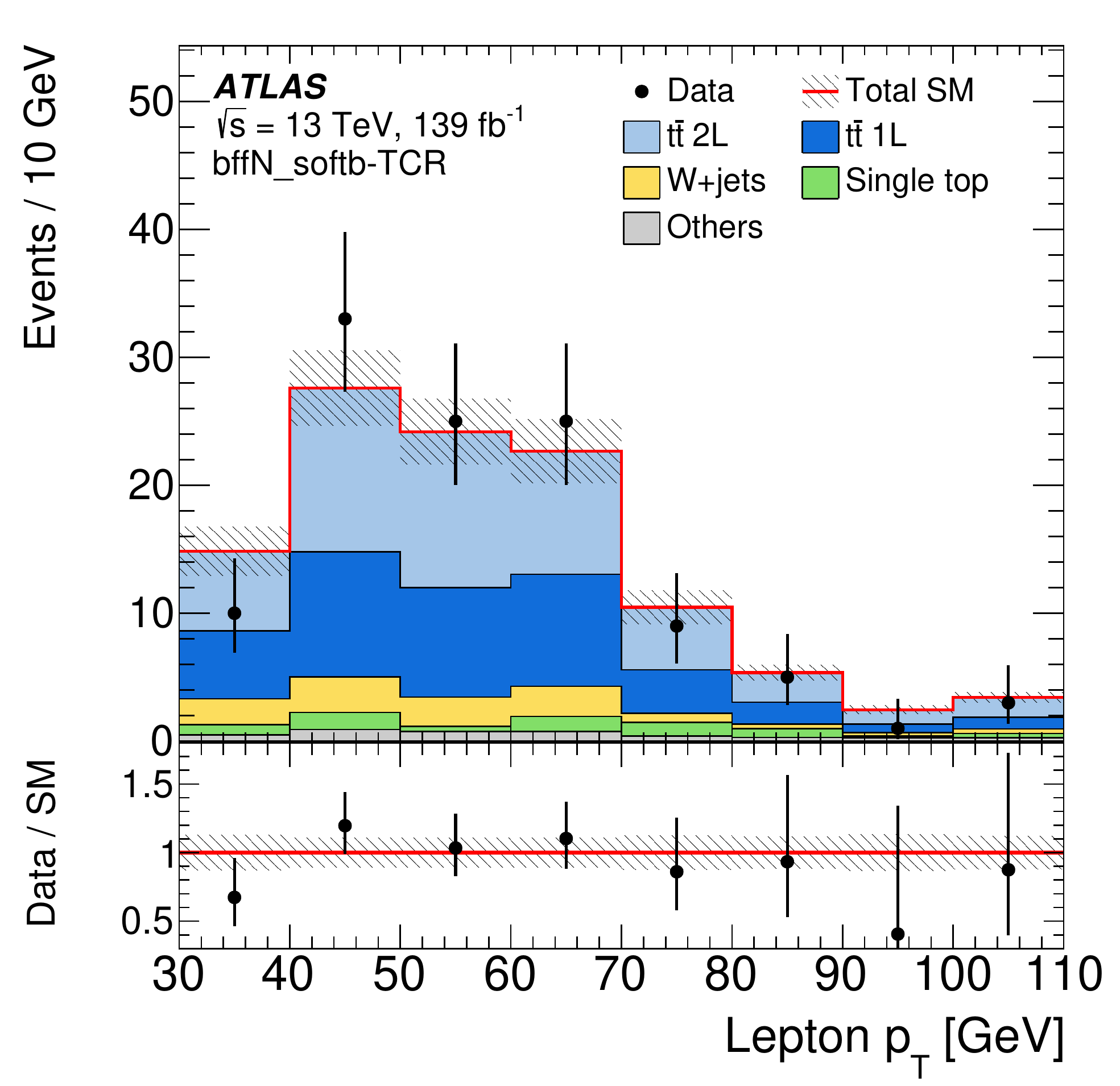}}
\subfigure[]{\includegraphics[width=0.45\textwidth]{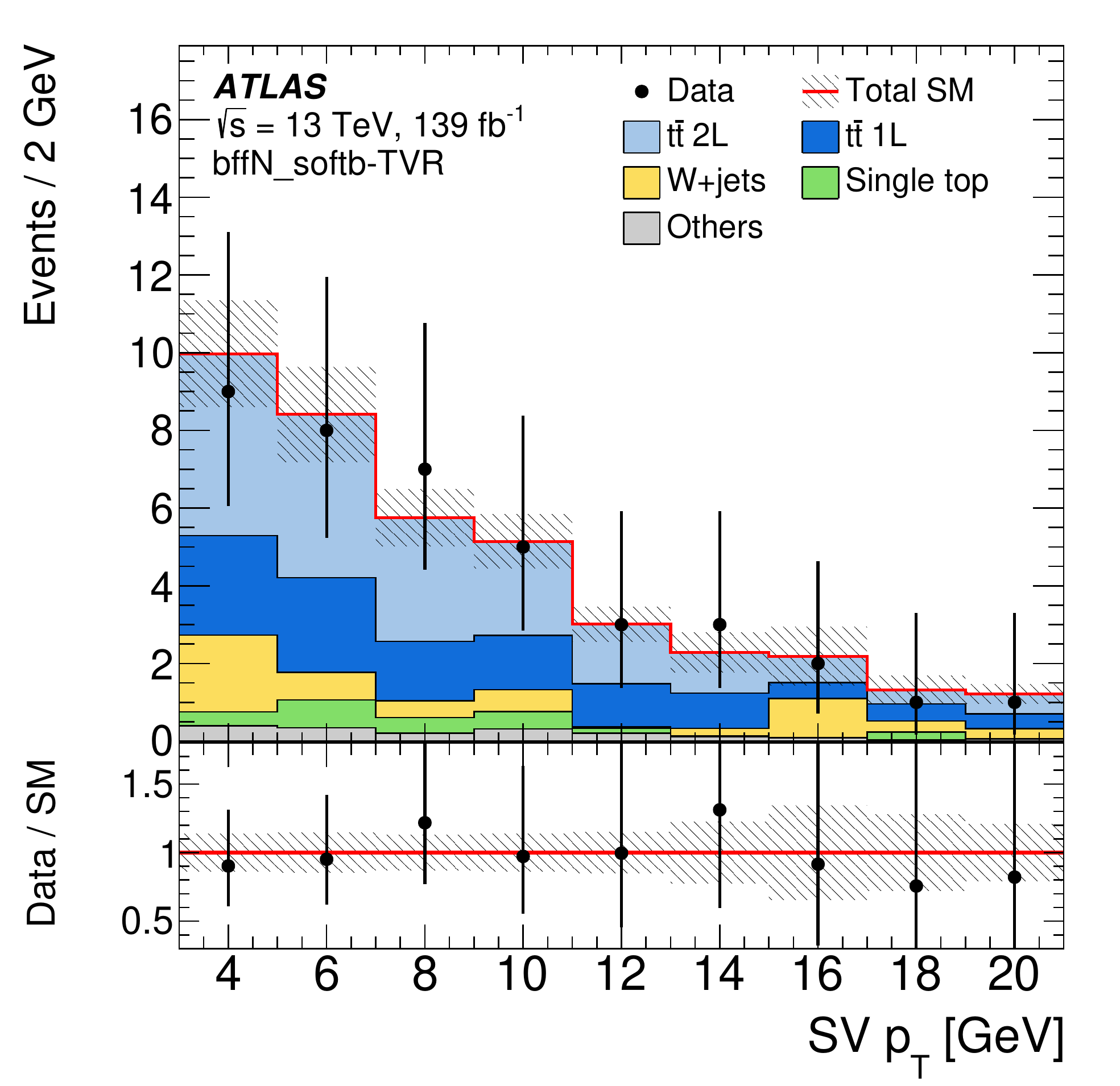}}
\subfigure[]{\includegraphics[width=0.45\textwidth]{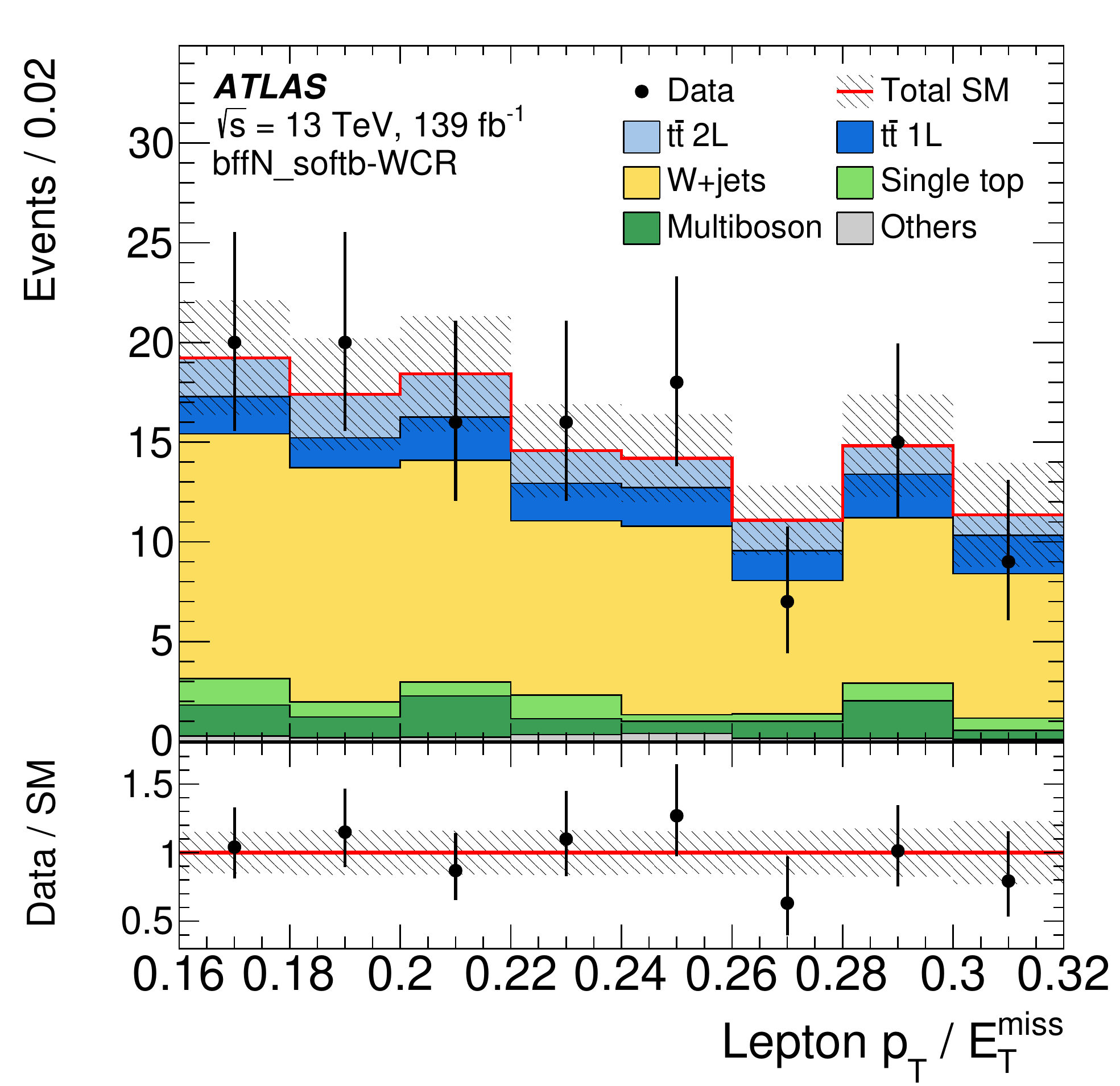}}
\subfigure[]{\includegraphics[width=0.45\textwidth]{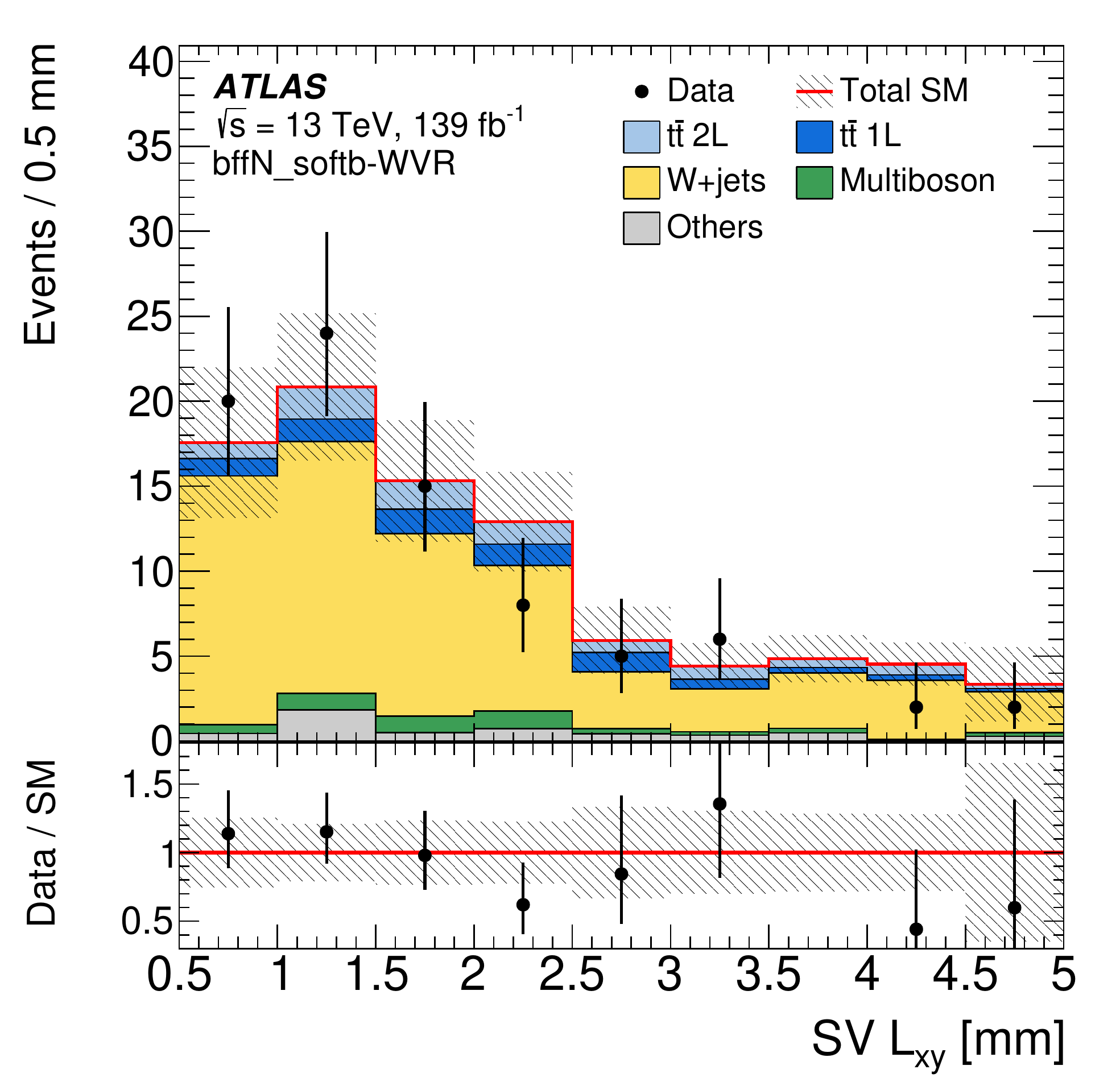}}
\caption{Selected kinematic distributions in \bffNsoftb\ CRs and VRs: (a) lepton \pT in the TCR, (b) transverse component of the total track momentum attached to the secondary vertex, SV \pT, in the TVR, (c) \lepPtoverMET\ in the WCR, (d) distance from the primary vertex to the secondary vertex in the transverse plane, SV $L_{xy}$, in the WVR. The distributions shown are post-fit, i.e.\ each background is scaled by a normalisation factor obtained from a background-only likelihood fit to the CRs (see Table~\ref{tab:yields}). The hatched area around the total SM prediction and the hatched band in the Data/SM ratio include all statistical and systematic uncertainties. The last (first) bin contains overflows (underflows).}
\label{fig:bkg:bffN_softb}
\end{figure}

\begin{figure}[ht] \centering
\subfigure[]{\includegraphics[width=0.45\textwidth]{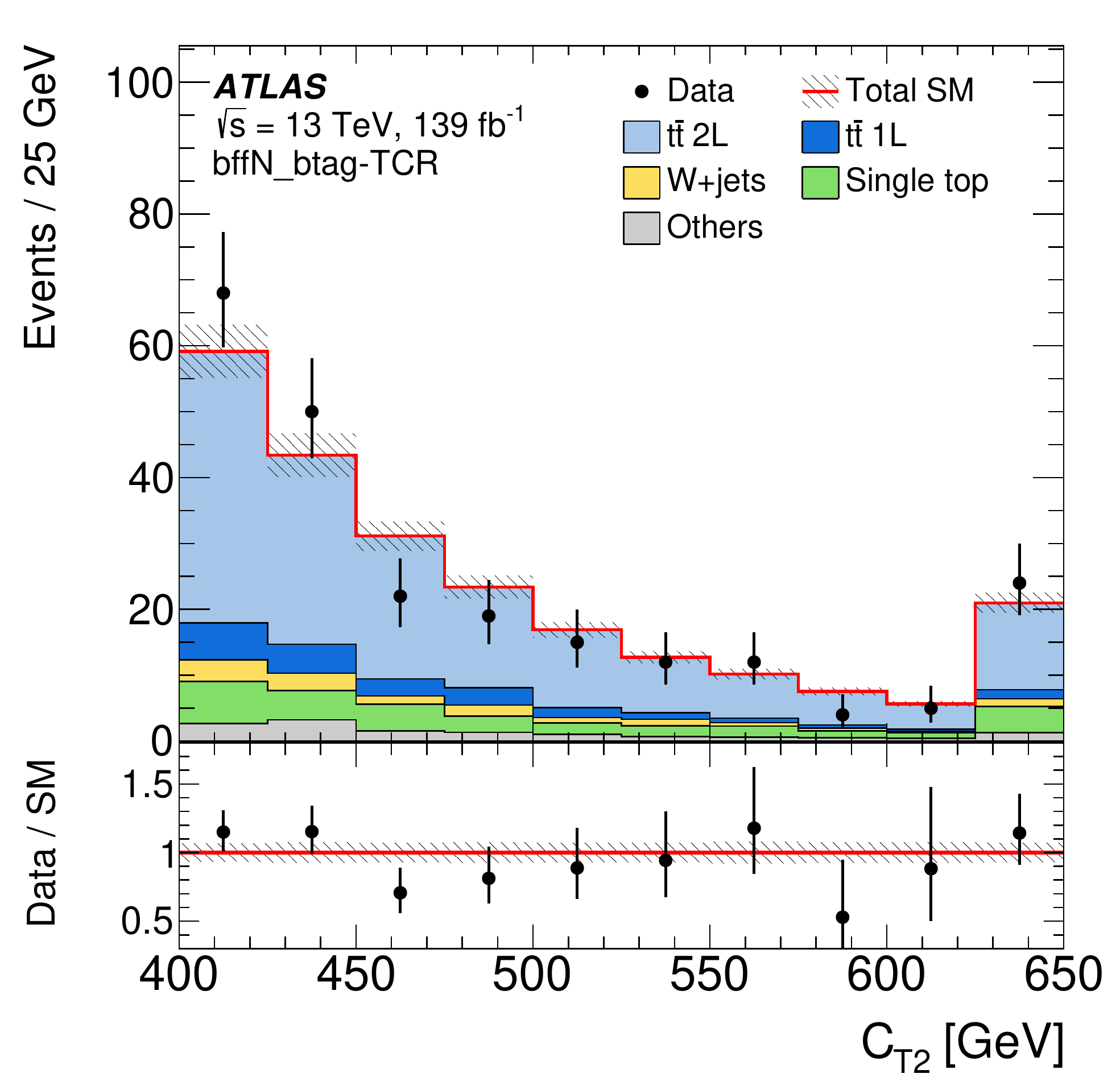}}
\subfigure[]{\includegraphics[width=0.45\textwidth]{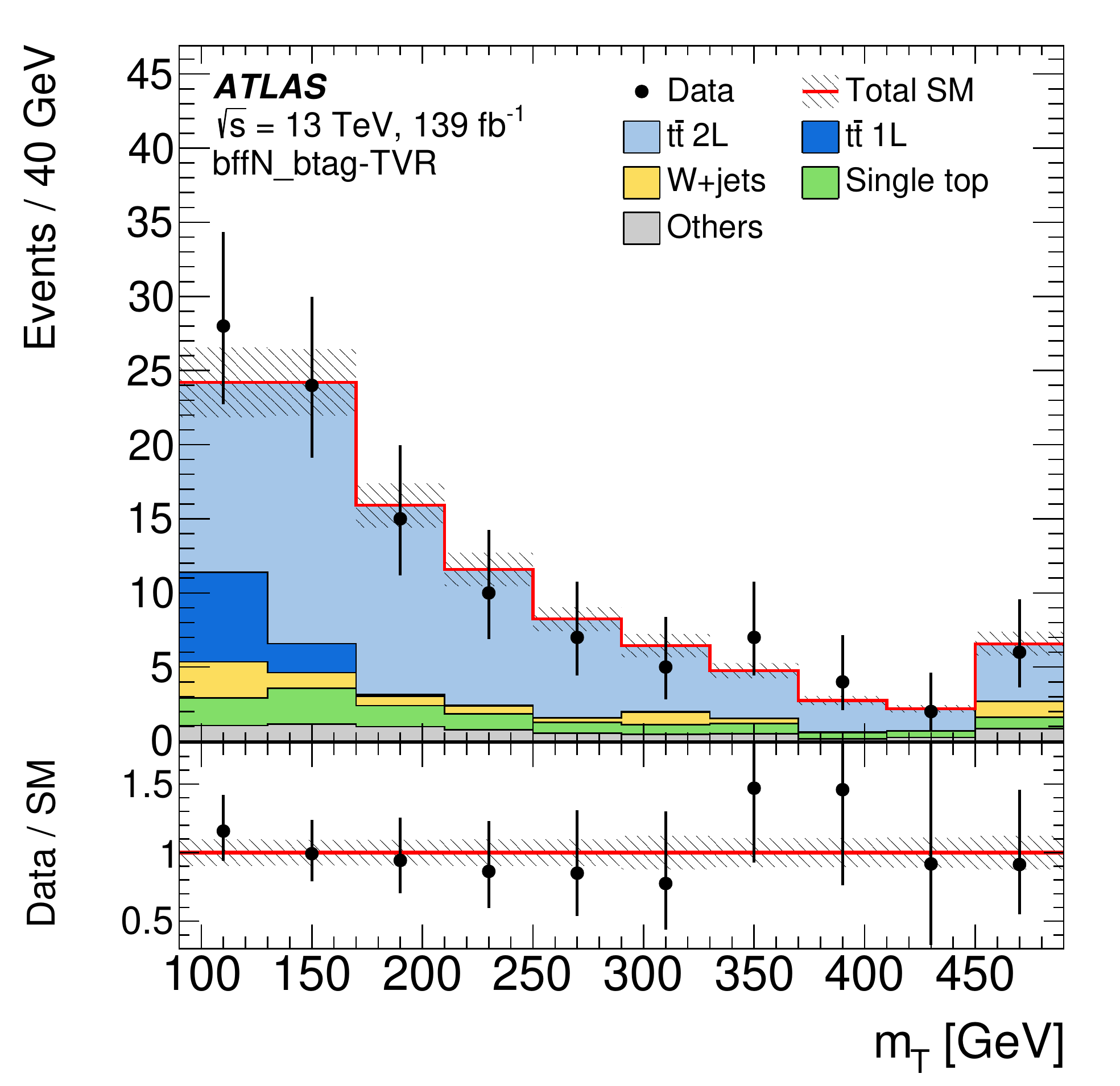}}
\subfigure[]{\includegraphics[width=0.45\textwidth]{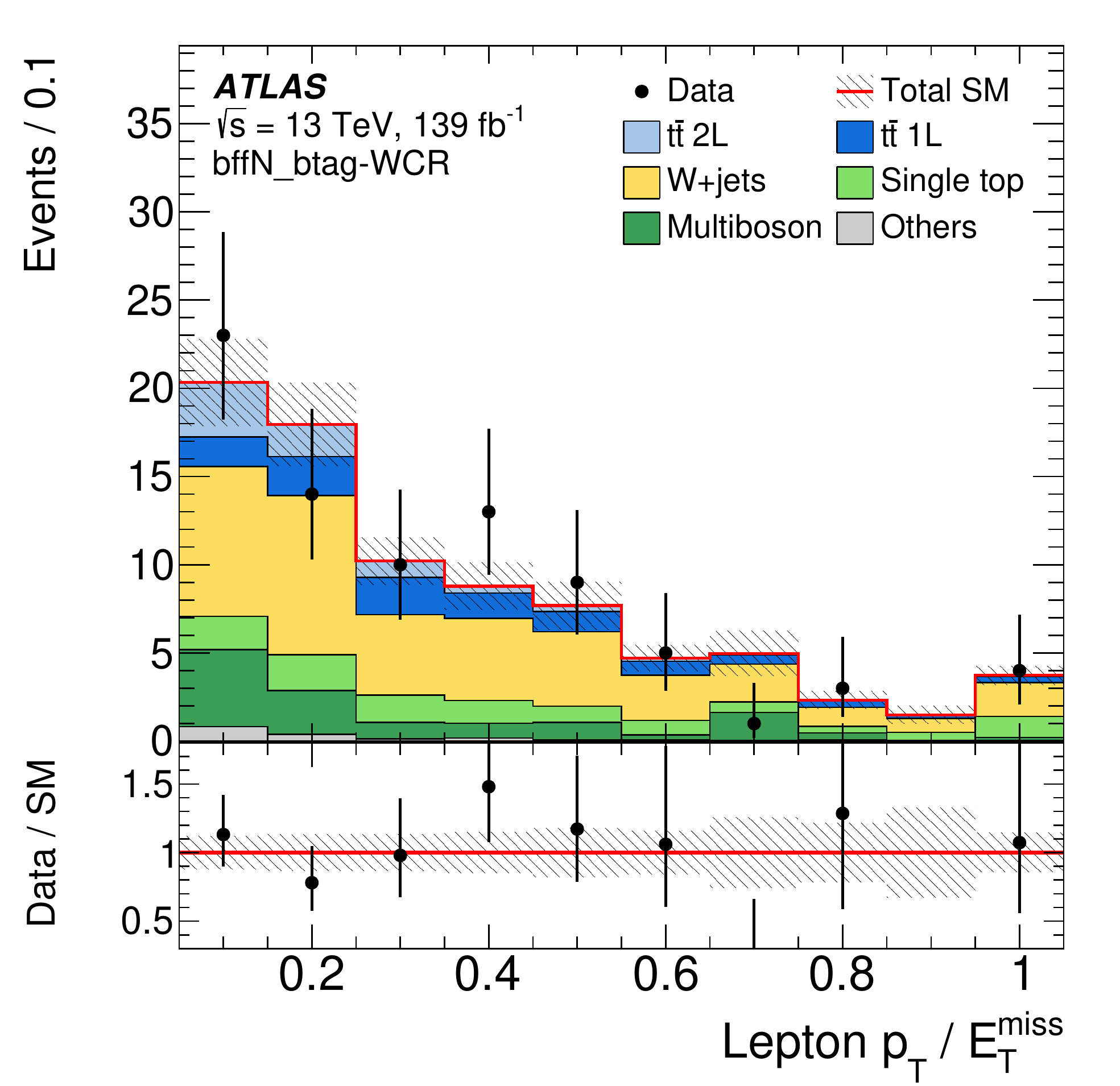}}
\subfigure[]{\includegraphics[width=0.45\textwidth]{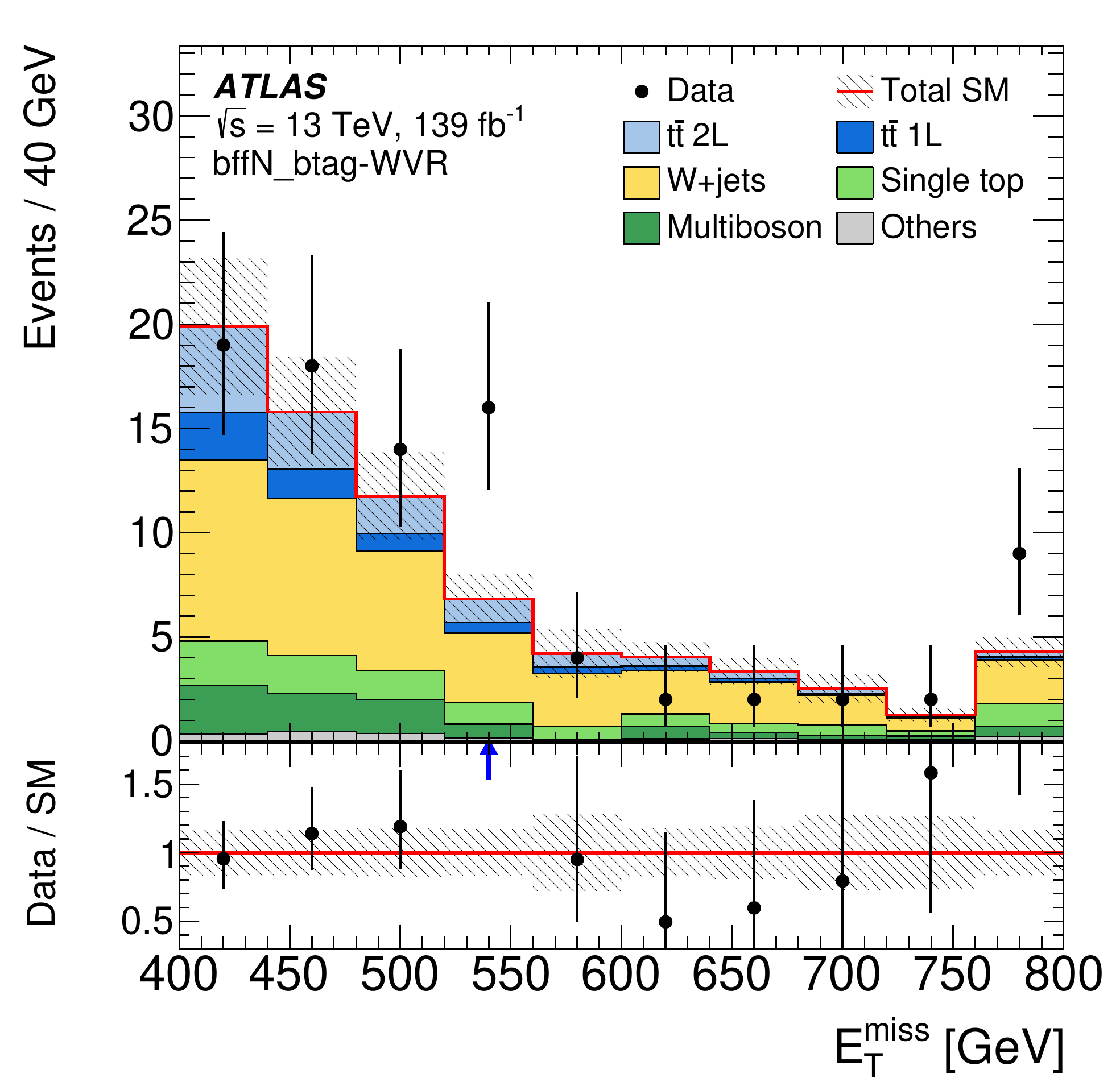}}
\caption{Selected kinematic distributions in \bffNbtag\ CRs and VRs: (a) \CTtwo\ in the TCR, (b) \mt  in the TVR, (c) \lepPtoverMET\ in the WCR, (d) \met in the WVR. The distributions shown are post-fit, i.e.\ each background is scaled by a normalisation factor obtained from a background-only likelihood fit to the CRs (see Table~\ref{tab:yields}). The hatched area around the total SM prediction and the hatched band in the Data/SM ratio include all statistical and systematic uncertainties. The last (first) bin contains overflows (underflows).}
\label{fig:bkg:bffN_btag}
\end{figure}

Figures~\ref{fig:bkg:bffN_softb} and~\ref{fig:bkg:bffN_btag} compare data and predictions in the \bffNsoftb\ and
\bffNbtag\ CRs and VRs. Good agreement is observed between data and prediction, within uncertainties.
 
\FloatBarrier

% End of text imported from the .//texfiles/backgrounds.tex input file
 
\clearpage
 
% The next lines are included from the .//texfiles/systematics.tex input file
\section{Systematic uncertainties}
\label{sec:systs}
 
The systematic uncertainties in the background estimates arise from multiple experimental and theoretical sources and can
enter the SR background yield either via direct predictions from theoretical cross-sections or from uncertainties in the
extrapolation from CRs to SRs. The sources of systematic uncertainties are grouped into categories whose labels are
defined in parentheses in the paragraphs below. Their effect on the background predictions in the SRs is summarised in
Table~\ref{tab:systematics:summary}. The systematic uncertainties are included as nuisance parameters constrained by Gaussian probability
distributions and profiled in the likelihood fits.
 
Experimental uncertainties arise from imperfect knowledge of the jet energy scale ({\em JES}), jet energy resolution
({\em JER})~\cite{PERF-2016-04}, scale and resolution of the \met soft term ({\em \met experimental})~\cite{PERF-2016-07},
as well as the modelling of the $b$-tagging or soft $b$-tagging efficiencies and mis-tag rates~\cite{FTAG-2018-01}
({\em $b$-tagging experimental}). Other experimental uncertainties arise from the modelling of the lepton
energy scales, energy resolutions, reconstruction and identification efficiencies ({\em Leptons experimental}). There
is also an experimental uncertainty arising from the reweighting of the simulation as a function of the number of
interactions per bunch crossing in data and the additional cuts applied to jets to ensure they arise from the hard-scatter
primary vertex ({\em Pile-up}).
 
Backgrounds such as dibosons and \Zjets, derived directly from a MC prediction and a theoretical cross-section,
have theoretical systematic uncertainties ({\em Theory}) arising from theoretical cross-section calculations, including those
related to parton distribution functions and factorisation and normalisation scales. The systematic uncertainty
on the integrated luminosity is also included in this category. As shown in Table~\ref{tab:CROverview},
the single-top, \ttZ\ and \Wjets\ backgrounds are also predicted directly from MC simulations for some SRs,
in which case the theory uncertainties apply also to those processes.
 
When the yield from a background such as \ttbar, single-top, \ttV\ or \Wjets\ is normalised using a CR, modelling
uncertainties affect the extrapolation from the control to the signal region, but not the overall normalisation. In
each of these cases, the background has a normalisation systematic uncertainty ({\em Normalisation}) from the fit,
arising from the statistical power of the CR for the given background and a modelling uncertainty ({\em Modelling})
that affects the extrapolation factor from the CR to the SR.
 
The uncertainties in the modelling of the \ttbar background include effects related to the MC event generator, the
hadronisation modelling and the amount of initial- and final-state radiation~\cite{ATL-PHYS-PUB-2017-007}. The MC
generator uncertainty is estimated by taking the full difference in event yields between \POWHEGBOXV{v2}\unskip+\PYTHIAV{8} and
\MGMCatNLOV{v2.6.0}\unskip+\PYTHIAV{8}. Events generated with \POWHEGBOXV{v2} are showered and subsequently hadronised with
either \PYTHIAV{8} or \HERWIGV{7.0} in order to estimate the effect from modelling of the hadronisation. The systematic
uncertainty from the amount of initial- and final-state radiation is derived by comparing \POWHEGBOX results obtained with
different shower radiation, NLO radiation and modified factorisation and renormalisation scales.
 
The single-top $Wt$ process modelling uncertainty is derived from the size of the interference between \ttbar and $Wt$
using the \ttbar, $Wt$ and $WWbb$ samples generated with \madgraph.
It is obtained by comparing $Wt$ with the difference between $WWbb$ and \ttbar.
The $Wt$ sample generated with \madgraph\ is found to be in
good agreement with the nominal samples generated with \POWHEGBOXV{v2}\unskip+\PYTHIA~8. For the \tNmed\ and \tNhigh\ SRs where STCR
is used, the $Wt$ modelling uncertainty enters via the ratio of the number of $Wt$ events in the signal region to the number in the STCR.
Given the potentially large modelling uncertainty in the interference between \ttbar and $Wt$, the modelling uncertainty
is also evaluated for the \DM\ SR by comparing the predicted single-top yield from $Wt$ with the difference between
$WWbb$ and \ttbar.

The modelling uncertainties considered for \ttZ\ are the renormalisation and factorisation scales, and the amount of initial- and
final-state radiation, obtained by considering the variation of the same parameters used for the \ttbar  initial- and final-state
radiation systematic uncertainties. The \Wjets\ modelling uncertainties include generator modelling, derived by considering an alternative
\Wjets\ sample generated with \madgraph\ as well as modified factorisation, renormalisation, resummation and parton matching scales.
 
Most of the SRs are binned in one or two variables in order to enhance sensitivity to a wider range of models for
exclusion limits. In this situation the normalisation factors to go from the CR to the SR are rederived specifically for
each bin of the SR. The modelling systematic uncertainties are also rederived following the scheme above but applied to the
normalisation factor from the CR to each specific bin of the SR.
 
The SUSY signal cross-section uncertainty is taken from an envelope of cross-section predictions using different PDF sets
and factorisation and renormalisation scales as described in Ref.~\cite{Borschensky:2014cia}. The uncertainty in the DM production
cross-section is derived from the scale variations and PDF choices. Dedicated uncertainties in the SUSY and DM signal acceptance
due to the modelling of additional radiation, factorisation, renormalisation and parton matching scales are considered.
The total systematic uncertainty for the SUSY models varies between 9\% and 35\%, increasing at higher stop mass and at lower
values of $\Delta m_{\stopone,\ninoone}$. For the spin-0 mediator signals the total systematic uncertainty is between 15\% and 18\%.
 
\begin{table}[bt]
\begin{center}
\caption{Summary of the dominant systematic uncertainties as a percentage of the total predicted background yields in the SRs, obtained from the background-only fits described in Section~\ref{sec:results}. }
\vspace{3mm}
\renewcommand{\arraystretch}{1.5}
{\scriptsize
\begin{tabular}{ l  S  S  S  S  S  S  S  S}
\toprule
SR Uncertainty [\%]            & \tNmed & \tNhigh & \tNdiaglow   & \tNdiaghigh & \bWN   & \bffNbtag & \bffNsoftb & \DM \\
\midrule
\ttbar normalisation      &   4.4  &    2.7  & 12.3         &  15.8       &  7.8   &   6.6     &   3.7     &  3.0 \\
\ttZ\ normalisation      &   9.0  &    6.8  & {--}         &  {--}       &  {--}  &  {--}     &  {--}     &  8.5 \\
\Wjets\ normalisation      &   3.0  &    4.8  & {--}         &  {--}       &  {--}  &   5.5     &  11.1     & {--} \\
$Wt$ normalisation      &   2.8  &    3.4  & {--}         &  {--}       &  {--}  &  {--}     &  {--}     & {--} \\
\midrule
\ttbar modelling      &   3.0  &    9.1  & 18.4         &  29.3       & 17.6   &    3.1    &   3.3     &  4.1 \\
\ttZ\ modelling      &   7.7  &    7.1  & {--}         &  {--}       &  {--}  &   {--}    &   {--}    &  7.4 \\
\Wjets\ modelling      &   2.3  &    3.8  & {--}         &  {--}       &  {--}  &    4.3    &   9.7     & {--} \\
$Wt$ modelling      &   0.5  &    0.8  & {--}         &  {--}       &  {--}  &   {--}    &   {--}    &  6.4 \\
\midrule
JER      &  10.9  &    5.1  &  4.1         &   5.0       &  6.1   &    1.8    &    7.6    &  6.5 \\
\met experimental      &   0.7  &    0.4  &  1.0         &   0.2       &  0.9   &    2.4    &    3.4    &  0.1 \\
$b$-tagging experimental      &   1.7  &    3.6  &  1.3         &   0.9       &  1.6   &    1.9    &    3.2    &  3.1 \\
JES      &   6.0  &    2.5  &  4.7         &   4.1       &  2.7   &    6.1    &   11.7    &  2.4 \\
Leptons experimental      &   1.0  &    1.6  &  1.3         &   0.3       &  0.1   &    2.3    &    4.9    &  0.6 \\
Pile-up      &   1.1  &    1.2  &  1.2         &   0.3       &  0.8   &    1.0    &    2.1    &  1.1 \\
\midrule
Theory      &   0.9  &    1.3  &  1.4         &   0.5       &  4.8   &    3.8    &    3.7    &  0.7 \\
MC statistics      &   4.1  &    6.6  &  5.8         &   3.5       &  3.1   &    4.9    &   17.2    &  3.2 \\
\midrule
Total      &  19    &   17    & 24           &  33         & 20     &   12      &   27      & 15  \\
\bottomrule
\end{tabular}
}
\label{tab:systematics:summary}
\end{center}
\end{table}

% End of text imported from the .//texfiles/systematics.tex input file
 
% The next lines are included from the .//texfiles/results.tex input file
\section{Results}
\label{sec:results}
 
To determine the SM background yields in the SRs, a background-only likelihood fit is performed for
each analysis. The fit does not use the signal region data, but only the dedicated CRs to normalise the
backgrounds.
 
\begin{figure}[h!]
\begin{center}
\includegraphics[width=1.0\textwidth]{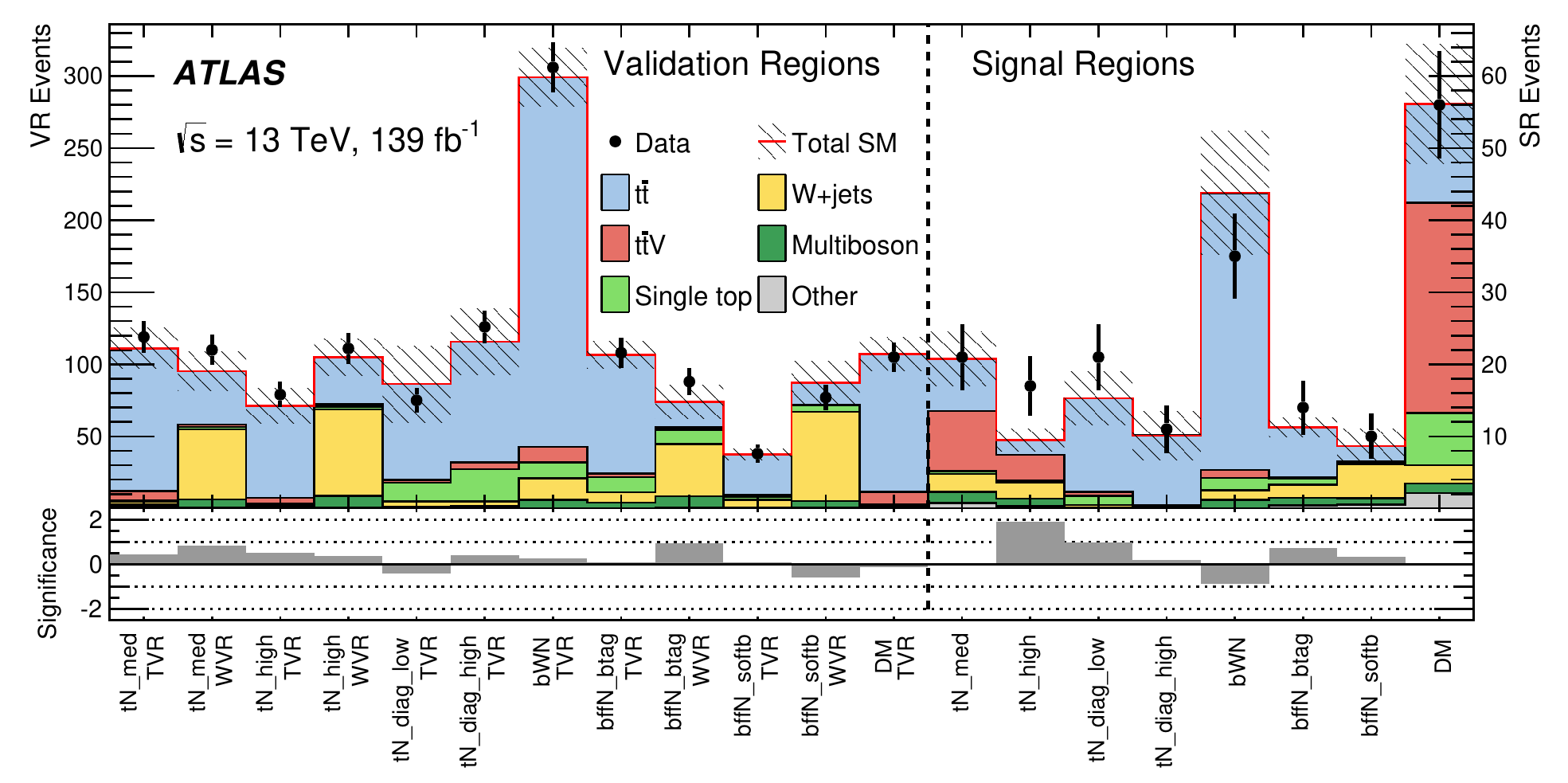}
\caption{The upper panel shows the comparison between the observed data ($n_{\textnormal{obs}}$) and
the predicted SM background ($n_{\textnormal{exp}}$) in all VRs and SRs.
The background predictions are obtained using the background-only fit, and the
hatched area around the SM prediction includes all uncertainties.
The bottom panel shows the $Z$ significance of the observed number of events given the
SM expectation.}
\label{fig:VR_PullPlot}
\end{center}
\end{figure}

The number of observed events and the predicted number of SM background events from the background-only
fits in all VRs and SRs are shown in Figure~\ref{fig:VR_PullPlot} together with the $Z$ significance of
the observation. The SRs are not mutually exclusive and are therefore not statistically independent. In
all SRs, the distributions indicate good agreement between the data and the SM background estimate. The
largest excess over the background-only hypothesis is $1.9 \sigma$ observed in the \tNhigh\ SR.
 
The number of observed events together with the predicted number of SM background events in all SRs
are summarised in Table~\ref{tab:yields}, showing the breakdown of the various backgrounds that contribute to
the SRs. The table also lists the results for the fit parameters that control the normalisation of the
main backgrounds (normalisation factors, NFs),\footnote{The \ttbar\ NFs in the \tNdiaglow, \bffNbtag,
and \bffNsoftb\ SRs are applied to semileptonic and dileptonic \ttbar\ events while all other SRs apply
the \ttbar\ NFs to the dileptonic component only.} together with the associated fit uncertainties including
the theoretical modelling uncertainties. To quantify the level of agreement of the SM background-only
hypothesis with the observations in the SRs, a profile-likelihood-ratio test~\cite{Cowan:2010js} is performed.
The resulting $p$-values ($p_{0}$) are also presented in the table together with the $Z$ significances.
For SRs with an observed number of events below the SM prediction, the $p_0$ values are capped at 0.5.
Model-independent upper limits on
beyond-the-SM contributions are derived for each SR. A generic signal model is assumed that contributes
only to the SR and for which neither experimental nor theoretical systematic uncertainties except for the
luminosity uncertainty are considered. All limits are calculated using the CL$_{\text{s}}$ prescription~\cite{Read:2002hq}.
The NFs are compatible with unity in most cases. One exception is for the single-top NFs in the \tNmed\ and \tNhigh\ SRs.
The single-top NFs are significantly below unity when using the DR scheme for the treatment of the
interference between the $Wt$ and \ttbar\ processes.
When changing to the DS scheme, the NFs become larger than unity but the predicted number of single-top
events in the signal regions after the fit does not change significantly. This is explained by the fact
that the DS and DR schemes give the same SR to STCR event yield ratio, within uncertainties.
The \ttbar\ NFs in \tNdiaghigh, \bffNbtag\ and \bffNsoftb\ are below unity,
which could potentially point to some mismodelling in this extreme region of phase-space.
But good agreement is seen in the \ttbar\ VRs, giving confidence in the \ttbar\ background estimates.
 
\begin{table}
\begin{center}
\caption{The number of observed events in the various SRs together with the expected numbers of background events and their uncertainties as predicted by the background-only fits, the normalisation factors for the background predictions obtained in the fit, the probabilities (represented by $p_0$ and $Z$ values) that the observed numbers of events are compatible with the background-only hypothesis, and the expected ($N_{\textnormal{non-SM}}^{\textnormal{limit}}$ exp.) and observed ($N_{\textnormal{non-SM}}^{\textnormal{limit}}$ obs.) 95\% CL upper limits on the number of beyond-SM events.
\label{tab:yields}}
\vspace{3mm}
\renewcommand{\arraystretch}{1.5}
\setlength\tabcolsep{3.0pt}
\sisetup{retain-explicit-plus}
\sisetup{round-mode = places}
{\scriptsize
\begin{tabular}{l c c c c c c c c}
\toprule
& tN\_med & tN\_high & tN\_diag\_low & tN\_diag\_high & bWN & bffN\_btag & bffN\_softb & DM\\
\midrule
Observed & $21$ & $17$ & $21$ & $11$ & $35$ & $14$ & $10$ & $56$\\
\midrule
Total SM & $21 \pm 4$ & $9.5 \pm 1.6$ & $15 \pm 4$ & $10.1 \pm 3.4$ & $44 \pm 9$ & $11.3 \pm 1.4$ & $8.7 \pm 2.3$ & $56 \pm 8$\\
\midrule
$t\bar{t}$ & $7.2 \pm 1.2$ & $2.0 \pm 1.0$ & $13.0 \pm 2.8$ & $9.6 \pm 2.6$ & $38 \pm 9$ & $6.9 \pm 1.1$ & $2.2 \pm 0.6$ & $14 \pm 4$\\
$t\bar{t}V$ & $8.3 \pm 2.5$ & $3.5 \pm 1.0$ & $0.55 \pm 0.17$ & $0.12 \pm 0.04$ & $1.1 \pm 1.1$ & $0.21 \pm 0.12$ & $0.06 \pm 0.04$ & $29 \pm 6$\\
Single top & $0.4^{+0.6}_{-0.4}$ & $0.27^{+0.34}_{-0.27}$ & $1.24 \pm 0.27$ & $0.26 \pm 0.06$ & $1.7 \pm 1.7$ & $0.8 \pm 0.5$ & $0.22 \pm 0.08$ & $7 \pm 4$\\
$W$+jets & $2.5 \pm 2.3$ & $2.3 \pm 1.0$ & $0.41 \pm 0.13$ & $0.080 \pm 0.020$ & $1.3 \pm 0.6$ & $1.9 \pm 0.8$ & $4.8 \pm 2.1$ & $2.56 \pm 0.24$\\
Multiboson & $1.49 \pm 0.21$ & $1.06 \pm 0.16$ & $0.070 \pm 0.020$ & $0.020 \pm 0.010$ & $1.22 \pm 0.30$ & $1.07 \pm 0.35$ & $0.89 \pm 0.32$ & $1.31 \pm 0.18$\\
Other & $0.78 \pm 0.06$ & $0.320 \pm 0.024$ & -- & -- & -- & $0.40 \pm 0.13$ & $0.52 \pm 0.19$ & $2.15 \pm 0.16$\\
\midrule
$t\bar{t}$ NF & $0.98^{+0.14}_{-0.12}$ & $0.90 \pm 0.12$ & $0.88^{+0.13}_{-0.12}$ & $0.73^{+0.14}_{-0.13}$ & $1.06 \pm 0.10$ & $0.80^{+0.09}_{-0.08}$ & $0.68 \pm 0.10$ & $1.12^{+0.15}_{-0.13}$\\
$t\bar{t}V$ NF & $0.95^{+0.22}_{-0.20}$ & $0.92 \pm 0.17$ & -- & -- & -- & -- & -- & $1.18^{+0.20}_{-0.18}$\\
Single top NF & $0.11^{+0.26}_{-0.11}$ & $0.12^{+0.22}_{-0.12}$ & -- & -- & -- & -- & -- & --\\
$W$+jets NF & $0.96^{+0.25}_{-0.23}$ & $0.86 \pm 0.17$ & -- & -- & -- & $0.83 \pm 0.28$ & $1.04^{+0.22}_{-0.20}$ & -\\
\midrule
$p_0\ (Z)$  & $0.49\ (0.03)$ & $0.01\ (2.20)$ & $0.17\ (0.95)$ & $0.31\ (0.50)$ & $0.50\ (0.00)$ & $0.20\ (0.84)$ & $0.26\ (0.64)$ & $0.50\ (0.00)$ \\
$N_{\textnormal{non-SM}}^{\textnormal{limit}}$ exp.  & $12.4^{+5.4}_{-2.6}$ & $9.8^{+2.8}_{-1.8}$ & $14.1^{+4.8}_{-3.3}$ & $9.7^{+3.7}_{-2.1}$ & $21.8^{+7.6}_{-7.9}$ & $8.6^{+3.6}_{-1.2}$ & $8.9^{+3.8}_{-2.0}$ & $27.4^{+7.6}_{-5.0}$ \\
$N_{\textnormal{non-SM}}^{\textnormal{limit}}$ obs.  & $12.9$ & $16.2$ & $17.7$ & $10.9$ & $15.3$ & $11.7$ & $10.5$ & $28.2$ \\
\bottomrule
\end{tabular}
}
\end{center}
\end{table}

\begin{figure}
\subfigure[]{\includegraphics[width=0.49\textwidth]{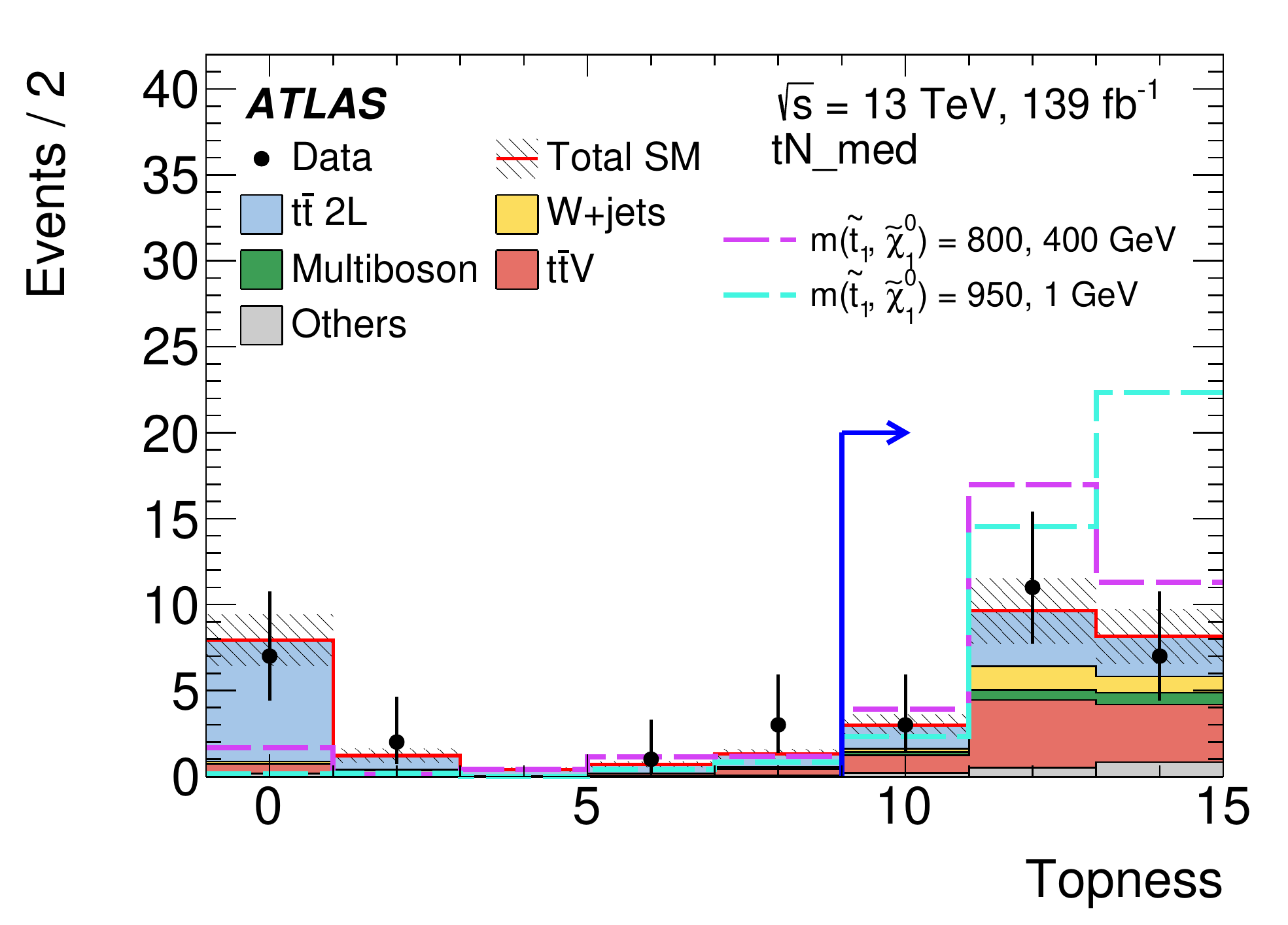}}
\subfigure[]{\includegraphics[width=0.49\textwidth]{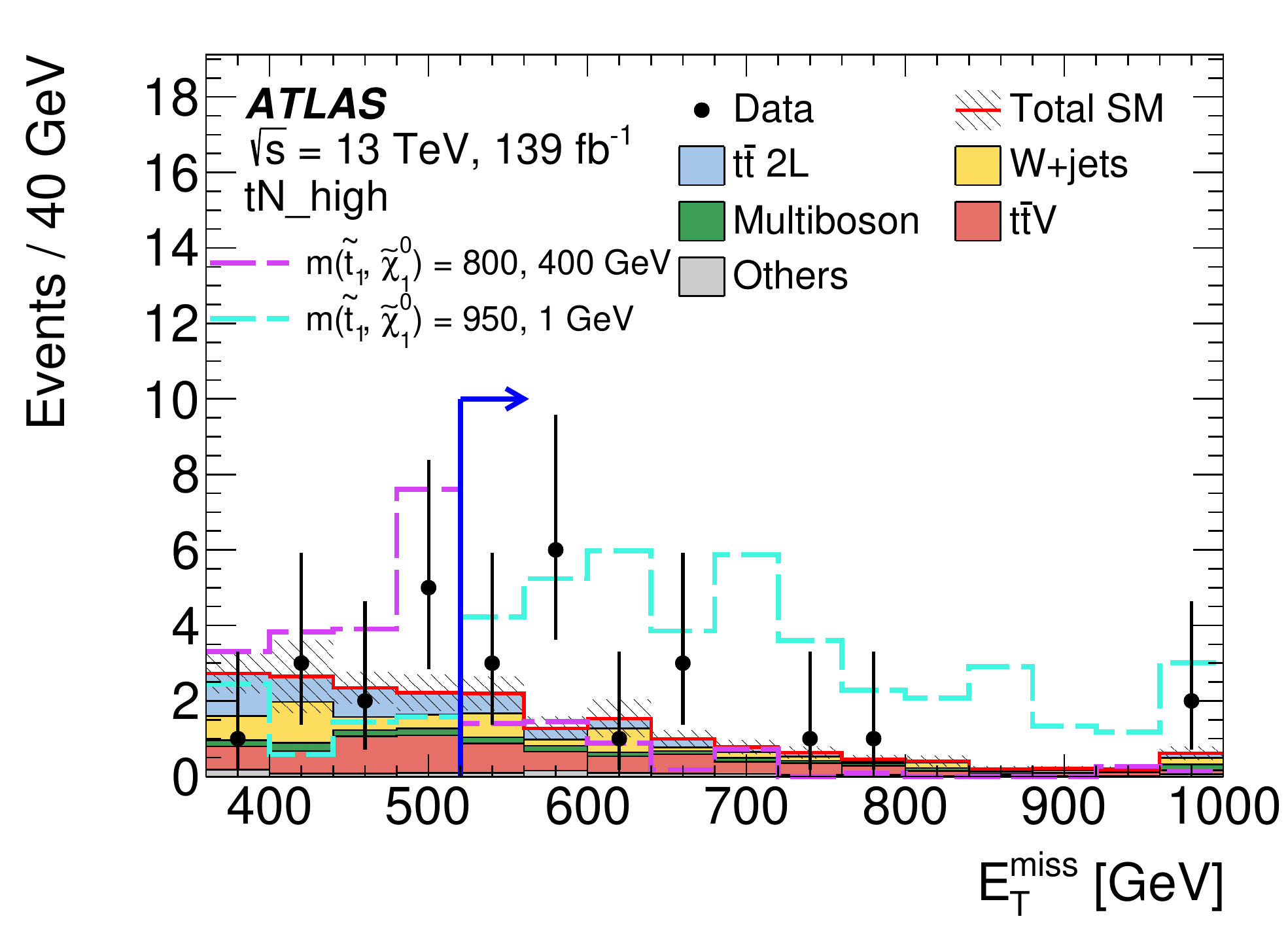}}
\subfigure[]{\includegraphics[width=0.49\textwidth]{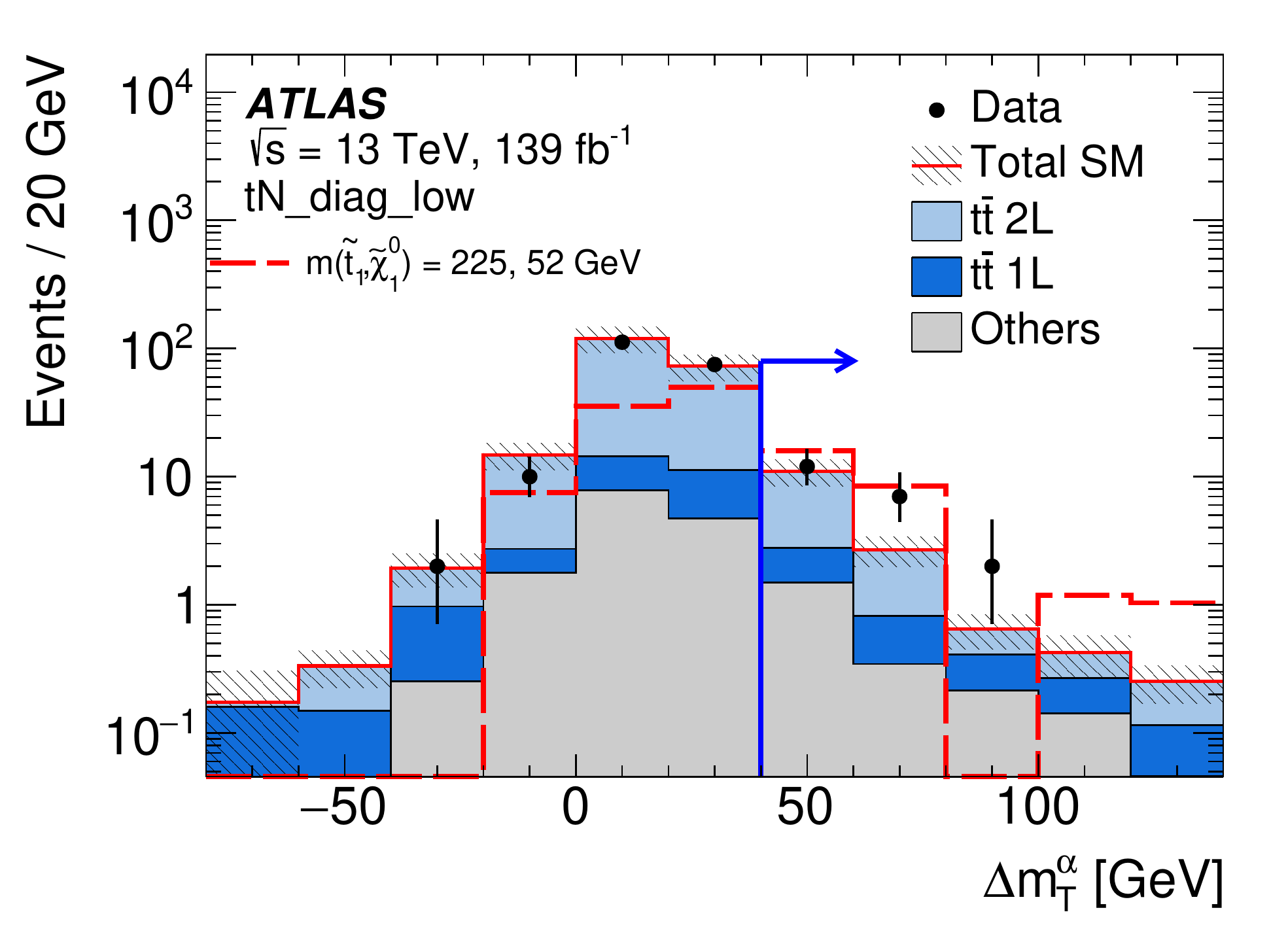}}
\subfigure[]{\includegraphics[width=0.49\textwidth]{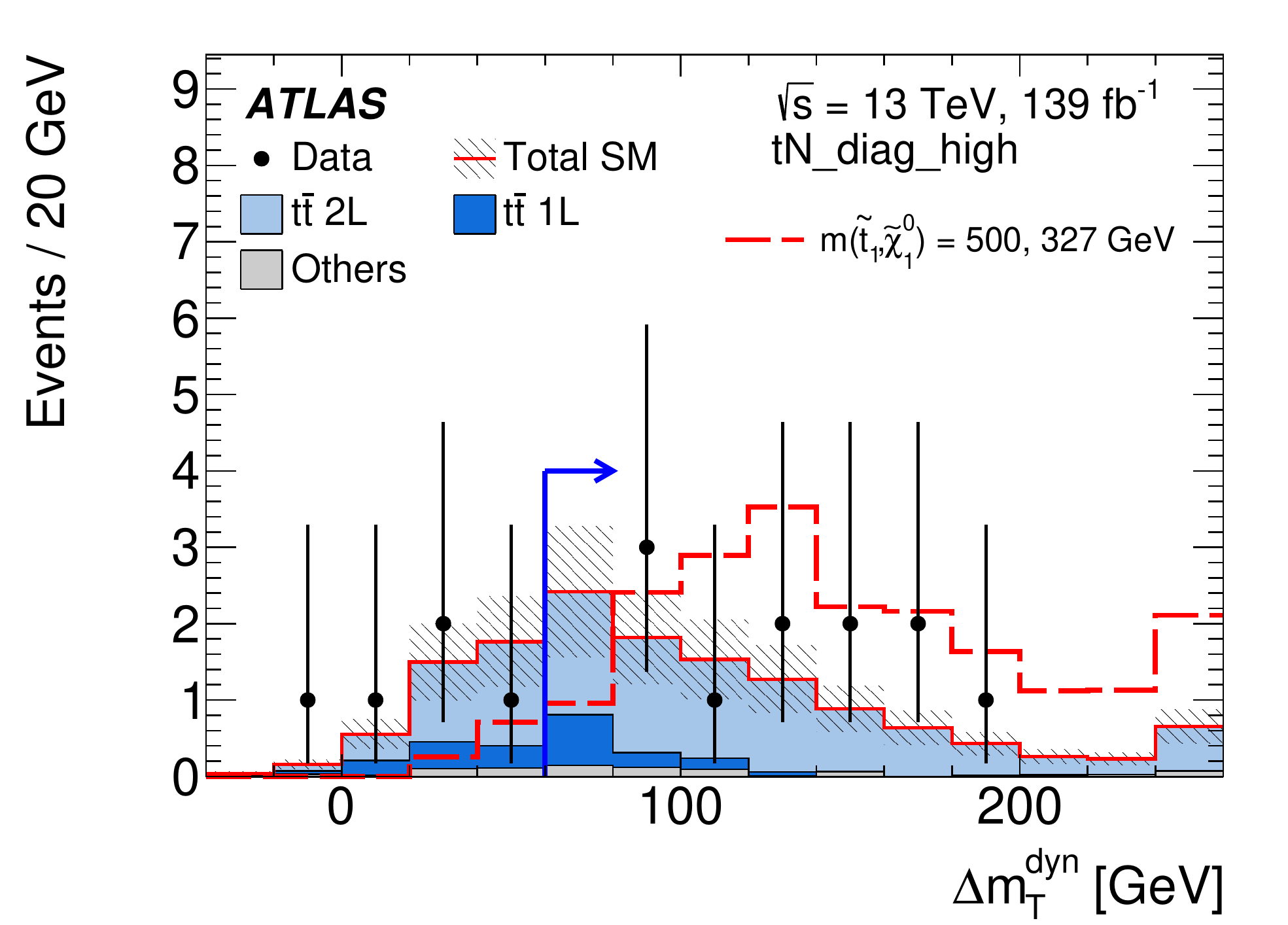}}
\caption{Kinematic distributions in the (a) \tNmed, (b) \tNhigh, (c) \tNdiaglow\ and (d) \tNdiaghigh\ SRs. The full event selection in the corresponding signal region is applied, except
for the requirement (indicated by an arrow) that is imposed on the variable being plotted. The distributions shown
are post-fit, i.e.\ each background is scaled by a normalisation factor obtained from a background-only likelihood
fit to the CRs (see Table~\ref{tab:yields}). In addition to the background prediction, a signal model is shown on
each plot. The hatched area around the total SM prediction includes statistical and experimental uncertainties. The
last (first) bin contains overflows (underflows).}
\label{fig:nminus1_1}
\end{figure}

\begin{figure}
\begin{center}
\subfigure[]{\includegraphics[width=0.49\textwidth]{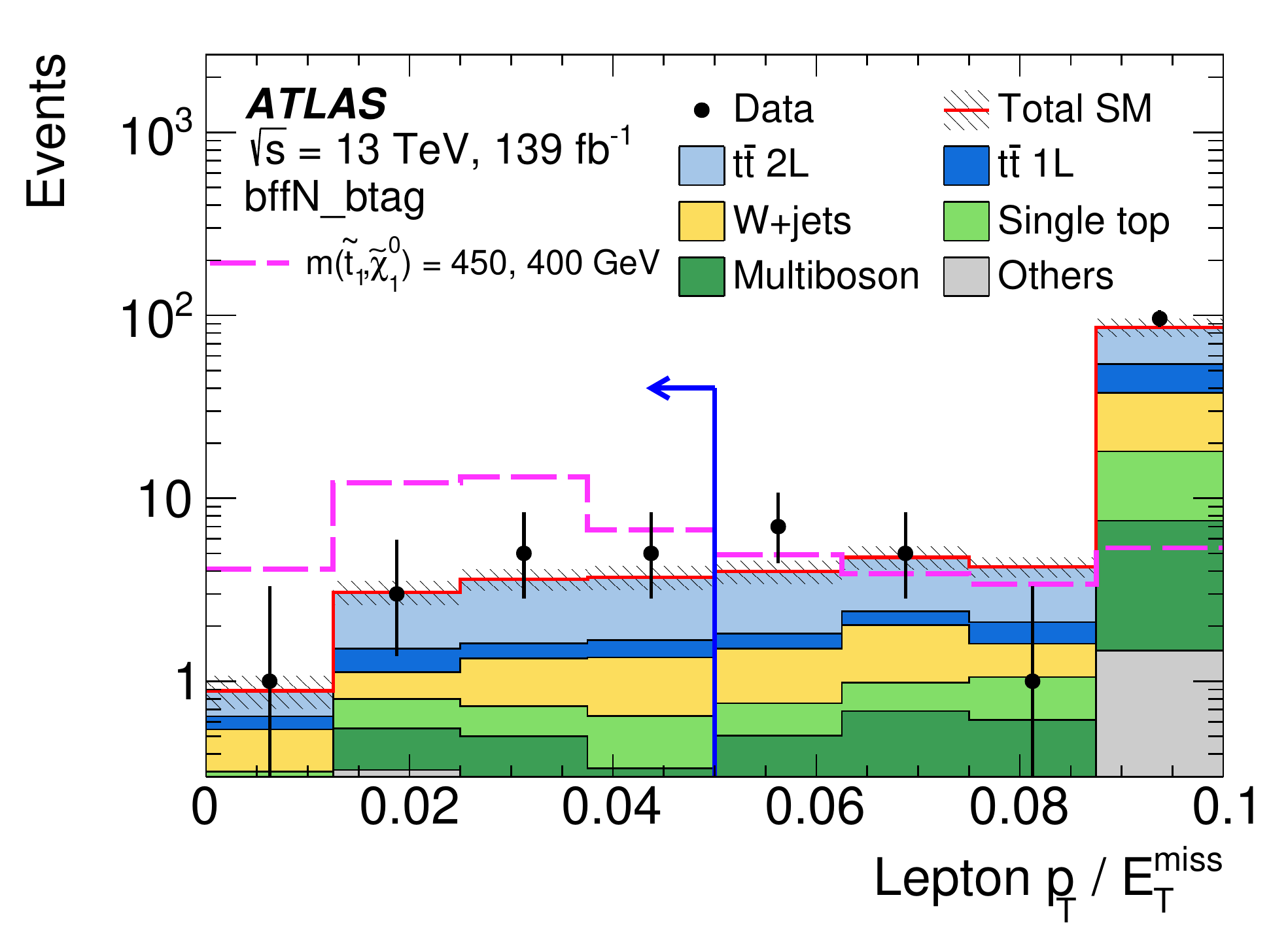}}
\subfigure[]{\includegraphics[width=0.49\textwidth]{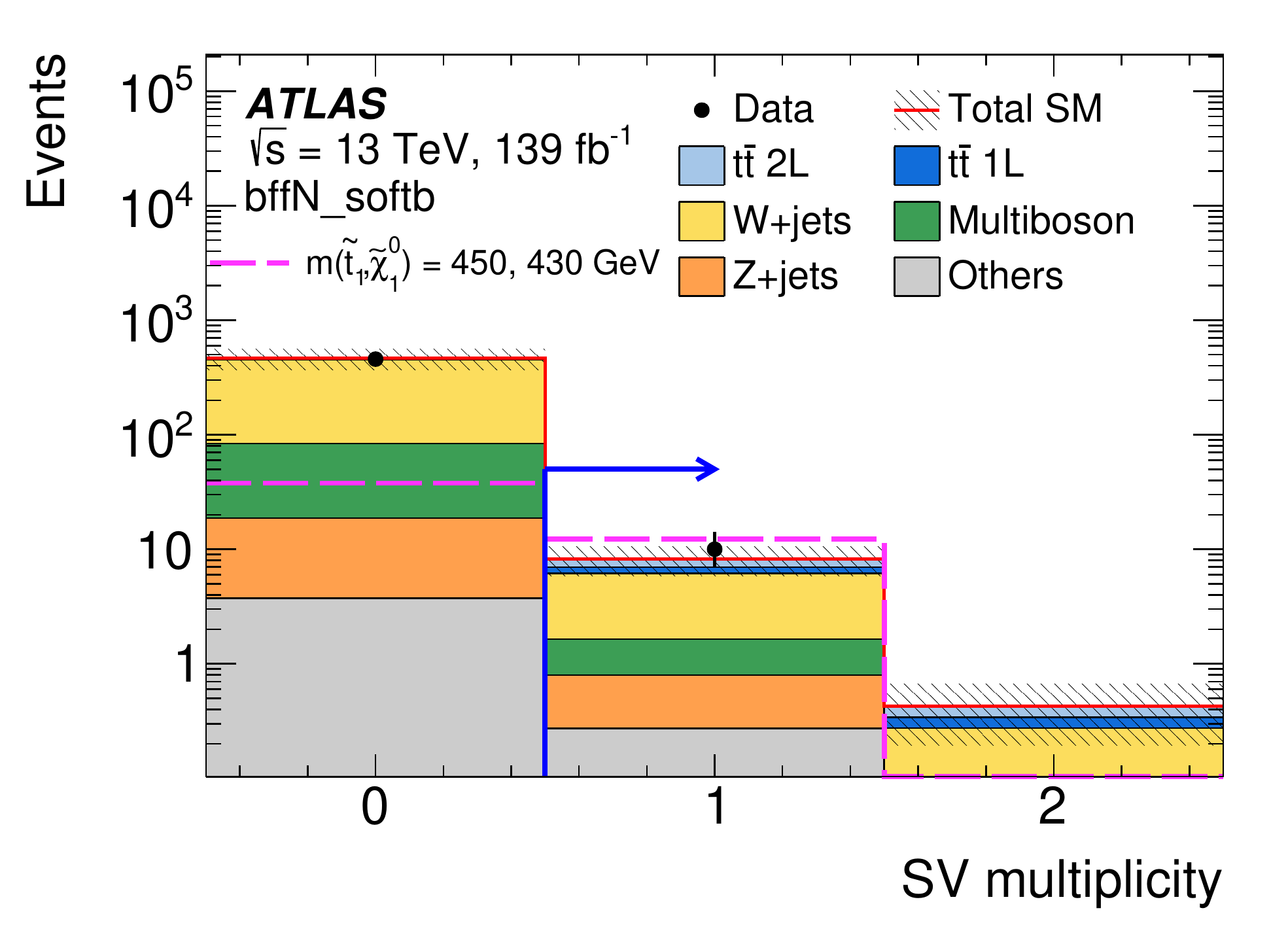}}
\subfigure[]{\includegraphics[width=0.49\textwidth]{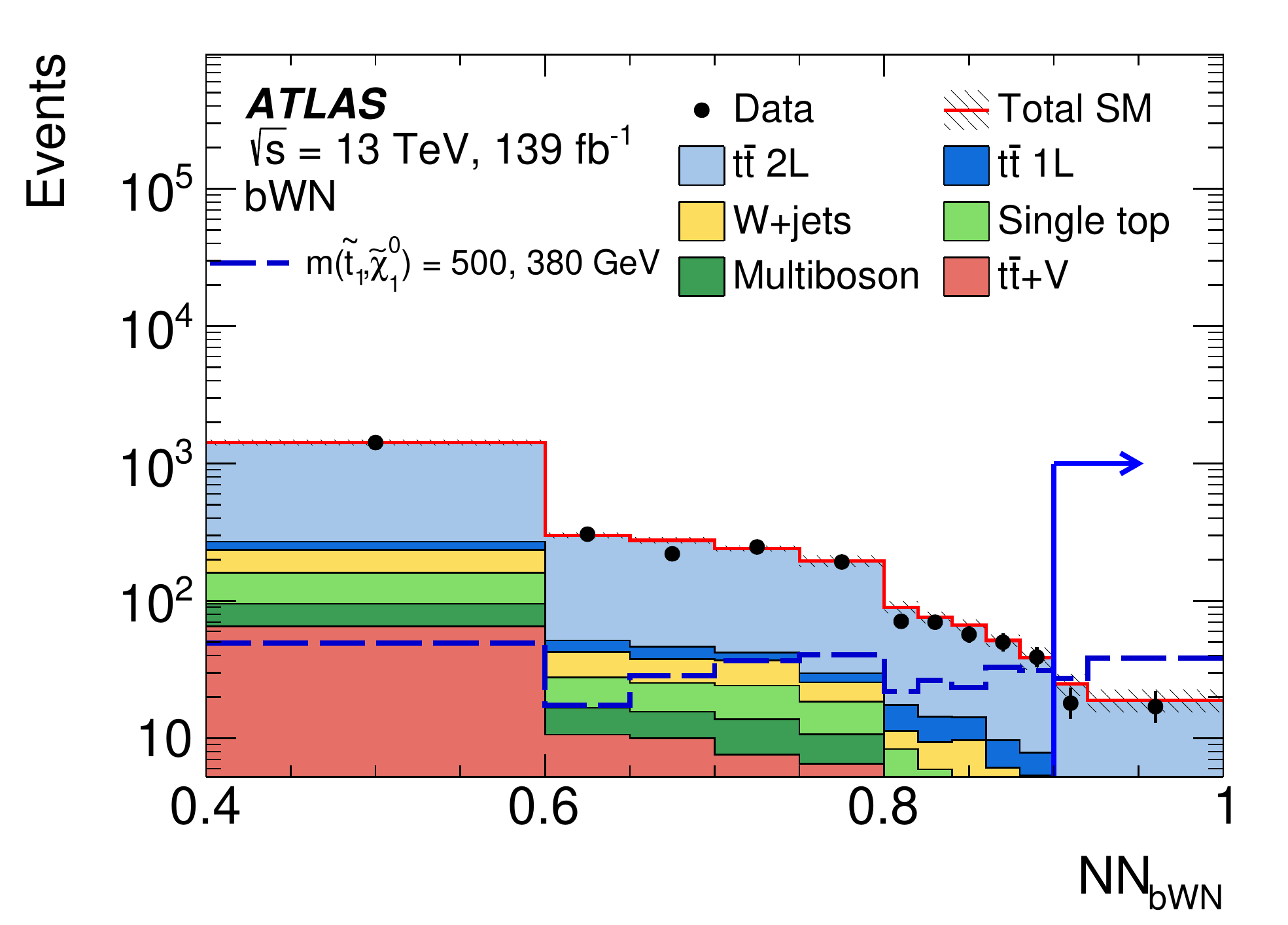}}
\subfigure[]{\includegraphics[width=0.49\textwidth]{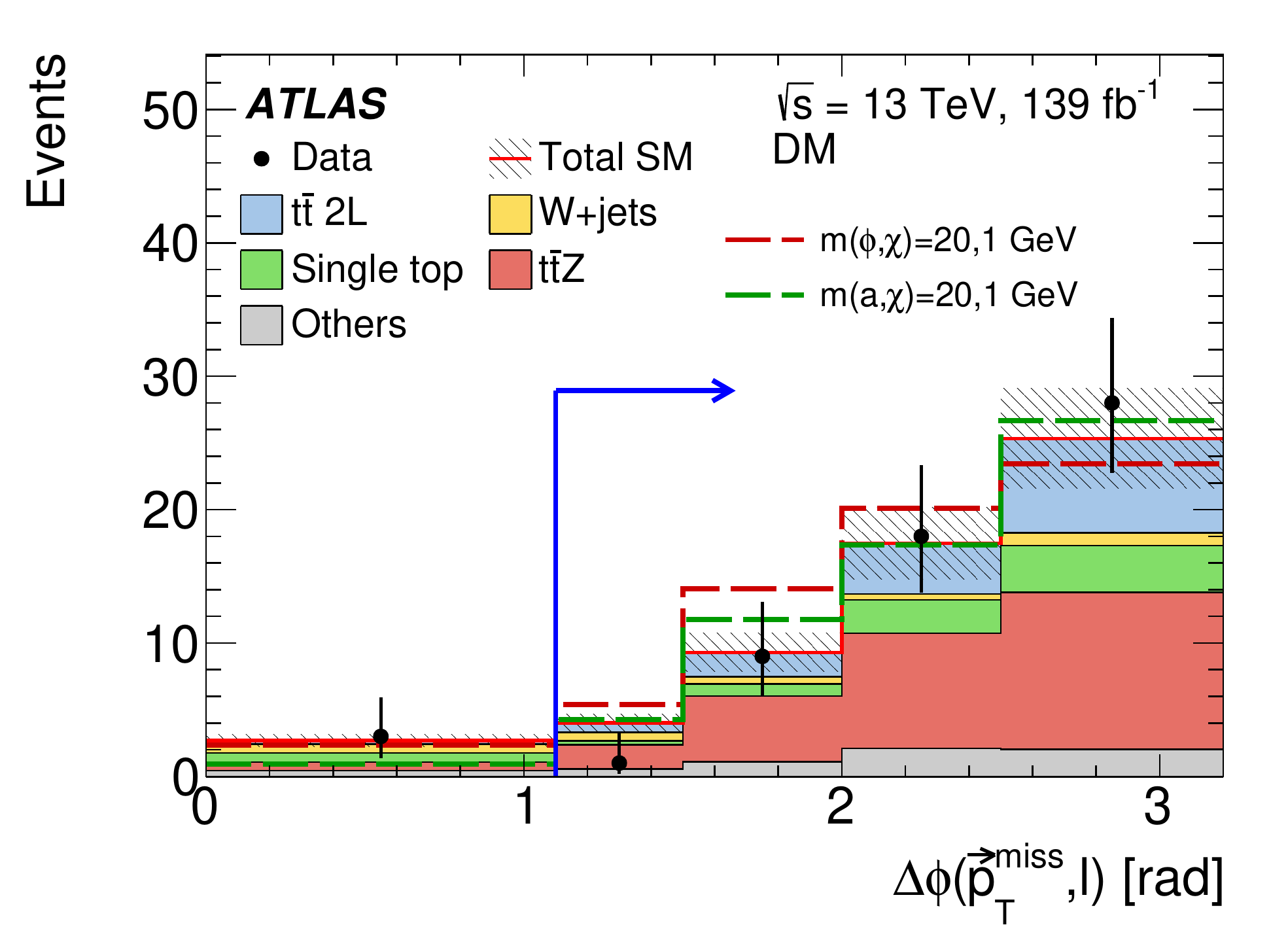}}
\end{center}
\caption{Kinematic distributions in the (a) \bffNbtag, (b) \bffNsoftb, (c) \bWN\ and (d) \DM\ SRs.
The full event selection in the corresponding signal region is applied, except for the requirement
(indicated by an arrow) that is imposed on the variable being plotted. In the \DM\ SR, the signal
is normalised under the assumption of the coupling strength $g$=1.0.
The distributions shown are post-fit, i.e.\ each background is scaled by a normalisation factor obtained
from a background-only likelihood fit to the CRs (see Table~\ref{tab:yields}). In addition to the
background prediction, a signal model is shown on each plot. The hatched area around the total SM
prediction includes statistical and experimental uncertainties. The last (first) bin contains
overflows (underflows).}
\label{fig:nminus1_2}
\end{figure}

Figures~\ref{fig:nminus1_1} and~\ref{fig:nminus1_2} show comparisons between the observed data and the SM background prediction
with all SR selections applied except the requirement on the plotted variable.
The expected distributions from representative signal benchmark models are overlaid.
 
\section{Interpretations}
\label{sec:results_limits}
 
No significant excess is observed, and exclusion limits based on profile-likelihood fits are set for the stop
pair production models and the spin-0 mediator models. Exclusion limits at 95\% confidence level (CL) are
obtained by selecting the signal region with the lowest expected CL$_{\text{s}}$ value for each signal model and the
exclusion contours are derived by interpolating in the CL$_{\text{s}}$ value. The signal uncertainties and potential
signal contributions to all regions are taken into account, and all uncertainties except those in the
theoretical signal cross-section are included in the fit. In all exclusion plots, the
$\pm 1 \sigma_{\textnormal{exp}}$ uncertainty band indicates how much the expected limit is affected by the
systematic and statistical uncertainties included in the fit. The $\pm 1 \sigma^{\textnormal{SUSY}}_{\textnormal{theory}}$
uncertainty lines around the observed limit illustrate the change in the observed limit as the nominal signal
cross-section is scaled up and down by the theoretical cross-section uncertainty.
 
\begin{figure}
\includegraphics[width=1.0\textwidth]{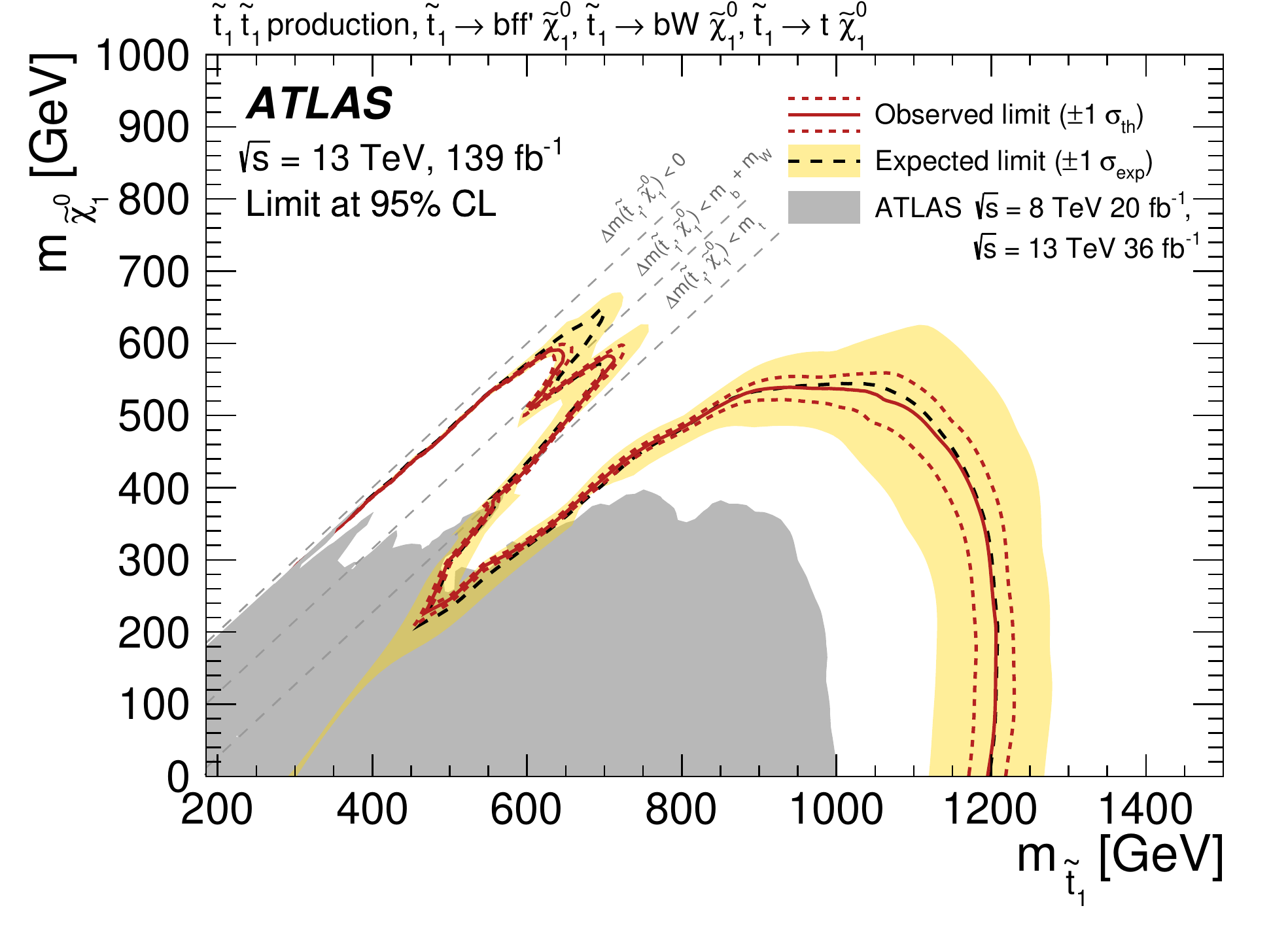}
\caption{Expected and observed 95\% CL excluded regions in the plane of $m_{\ninoone}$ and $m_{\stopone}$
for direct stop pair production assuming either a \topLSP, \threeBody\ or \fourBody\ decay with a branching ratio of 100\%.
The excluded regions from previous publications~\cite{SUSY-2016-16,SUSY-2016-15,SUSY-2016-17,EXOT-2016-27} are shown by the shaded area and include additional topologies. The diagonal dashed lines indicate the kinematical border of the stop decay modes.}
\label{fig:stop_exclusion}
\end{figure}

\begin{figure}
\includegraphics[width=1.0\textwidth]{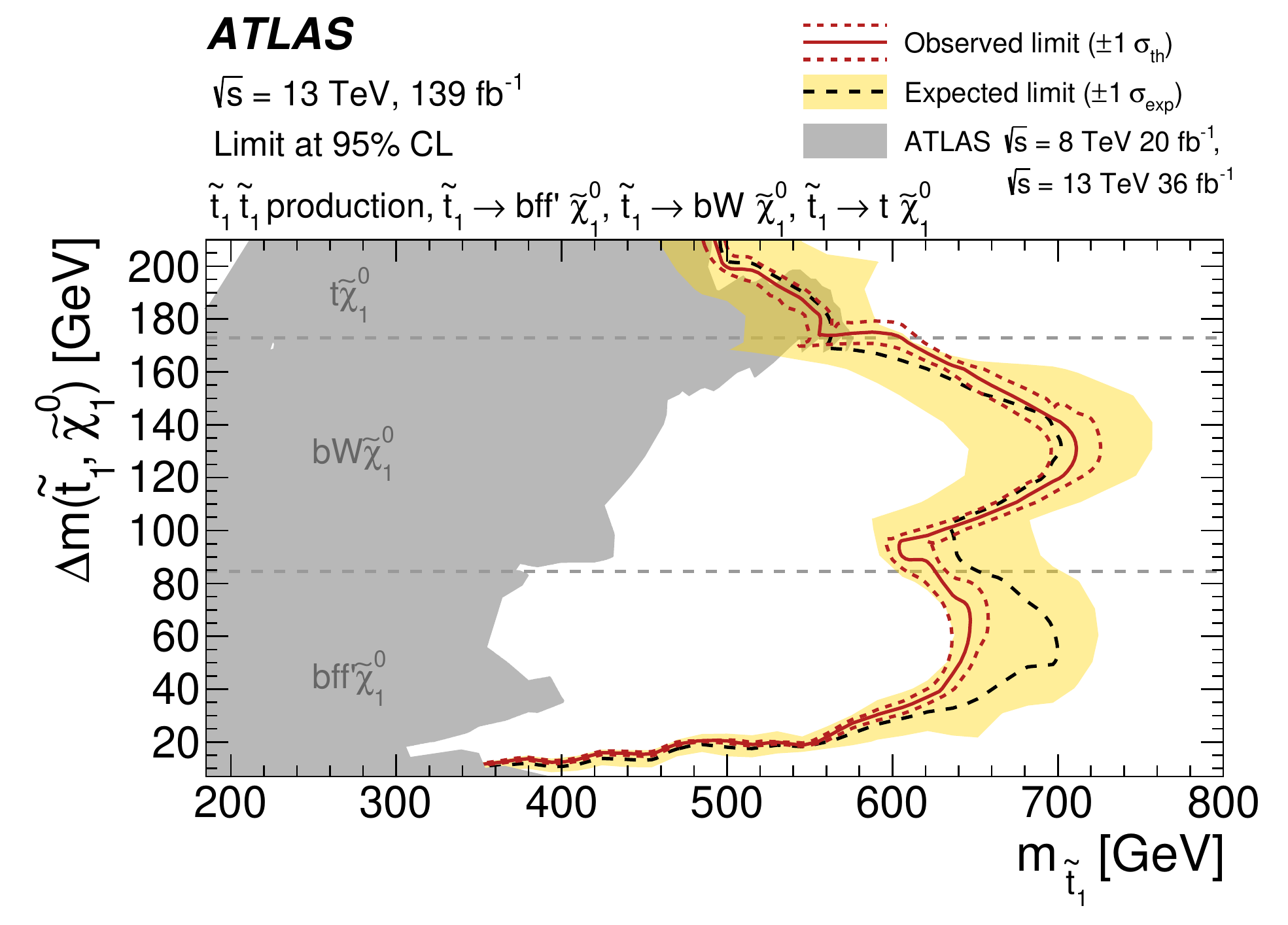}
\caption{Expected and observed  95\% CL excluded regions in the plane of $\Delta(m_{\stopone},m_{\ninoone})$ and $m_{\stopone}$
for direct stop pair production assuming either a \topLSP, \threeBody\ or \fourBody\ decay with a branching ratio of 100\%. The excluded regions
from previous publications~\cite{SUSY-2016-16,SUSY-2016-15,SUSY-2016-17,EXOT-2016-27} are shown by the shaded area and include additional topologies.
The horizontal dashed lines indicate the kinematical border of the stop decay modes.}
\label{fig:stop_exclusion_deltaM}
\end{figure}

Figures~\ref{fig:stop_exclusion} and~\ref{fig:stop_exclusion_deltaM} show the expected and observed exclusion
contours as a function of the stop mass, the neutralino mass and the mass difference between the stop and the
neutralino, for the \topLSP, \threeBody\ and \fourBody\ scenarios.
In models with a massless neutralino, stop masses up to 1200~\GeV are excluded at 95\% CL.
In the diagonal region, where the mass difference between the stop and the neutralino coincides with the mass of the top quark, stop masses up to 600 GeV are excluded, which covers the previously unexcluded diagonal region between 210 GeV and 250 GeV in stop mass.
In the three-body (four-body) region, stop masses up to 710~\GeV (640~\GeV) are excluded for a neutralino mass of approximately 580~\GeV.
The small excess observed in \tNhigh\ does not appear because
the exclusion limits are obtained from the shape-fit in the \tNmed\ signal region (Table~\ref{tab:tnbulk_sr}). The
\perpmet\ requirement applied in \tNmed\ but not in \tNhigh\ removes most of the excess. The shape-fit
is designed to have better expected sensitivity than the single-bin SRs over the whole \topLSP\ parameter space.
 
\begin{figure}
\subfigure[]{\includegraphics[width=0.49\textwidth]{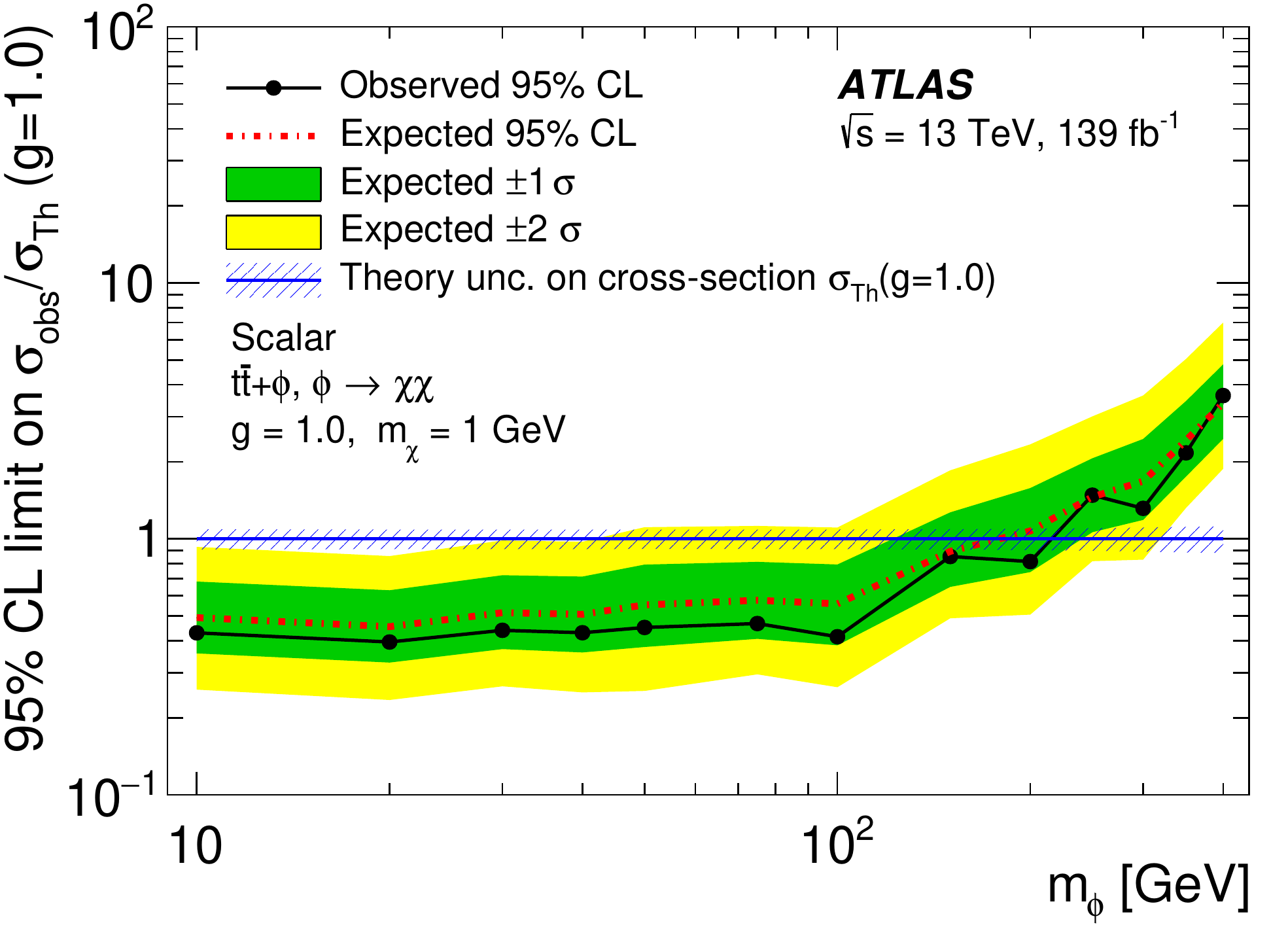}}
\subfigure[]{\includegraphics[width=0.49\textwidth]{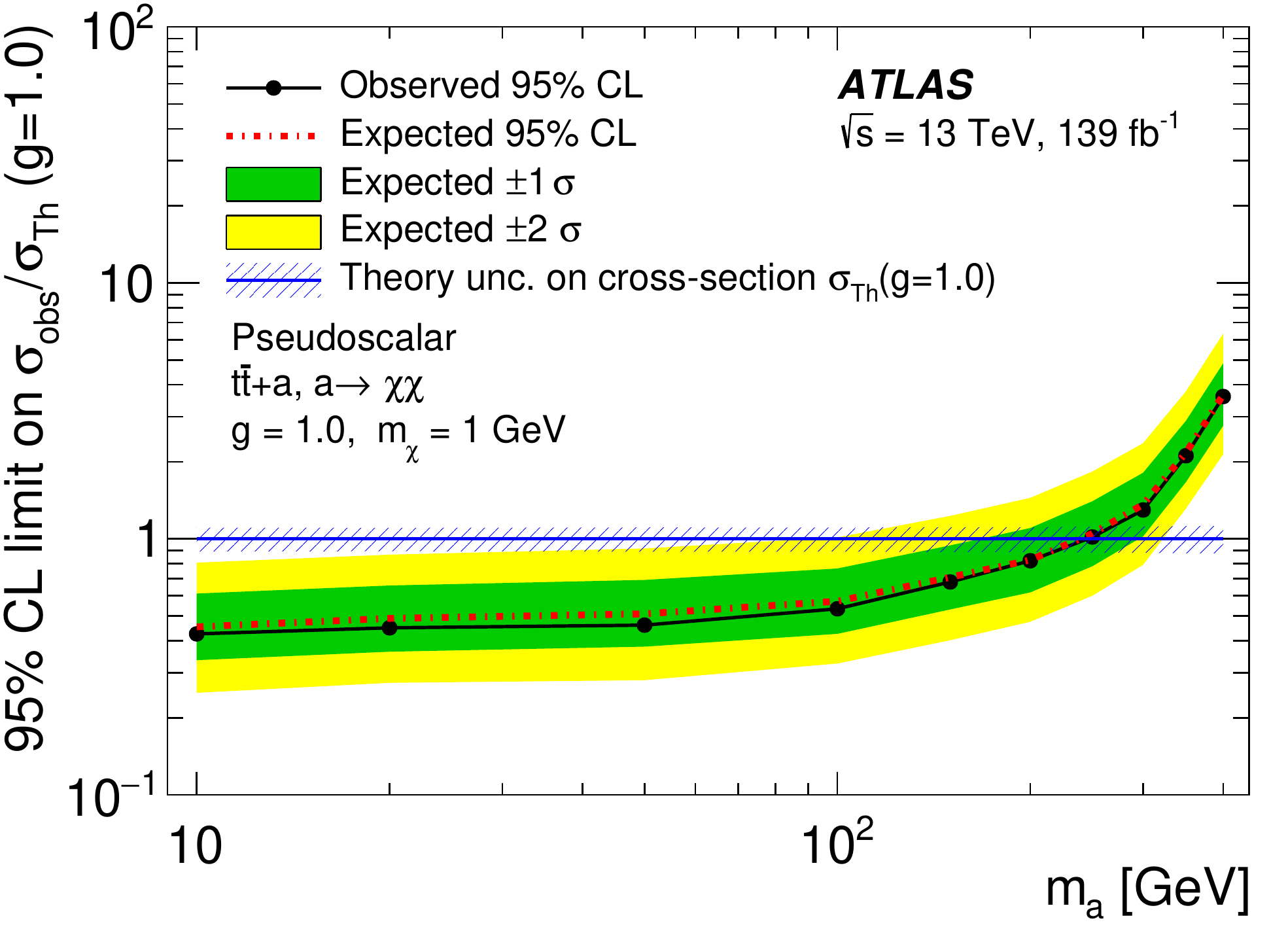}}
\caption{Upper limit on the ratio of the production cross-section for the spin-0 mediator model
to the theoretical cross-section under the hypothesis of (a) a scalar or (b) a pseudoscalar mediator. The limit is
shown as a function of the mediator mass for a fixed mass of the DM candidate of 1~\GeV. The coupling of the mediator to SM
and DM particles is assumed to be $g = 1$.}
\label{fig:DM_exclusion}
\end{figure}

Figure~\ref{fig:DM_exclusion} shows the upper limit on the ratio of the production cross-section for
the spin-0 mediator model to the theoretical cross-section.
Limits are shown under the hypothesis of a scalar or pseudoscalar mediator for a
fixed DM candidate mass.
Scalar and pseudoscalar mediator masses up to approximately 200~\GeV are excluded at 95\% CL,
assuming a 1~\GeV dark-matter particle mass and a common
coupling of $g = 1$ to SM and dark-matter particles.
With the common coupling reduced to $g = 0.8$, mediator masses up to approximately 100~\GeV are excluded.
Models with a mediator mass of 10~\GeV and a dark-matter particle mass of
1~\GeV are excluded down to a coupling of approximately $g= 0.7$.
 
% End of text imported from the .//texfiles/results.tex input file
 
\FloatBarrier
 
\clearpage
% The next lines are included from the .//texfiles/conclusion.tex input file
\section{Conclusion}
\label{sec:conclusion}
 
This paper presents searches for direct stop pair production covering various regions of SUSY phase space
and searches for a spin-0 mediator decaying into pair-produced dark-matter particles. The searches use
the final state with one isolated lepton, jets, and \met.
 
The analysis uses \ourLumi\ of $pp$ collision data collected with the ATLAS detector at the LHC at a
centre-of-mass energy of $\sqrt{s}=13$~\TeV. The largest excess over the background-only hypothesis is
$1.9 \sigma$ in the \tNhigh\, signal region. As no significant deviation from the Standard Model
expectation is observed, exclusion limits at 95\% confidence level are derived for the models considered.
Stops are excluded up to 1200~\GeV (710~\GeV) in the two-body (three-body) decay scenario, extended from
about 1000~\GeV (400-600~\GeV) in previous results.
The introduction of ML techniques contributes to the significantly improved sensitivity in the challenging
three-body decay scenario.
In the four-body scenario, the exclusion of stops is
extended from about 400~\GeV in earlier results to up to 640 GeV for a stop–neutralino mass difference of
60~\GeV. The introduction of the soft $b$-tagging algorithm contributes to the significantly improved sensitivity
at small mass differences between the stop and the lightest neutralino.
Scalar and pseudoscalar dark-matter mediators are excluded up to 200~\GeV for a common coupling
of $g = 1$ to Standard Model and dark-matter particles, from 100 GeV in earlier results. The introduction
of a shape-fit in the DM signal region contributes to this improvement.
% End of text imported from the .//texfiles/conclusion.tex input file

\section*{Acknowledgements}
 
% The next lines are included from the .//Acknowledgements/Acknowledgements.tex input file

We thank CERN for the very successful operation of the LHC, as well as the
support staff from our institutions without whom ATLAS could not be
operated efficiently.
 
We acknowledge the support of ANPCyT, Argentina; YerPhI, Armenia; ARC, Australia; BMWFW and FWF, Austria; ANAS, Azerbaijan; SSTC, Belarus; CNPq and FAPESP, Brazil; NSERC, NRC and CFI, Canada; CERN; ANID, Chile; CAS, MOST and NSFC, China; COLCIENCIAS, Colombia; MSMT CR, MPO CR and VSC CR, Czech Republic; DNRF and DNSRC, Denmark; IN2P3-CNRS and CEA-DRF/IRFU, France; SRNSFG, Georgia; BMBF, HGF and MPG, Germany; GSRT, Greece; RGC and Hong Kong SAR, China; ISF and Benoziyo Center, Israel; INFN, Italy; MEXT and JSPS, Japan; CNRST, Morocco; NWO, Netherlands; RCN, Norway; MNiSW and NCN, Poland; FCT, Portugal; MNE/IFA, Romania; MES of Russia and NRC KI, Russia Federation; JINR; MESTD, Serbia; MSSR, Slovakia; ARRS and MIZ\v{S}, Slovenia; DST/NRF, South Africa; MICINN, Spain; SRC and Wallenberg Foundation, Sweden; SERI, SNSF and Cantons of Bern and Geneva, Switzerland; MOST, Taiwan; TAEK, Turkey; STFC, United Kingdom; DOE and NSF, United States of America. In addition, individual groups and members have received support from BCKDF, CANARIE, Compute Canada and CRC, Canada; ERC, ERDF, Horizon 2020, Marie Sk{\l}odowska-Curie Actions and COST, European Union; Investissements d'Avenir Labex, Investissements d'Avenir Idex and ANR, France; DFG and AvH Foundation, Germany; Herakleitos, Thales and Aristeia programmes co-financed by EU-ESF and the Greek NSRF, Greece; BSF-NSF and GIF, Israel; La Caixa Banking Foundation, CERCA Programme Generalitat de Catalunya and PROMETEO and GenT Programmes Generalitat Valenciana, Spain; G\"{o}ran Gustafssons Stiftelse, Sweden; The Royal Society and Leverhulme Trust, United Kingdom.
 
The crucial computing support from all WLCG partners is acknowledged gratefully, in particular from CERN, the ATLAS Tier-1 facilities at TRIUMF (Canada), NDGF (Denmark, Norway, Sweden), CC-IN2P3 (France), KIT/GridKA (Germany), INFN-CNAF (Italy), NL-T1 (Netherlands), PIC (Spain), ASGC (Taiwan), RAL (UK) and BNL (USA), the Tier-2 facilities worldwide and large non-WLCG resource providers. Major contributors of computing resources are listed in Ref.~\cite{ATL-SOFT-PUB-2020-001}.
 
% End of text imported from the .//Acknowledgements/Acknowledgements.tex input file

\printbibliography

\clearpage
 
% ATLAS Collaboration author list
% Reference date of SUSY-2018-07 is 2020-01-29
% Author list last updated on date 19-FEB-21
% Data extracted on 19-Feb-2021 for paper reference SUSY-2018-07
% at 1:14pm
 
\begin{flushleft}
\hypersetup{urlcolor=black}
{\Large The ATLAS Collaboration}

\bigskip

\AtlasOrcid[0000-0002-6665-4934]{G.~Aad}$^\textrm{\scriptsize 102}$,    
\AtlasOrcid[0000-0002-5888-2734]{B.~Abbott}$^\textrm{\scriptsize 128}$,    
\AtlasOrcid[0000-0002-7248-3203]{D.C.~Abbott}$^\textrm{\scriptsize 103}$,    
\AtlasOrcid[0000-0002-2788-3822]{A.~Abed~Abud}$^\textrm{\scriptsize 36}$,    
\AtlasOrcid[0000-0002-1002-1652]{K.~Abeling}$^\textrm{\scriptsize 53}$,    
\AtlasOrcid[0000-0002-2987-4006]{D.K.~Abhayasinghe}$^\textrm{\scriptsize 94}$,    
\AtlasOrcid[0000-0002-8496-9294]{S.H.~Abidi}$^\textrm{\scriptsize 167}$,    
\AtlasOrcid[0000-0002-8279-9324]{O.S.~AbouZeid}$^\textrm{\scriptsize 40}$,    
\AtlasOrcid{N.L.~Abraham}$^\textrm{\scriptsize 156}$,    
\AtlasOrcid[0000-0001-5329-6640]{H.~Abramowicz}$^\textrm{\scriptsize 161}$
% ...
\bigskip
\\

% ...
$^{36}$CERN, Geneva; Switzerland.\\
% ...
$^{40}$Niels Bohr Institute, University of Copenhagen, Copenhagen; Denmark.\\
% ...
$^{53}$II. Physikalisches Institut, Georg-August-Universit\"{a}t G\"ottingen, G\"ottingen; Germany.\\
% ...
$^{94}$Department of Physics, Royal Holloway University of London, Egham; United Kingdom.\\
% ...
$^{102}$CPPM, Aix-Marseille Universit\'e, CNRS/IN2P3, Marseille; France.\\
$^{103}$Department of Physics, University of Massachusetts, Amherst MA; United States of America.\\
% ...
$^{128}$Homer L. Dodge Department of Physics and Astronomy, University of Oklahoma, Norman OK; United States of America.\\
% ...
$^{156}$Department of Physics and Astronomy, University of Sussex, Brighton; United Kingdom.\\
% ...
$^{161}$Raymond and Beverly Sackler School of Physics and Astronomy, Tel Aviv University, Tel Aviv; Israel.\\
% ...
$^{167}$Department of Physics, University of Toronto, Toronto ON; Canada.\\
% ...

\end{flushleft}

% Created with Glance <Atlas.Glance@cern.ch>
 
\end{document}